      [1999/12/01 v1.4c Il Nuovo Cimento]
\documentclass{cimento}
\usepackage{graphicx,epsfig}
\def\sus#1{\sin^{#1}i}
\def\jw2{w_{.2}}
\def\ecn#1{(1-e^{2})^{#1}}
\def\rtr#1{\left(\frac{R}{a}\right)^{#1}}
\def\jn2{\dot\Omega_{.2}}
\def\jo2{\dot\omega_{.2}}
\def\ekn#1#2{(1+\frac{#1}{#2}e^{2})}
\def\djn{\dot\Omega_{.2n}\equiv\derp{{\dot\Omega_{\rm class}}}{(J_{2n})}}
\def\djo{\dot\omega_{.2n}\equiv\derp{{\dot\omega_{\rm class}}}{(J_{2n})}}
\def\jn#1{\dot\Omega_{.#1}}
\def\jo#1{\dot\omega_{.#1}}
\def\eqiaz{\begin{eqnarray*}}
\def\eqfaz{\end{eqnarray*}}
\def\eqia{\begin{eqnarray}}
\def\eqfa{\end{eqnarray}}
\def\btab{\begin{tabular}}
\def\etab{\end{tabular}}
\def\bar{\begin{array}}
\def\ear{\end{array}}

\def\sat{satellite}
\def\nvd{4\b-\ga-3}
\def\ppn{\rp{2+2\ga-\b}{3}}

\def\nd{node}
\def\pg{perigee}

\def\grc{gravitomagnetic}

\def\btab{\begin{tabular}}
\def\etab{\end{tabular}}
\def\rfr#1{eq. (\ref{#1})}
\def\rfrs#1#2{eqs. (\ref{#1})-(\ref{#2})}
\def\Rfr#1{Eq. (\ref{#1})}
\def\Rfrs#1#2{Eqs. (\ref{#1})-(\ref{#2})}
\def\dfa{\derp{\mtc{R}}{a}}
\def\dfm{\derp{\mtc{R}}{\mtc{M}}}
\def\dfog{\derp{\mtc{R}}{\og}}
\def\dfi{\derp{\mtc{R}}{i}}
\def\dfe{\derp{\mtc{R}}{e}}
\def\dfo{\derp{\mtc{R}}{\O}}
\def\cu{\rp{2}{na}}
\def\cd{\rp{1-e^2}{na^{2} e}}
\def\ctr{\rp{(1-e^2)^{1/2}}{na^{2} e}}
\def\cq{\rp{1}{na^{2}(1-e^2)^{1/2}\si}}

\def\bb{\bibitem}
\def\bar{\begin{array}}
\def\ear{\end{array}}
\def\eqi{\begin{equation}}
\def\eqf{\end{equation}}
\def\mc#1#2{\left[\matrix{#1 \cr #2\cr}\right]}

\def\mtc#1{\mathcal{#1}}

\def\b{\beta}
\def\ga{\gamma}
\def\d{\delta}
\def\ve{\varepsilon}

\def\z{\zeta}
\def\et{\eta}
\def\th{\theta}
\def\vth{\vartheta}

\def\k{\kappa}
\def\l{\lambda}
\def\m{\mu}
\def\n{\nu}
\def\x{\xi}

\def\p{\pi}
\def\r{\rho}

\def\s{\sigma}

\def\t{\tau}

\def\f{\phi}

\def\ps{\psi}
\def\og{\omega}
\def\G{\Gamma}
\def\D{\Delta}

\def\S{\Sigma}

\def\O{\Omega}
\def\cO{\cos{\O}}
\def\sO{\sin{\O}}
\def\cog{\cos{\og}}
\def\sog{\sin{\og}}
\def\ci{\cos{i}}
\def\si{\sin{i}}

\def\vass#1{\left\vert\ #1 \right\vert}
\def\rp#1#2{{#1\over#2}}
\def\ol#1{\overline{#1}}
\def\km{\left(\matrix{\cO\cog-\sO\ci\sog & -\cO\sog-\sO\ci\cog & \sO\si
\cr \sO\cog+\cO\ci\sog & -\sO\sog+\cO\ci\cog & -\cO\si
\cr \si\sog & \si\cog & \ci \cr}\right)}
\def\gr{General Relativity}
\def\derp#1#2{\rp{\partial{#1}}{\partial{#2}}}
\def\dert#1#2{\frac{{{d}}{#1}}{{{d}}{#2}}}

\def\somma#1#2{\sum_{#1}^{#2}}
\def\ct#1{\cite{#1}}
\def\lb#1{\label{#1}}

\def\orot{\og_{\oplus}}

\def\cd{\cos{\d}}

\def\gr{General Relativity}
\def\derp#1#2{\rp{\partial{#1}}{\partial{#2}}}

\def\somma#1#2{\sum_{#1}^{#2}}
\def\ct#1{\cite{#1}}
\def\lb#1{\label{#1}}

\def\orot{\og_{\oplus}}

\def\cd{\cos{\d}}

\def\se{\sin{\ve}}
\def\ci{\cos{i}}
\def\si{\sin{i}}

\def\sf{\sin{f}}

\def\med#1{\left\langle{#1}\right\rangle_{2\p}}

\def\se{\sin{\ve}}
\def\ci{\cos{i}}
\def\si{\sin{i}}

\def\sf{\sin{f}}

\def\sa{semimajor axis}

\def\far{{Farinella}}

\def\dman2{-\sqrt{1-e^2}\ (\dert{\og}{t}+\ci\dert{\O}{t})}

\def\lg{{{\rm LAGEOS}}}
\def\lgg{{\rm LAGEOS}\ {\rm II}}
\def\jgr{{J. Geophys. Res.}}

\def\lt{Lense-Thirring\ }

\def\nd{node}
\def\pg{perigee}
\def\btab{\begin{tabular}}
\def\etab{\end{tabular}}
\def\dfa{\derp{\mtc{R}}{a}}
\def\dfm{\derp{\mtc{R}}{\mtc{M}}}
\def\dfog{\derp{\mtc{R}}{\og}}
\def\dfi{\derp{\mtc{R}}{i}}
\def\dfe{\derp{\mtc{R}}{e}}
\def\dfo{\derp{\mtc{R}}{\O}}
\def\cu{\rp{2}{na}}
\def\cd{\rp{1-e^2}{na^{2} e}}
\def\ctr{\rp{(1-e^2)^{1/2}}{na^{2} e}}
\def\cq{\rp{1}{na^{2}(1-e^2)^{1/2}\si}}
\def\kep{Keplerian orbital elements}
\def\btab{\begin{tabular}}
\def\etab{\end{tabular}}
\def\bar{\begin{array}}
\def\ear{\end{array}}
\def\dert#1#2{\rp{{d}{#1}}{{d}{#2}}}

\def\grl{general relativistic}

\def\leti{Lense-Thirring}
\def\grc{gravitomagnetic}

\def\se{systematic error}
\def\er{error}
\def\zh{even zonal harmonics}

\def\gp{geopotential}

\def\lg{{\rm LAGEOS}}
\def\lgg{{\rm LAGEOS} II}

\def\lb#1{\label{#1}}
\def\pc{precession}
\def\nd{node}
\def\pg{perigee}
\def\nl{nodal}

\def\sa{semimajor axis}

\def\ic{inclination}

\def\rt{Earth}

\def\dt#1{\dot{#1}}
\def\mlt{{\rm \mu_{LT}}}
\def\dmu{\frac{\delta\mlt}{\mlt}}
\def\zs{{\rm zonals}}
\def\st{satellite}

\def\rp#1#2{{#1\over#2}}
\def\msy{{\rm mas/y}}

\title{Recent Developments in Testing General Relativity with Satellite Laser Ranging}
\author{L.~Iorio\from{ins:x}\thanks{E--mail: Lorenzo.Iorio@ba.infn.it}}
\instlist{\inst{ins:x} Dipartimento di Fisica
dell'Universit${\grave{a}}$ di Bari, Via Amendola 173, 70126,
Bari, Italy.
  }
\begin{document}

\maketitle


\section{Introduction}
The last decade can be characterized by an impressive diversity of
techniques monitoring the artificial and natural satellite
dynamics, as well as the Earth rotation: improved laser
technology, renewed Doppler techniques, satellite radar altimetry,
massive usage of the Global Positioning System (GPS), etc. Each of
these techniques is optimally tailored to a specific type of
application or scientific problem. For example, it appears that
laser tracking (SLR: see on the WEB: {\bf
http://ilrs.gsfc.nasa.gov}) of passive geodynamics satellites
(\lg, \lgg, Starlette, Stella, Ajisai, Etalon I and II) over
relatively long time intervals provides an excellent method for
determining the long-term variations of the geopotential
\ct{ref:kau} (including tidal effects) and many small
non-gravitational phenomena \ct{ref:mil}. It turns out that the
precision reached in the latest years by such technique in
measuring the position of the passive laser-ranged geodetic
satellites \lg\ and \lgg\ amounts to 1 cm or better\footnote{In
August 2001 the single-shot accuracy in tracking LAGEOS at Matera
amounts to 5 mm}. See, e.g., {\bf
http://earth.agu.org/revgeophys/marsha01/node1.html}

Such astonishing levels of accuracy have disclosed an unexpectedly
wide field of research in space geodesy, geophysics and
fundamental physics. E.g., now it is possible to plan
satellite-based experiments devoted to the experimental control
even of some tiny post-Newtonian features of the
Earth'gravitational field predicted by Einstein' s General
Relativity \ct{ref:lamz}. They are usually expressed in terms of
certain solve-for least squares fits' parameters. In evaluating
the precision of the results of these experiments it must be
considered that, in general, the main error does not consist of
the standard statistical error of the fits but of the various
systematic errors. They are induced by a complete set of other
physical effects acting upon the satellites to be employed
\ct{ref:mon}. Such perturbations are often quite larger than the
relativistic effects investigated and induce systematic errors to
be correctly evaluated and assessed. Many of these aliasing
effects are traditionally investigated  by geophysics and space
geodesy, so that such experiments are multidisciplinary efforts
which cover many scientific areas until now separated
\ct{ref:sof,ref:gra}. Indeed, the expansion of the SLR network,
together with improved system accuracies, has enabled the laser
data to contribute directly to improving orbit force models. It
allows one to calculate more accurately, among other things, the
error budget of many space-based general relativistic experiments:
this is the main topic of the present work. It does not treat the
measurement modelling errors specifically related to the
laser-ranging technique.
\subsection{The Lense-Thirring drag of inertial frames}
One of the most interesting topic in General Relativity is the
structure of the spacetime around a spherically symmetric rotating
mass-energy distribution. Indeed, in the slow-motion and
weak-field approximation, it exhibits the characteristic feature
of exerting a non-central force on a test particle due to the
total angular momentum of the central object, contrary to
Newtonian mechanics in which the gravitational action of a body is
caused only by its mass, regardless to its rotational motion.
Because of the formal analogies with the electrodynamics, this
effect, deduced from the equations of Einstein for the first time
by Lense and Thirring in 1918 \ct{ref:len}, is also defined as
gravitomagnetism. A comprehensive review of its properties can be
found in \ct{ref:ciuwhe,ref:masetal,ref:mas}.


In the past few years we have seen increasing efforts, both from a
theoretical and an experimental point of view, devoted to the
measurement of the Lense-Thirring effect in the weak gravitational
field of the Earth by means of artificial satellites. At present,
there are two main proposals which point towards the
implementation of this goal: the Gravity Probe-B mission
\ct{ref:eve}, and the approach proposed in \ct{ref:ciu86} which
consists in using the actually orbiting LAGEOS laser-ranged
satellite and launching another satellite of LAGEOS type, the
LARES, with the same orbital parameters of LAGEOS except for the
inclination, which should be supplementary with respect to it, the
eccentricity which should be one order of magnitude larger, and
the area-to-mass ratio which should be smaller so to reduce the
impact of the non-gravitational orbital perturbations (see
Appendix A for the LAGEOS type satellites' data). The observable
would be a secular linear trend built up with the sum of the
residuals of the longitudes of the ascending nodes $\Omega$ of
LARES and LAGEOS (see section 2 and fig. \ref{kpl} for an
explanation of the Keplerian orbital elements in terms of which
the orbit of a satellite can be parameterized). The proposed
orbital geometry would allow one to minimize the impact of the
aliasing trend due to the mismodelling in the classical nodal
precessions generated by the oblateness of the Earth, which would
represent the main systematic error.

The GP-B mission is aimed to the detection of the motion of a
spinning particle. At present, both the GP-B and the LARES have
not yet been launched: however, while the GP-B is scheduled to fly
in fall 2002, the fate of LARES project is still uncertain.

Recently, Ciufolini, in ref.\ct{ref:ciu96}, has put forward an
alternative strategy based on the utilization of the already
existing LAGEOS and LAGEOS\ II
which allowed the detection of the Lense-Thirring drag at a
precision level of the order of $20\%$ \ct{ref:ciu98,ref:ciu00}.
While the GP-B mission is focused on the gravitomagnetic dragging
of the spin of a freely falling body, in the LAGEOS experiment it
is the entire orbit of the satellite which is considered to
undergo the secular Lense-Thirring precession. More exactly, among
the \kep\ \ct{ref:ste,ref:all}, the node $\Omega$ and the perigee
$\omega$ are affected by the gravitomagnetic perturbation. For
\lg\ and \lgg\ it amounts to\begin{eqnarray} \dot
\O_{{\rm LT}}^{{\rm LAGEOS}}&\simeq& 31 \ \textrm{mas/y},\\
\dot
\O_{{\rm LT}}^{{\rm LAGEOS\ II}}&\simeq& 31.5 \ \textrm{mas/y}, \\
\dot \og_{{\rm LT}}^{{\rm LAGEOS}}&\simeq& 31.6\
\textrm{mas/y},\\
 \dot \og_{{\rm LT}}^{{\rm LAGEOS\ II}}&\simeq&
-57\ \textrm{mas/y},
\end{eqnarray} where mas/y stands for
milliarcseconds per year.

In section 2 a brief derivation of such results is presented. It
follows ordinary linear satellite perturbation theory. Indeed, in
the slow-motion and weak-field approximation, the \grc\ potential
is treated as a classical disturbing term with respect to the
Newtonian gravitoelectric monopole.

The observable quantity proposed in \ct{ref:ciu96} is a suitable
combination of the orbital residuals of the nodes of \lg\ and
\lgg\ and the \pg\ of \lgg \eqi\dot y\equiv\d\dot\O^{{\rm I}}_{\rm
exp}+c_1 \times\d\dot\O^{{\rm II}}_{\rm
exp}+c_2\times\d\dot\og^{{\rm II}}_{\rm exp}\simeq\m_{\rm LT}
\times 60.2.\lb{unolt}\eqf In it $c_1=0.295$, $c_2=-0.35$,
$\m_{\rm LT}$ is the scaling, solve-for parameter, equal to 1 in
General Relativity and 0 in classical mechanics, to be determined
and $\d\O^{{\rm I}}_{\rm exp},\ \d\O^{{\rm II}}_{\rm exp},\
\d\og^{{\rm II}}_{\rm exp}$ are the orbital residuals, in mas,
calculated with the aid of some orbit determination software like
UTOPIA or GEODYN, of the nodes of LAGEOS and LAGEOS\ II and the
perigee of LAGEOS\ II. The residuals account for any unmodelled or
mismodelled physical phenomena acting on the observable analyzed.
By dealing with the gravitomagnetism as an unmodelled physical
effect, General Relativity predicts a linear trend with a slope of
60.2 mas/y\eqi \dot y_{{\rm LT}}\equiv (31\
\textrm{mas/y})+c_1\times (31.5\ \textrm{mas/y})+c_2\times(-57\
\textrm{mas/y})\simeq 60.2\ \textrm{mas/y}.\lb{duelt}\eqf The
coefficients $c_1$ and $c_2$ have been obtained by solving for the
three unknowns $\delta J_2$, $\delta J_4$ and $\mu_{\rm LT}$ a
nonhomogeneous algebraic linear system of three equations
expressing the observed mismodelled classical precessions of the
nodes of LAGEOS and LAGEOS II and the perigee of LAGEOS II (see
ref.\ct{ref:ciu96}). They depend on the orbital parameters of
LAGEOS and LAGEOS II  and are built up so to cancel out all the
static and dynamical contributions of degree $l=2,4$ and order
$m=0$ of the Earth' s gravitational field. This cancellation is
required to reduce especially the impact of the first two even
($l=2,4$) zonal ($m=0$) harmonics of the static geopotential.
Indeed, at the present level of accuracy of the terrestrial
gravitational field \ct{ref:lem}, the mismodelled parts of the
classical orbital precessions of the node and the perigee, caused
by the Earth's oblateness and parameterized with the first two
even zonal coefficients $J_2$ and $J_4$ \ct{ref:kau}, for a single
satellite amount to a significative part of the corresponding \lt\
effect itself on the considered orbital element.
The combined residuals of \rfr{unolt} are affected by a complete
set of gravitational and non-gravitational perturbations; among
the former ones those generated by the solid Earth and ocean tides
\ct{ref:mel,ref:car00} play an important role.

In section 3 the amplitudes and the periods of the most relevant
tidal perturbations acting upon the nodes of \lg\ and \lgg\ and
the perigee of LAGEOS II have been worked out \ct{ref:ior01}. This
analysis covers both the solid $l=2$ and the ocean $l=2,3,4$ part
of the tidal spectrum, includes the first order contributions, in
the sense of the Kaula' s theory of perturbations \ct{ref:kau},
and also a calculation of the mismodelling in the first order
perturbative amplitudes compared to the \lt\ drag over 4 years.

Section 4 is devoted to the study of the impact of the solid Earth
and ocean tidal perturbations on \rfr{unolt} with particular care
to the systematic errors induced by them on the measurement of
$\m_{{\rm LT}}$ \ct{ref:iorpav,ref:pavior}. The left hand side of
\rfr{unolt} has been calculated by adopting the nominal values of
the perturbative amplitudes worked out in section 3 in order to
test if the first two even degree zonal tidal lines do not really
affect the combined residuals. This is particularly important for
the 18.6-year tidal constituent whose period is that of the
longitude of the ascending node of the Moon, i.e. 18.6 years.
Indeed, over observational time spans of few years it could
resemble a trend as well and its mismodelled part, accounted for
by the residuals combination, could corrupt the measurement of
$\mu_{\rm LT}$. The results obtained in section 3 have been
employed also in order to obtain a simulated residual curve on
which several tests have been performed for the various tesseral
($m=1$) and sectorial ($m=2$) tidal constituents, not cancelled
out by the combined residuals, with the aim of assessing their
contribution on the systematic uncertainty in the measurement of
$\m_{\rm LT}$.

\subsection{The gravitoelectric pericenter shift}
In section 5 the experience gained in the Lense-Thirring LAGEOS
experiment is used to propose a possible experiment for measuring
the general relativistic pericenter shift of a test body due to
the Schwarzschild's gravitoelectric part of the terrestrial
gravitational field by means of the SLR data to \lg\ and \lgg. The
approach is similar to that of the current Lense-Thirring LAGEOS
experiment and it is based on another suitable combination of
orbital residuals of the nodes of LAGEOS and LAGEOS II and the
perigee of LAGEOS II. The possibility of adopting different
observables is examined as well. A discussion on the obtainable
values for the Parameterized Post Newtonian parameters $\beta$ and
$\gamma$ is also included.
\subsection{The LARES mission}
In section 6 we reanalyze the original LARES mission from the
point of view of the sensitivity of the related observable to the
departures of the LARES orbital parameters from their nominal
values due to the orbital injection errors. Moreover, we propose
an alternative combination of orbital residuals involving also the
LAGEOS II which would yield to a smaller and more stable value of
the error due to the mismodelling in the even zonal harmonics of
the geopotential which turns out to be the main source of
systematic error.
\subsection{Conclusions}
Section 7 is devoted to the conclusions and to the recommendations
for future work.
\section{The  Lense-Thirring effect on the orbit of a test body}
\subsection{Introduction}

In this section  an alternative strategy is presented in order to
derive the \lt\ effect \ct{ref:iorcim}. It reveals itself useful
especially in the prediction of the behavior of all the Keplerian
orbital elements of the test body in the gravitational field of a
rotating source. Our calculations are based on the Lagrangian
planetary equations \ct{ref:kau,ref:berfar,ref:vin} and a
non-central hamiltonian term whose existence, if from one hand can
be rigorously deduced, from the other hand can be intuitively
guessed by analogy from the corresponding term in electrodynamics
for the Lagrangian of a charged particle acted upon by  electric
and magnetic fields.

For a perfectly spherical source the well known \lt\ rate
equations for the secular precessions of the node and the perigee
are reobtained.
\subsection{The gravitomagnetic potential}
In general, it can be proved \ct{ref:masetal} that the general
relativistic equations of motion of a test particle of
 mass $m$ freely falling in a stationary
gravitational field, in the weak-field and slow-motion
approximation, are  given by \eqi m\rp{{ d}^2 {\bf r}}{{ d}t^2}\simeq
m\left({\bf E}_{g}+\frac{\textbf{v}}{c}\times {\bf B}_{g}\right).
\label{cif}\eqf In \rfr{cif} {\bf E}$_{g}$ and {\bf B}$_{g}$ are
the gravitoelectric and gravitomagnetic fields, respectively. If a
perfectly spherically symmetric rotating body is assumed as
gravitational source, in eq. (\ref{cif}) ${\bf E}_{g}=-G{M}{\bf
i}_{r}/{r^2}+{\mathcal{O}}(c^{-2})$ is the Newtonian gravitational
 field of a spherical body, with $M$ its mass and $G$ the Newtonian
gravitational constant\footnote{The terms of order
${\mathcal{O}(c^{-2})}$ are the post-Newtonian corrections due to
the static Schwarzschild part of the metric.}, while ${\bf B}_{g}$
is given by \ct{ref:masetal} \eqi {\bf B}_{g}= \nabla \times {\bf
A}_g\simeq 2\rp{G}{c} \left[\rp{ {\bf J}-3({\bf J}\cdot {{\bf i}_
{r}}){{\bf i}_{r}}}{r^3}\right]\lb{acca}, \eqf in which \eqi {\bf
A}_{g}({\bf r})\simeq -2\rp{G}{c}\rp{{\bf J} \times{\bf
r}}{r^3}.\label{acchina}\eqf In \rfr{acchina} ${\bf J}$ is the
angular momentum of the central body. The field ${\bf A}_{g}
\equiv (h_{01},h_{02},h_{03})$, named gravitomagnetic potential,
is due to the off-diagonal components of the spacetime metric
tensor \eqi g_{\m\n}=\et_{\m\n}+h_{\m\n},\ \m,\n=0,1,2,3,\eqf
where $\et_{\m\n}$ is the Minkowski metric tensor.  In obtaining
\rfr{acchina} a non-rotating
 reference frame $K\{x,\ y,\ z\}$
with the $z$ axis directed along the intrinsic angular momentum of the
central spinning body {\bf J} and the $\{x,\ y\}$
plane coinciding with its  equatorial plane  has been used. The origin is
located at the center of mass of the central body.
\Rfr{cif} holds if ${\bf A}_g$ is not time-varying.

 The gravitomagnetic potential
generates a non-central gravitational contribute due uniquely to
the angular momentum of the gravitational source
 that the Newtonian mechanics does not predict, though the conditions of
validity of eq. (\ref{cif}) are the same for which the latter
holds as well.\footnote{Incidentally, it may be interesting to
notice that eq. (\ref{cif}) are consistent with the fundamental
Einstein assumption \ct{ref:ein} that a non-accelerated reference
frame with a gravitational field is equivalent, within certain
limits, to an accelerated one without any gravitational field.
Indeed, if a reference frame solidal with the rotating body is
assumed, the equations of motion for a test particle are formally
identical to eq. (\ref{cif}): the gravitomagnetic force term
$\frac{m}{c} (\textbf{v}\times{\bf B}_{g})$ is  substituted by the
Coriolis force term $2m(\textbf{v}\times{\bf \O})$ where ${\bf
\O}$ is the angular velocity vector of the rotating body
\ct{ref:vla}. See also \ct{ref:mas93}.} So it is possible to speak
of mass-energy currents whose motion exerts a non-central
 gravitational force on a test
massive body analogous to the  Lorentz force felt by a charged
particle when it is moving in a magnetic field. Indeed, its
equations of motions \eqi m\rp{{d}^2 {\bf r}}{{d}t^2}\simeq
q\left({\bf E}+\rp{\textbf{v}}{c} \times {\bf B}\right)
\label{lor}\eqf are formally analogous to eq. (\ref{cif}). Eq.
(\ref{lor}) can be derived by means of the Lagrangian \eqi
{\mtc{L}}_{e.m.}=\rp{1}{2}mv^2-qV+ \rp{q}{c}({\bf v}\cdot{\bf
A}).\lb{ele}\eqf where $\textbf{v}$ is the velocity of the
particle and $V$ is its scalar potential.

Since one of the most promising ways to detect the gravitomagnetic
precession consists of employing artificial Earth satellites, it
would be helpful to derive the rate equations for the change in
the parameters that characterize their orbits. To this aim one
could
 introduce ``by hand'' a perturbative
 term $k\ ({\bf v}\cdot{\bf A}_{g})$ in the gravitational Lagrangian of the
particle and use it in
some particular perturbative scheme;
 the  constant $k$ would be determined by means
of dimensional considerations and taking into account that it
should be built up of universal constants. In fact it is possible
to show that a non-central term analogous to $\rp{q}{c}({\bf
v}\cdot{\bf A})$ can be rigorously deduced in the Lagrangian of  a
test body in the gravitational field of a spinning mass-energy
distribution, and that it can be exploited in deriving
straightforwardly the effect of the gravitomagnetic potential on
the Keplerian orbital elements of the test body.
\subsection{The rate equations for the
Keplerian orbital elements} The relativistic Lagrangian for a free
particle in a gravitational field  can be cast into the form:
\eqi{\mtc{L}}={\mtc{L}}^{(0)}+{\mtc{L}}^{(1)}\lb{lag2}.\eqf In
\rfr{lag2} the term ${\mtc{L}}^{(1)}$ is to be intended as
containing  the contributions of the off-diagonal terms of the
metric \eqi {\mtc{L}}^{(1)}
=\rp{m}{c}g_{0k}\dot{x^{0}}\dot{x^{k}}.\lb{perini}\eqf In this
case, recalling that the slow motion approximation is used, the
\rfr{perini} becomes \eqi
{\mtc{L}}^{(1)}\equiv{\mtc{L}}_{gm}=\rp{m}{c}({\bf A}_{g}\cdot{\bf
v}).\eqf For different derivations of ${\mtc{L}}_{gm}$ see also
\ct{ref:lan,ref:mis,ref:berfar}. In this section it is proposed to
adopt ${\mtc{L}}_{gm}$ in order to deriving the Lense-Thirring
effect on the orbital elements of a particle in the field of a
rotating gravitational source.

\begin{figure}[ht!] \begin{center}
\includegraphics*[width=10cm,height=7cm]{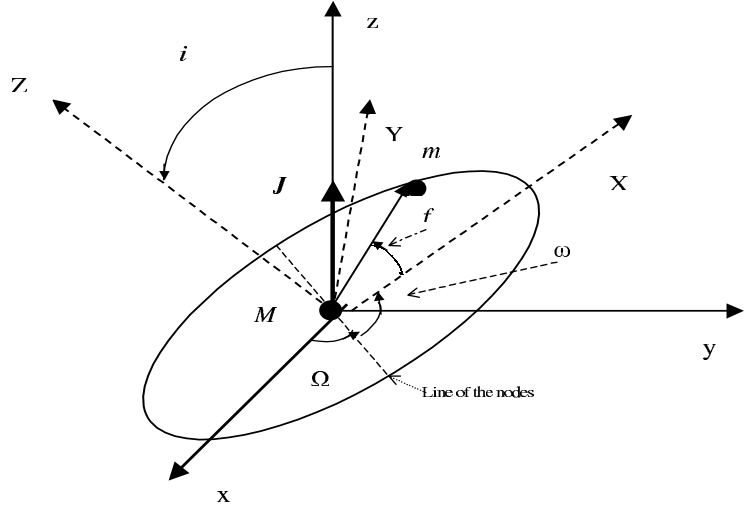}
\end{center}
\caption[Keplerian orbital elements for an elliptical
orbit]{\footnotesize Orbital elements for a general Keplerian
osculating ellipse.} \label{kpl}
\end{figure}

To this aim it must be assumed that under the gravitomagnetic
force the departures of the test body' s trajectory from the
unperturbed Keplerian ellipse are very small in time. This allows
to introduce the concept of osculating ellipse. It means that, at
a given instant of time, the particle may be assumed to lie on the
Keplerian ellipse determined by the position and velocity at that
instant thought as initial conditions for an unperturbed motion;
at the next instant of time the particle will be at a point of
another Keplerian ellipse, slightly different with respect to the
previous one and determined by the real position and velocity of
the test body at the new instant of time thought as new initial
conditions for an unperturbed Keplerian orbit. In other words, the
real trajectory of the test body at every instant may be
approximated by an osculating Keplerian ellipse. So the perturbed
motion can be described in terms of unperturbed Keplerian elements
varying in time. See fig. \ref{kpl}.

Consider the frame $K\{x,\ y,\ z\}$ previously defined and a frame
 $K^{'}\{X,
\ Y,\ Z\}$ with the $Z$ axis directed along the orbital angular
momentum ${\bf l}$ of the test body, the plane $\{X, \ Y\}$
coinciding with the orbital plane of the test particle and the $X$
axis directed toward the pericenter. $K$ and $K^{'}$ have the same
origin located in the center of mass of the central body. The
Keplerian orbital elements are \ct{ref:kau}

$\bullet\ \  a,\ e$, the semimajor axis and the ellipticity which
define the size and the shape of the orbit in its plane

$\bullet\ \  \O,\ i$, the longitude of the ascending node and the
inclination which fix the orientation of the orbit in the space,
i.e. of $K^{'}$ with respect to $K$. The
 longitude of the ascending node $\O$ is the angle in the equatorial plane
of $K\{x,\ y,\ z\}$ between the $x$ axis and the line of nodes in
which the orbital plane intersects the equatorial plane. The
inclination $i$ is the angle between the $z$ and $Z$ axis

$\bullet\ \  \og,\ \mtc{M}$, the argument of pericenter and the
mean anomaly. The argument of pericenter $\og$ is the angle in the
orbital plane between the line of nodes and the $X$ axis; it
defines the orientation of the orbit in its plane. The mean
anomaly $\mtc{M}$ specifies the motion of the test particle on its
orbit. It is related to the mean motion $n=(GM)^{1/2}a^{-3/2}$,
where $M$ is the mass of the gravitating central source, through
${\cal{M}}=n(t-t_{p})$ in which $t_{p}$ is the time of pericenter
passage

It is customary to define also

$\bullet\
 \varpi=\O+\og$, the longitude of pericenter

$\bullet\ u=\og+f$, the argument of latitude where $f$ is the
angle, called true anomaly, which in the orbital plane determines
the position of the test particle with respect to the pericenter

$\bullet\
 \varepsilon=\varpi+n(t_0-t_p)$, the mean longitude at the epoch $t_0$. If $t_0=0$, it is customary to write
$\varepsilon$ as $L_0=\varpi-n t_p$.

The matrix ${\bf R}_{xX}$ for
 the change of coordinates from
$K^{'}$ to $K$ is \ct{ref:kau} \eqi \km \lb{mat}.\eqf Using eq.
(\ref{mat}) and $X=r\cos{f}$, \ \ $Y=r\sin{f}$, \ \ $Z=0$ it is
possible to express the geocentric rectangular Cartesian
coordinates of the orbiter in terms of its Keplerian
elements\eqi\left\{\begin{array}{l}x=r (\cos{u}\cO-
\sin{u}\ci\sO)\\
y=r(\cos{u}\sO+
\sin{u}\ci\cO)\\
z=r\sin{u}\si.   \lb{z} \end{array}\right.\eqf Redefining suitably
the origin of the angle $\O$ so that $\cO=1$, $\sO=0$, the
previous equations become
\eqi\left\{\begin{array}{l}x=r\cos{u}\\y=r\sin{u}\ci\\z=
r\sin{u}\si.\lb{xx}\end{array}\right.\eqf Considering for the test
particle  the total mechanical energy with the sign reversed,
according to \ct{ref:vin}, ${\mtc{F}} \equiv
 -{\mathcal{E}}_{tot}=-(\mathcal{T}+
\mathcal{U})$, where $\mathcal{T}$ and $\mathcal{U}$ are the
kinetic and potential energies per unit mass, it is possible to
work out the analytical expressions for the rate of changes of
$a,\ e,\ i,\ \O,\ \og,\ {\mtc{M}}$ due to any non-central
gravitational contribution. To this aim it is useful to isolate in
$\mathcal{U}$ the central part $\mathcal{-C}$ of the gravitational
field from the terms $\mathcal{-R}$ which may cause the Keplerian
orbital elements to change in time:
${\mathcal{U}}=-{\mtc{C}}-{\mathcal{R}}$.  For a spherically
symmetric body, $\mtc{F}$ becomes \eqi
{\mtc{F}}=\rp{GM}{r}+{\mathcal{R}}-{\mathcal{T}}=\rp{GM}{2a}+
{\mathcal{R}}.\lb{kau}\eqf Concerning the perturbative scheme to
be employed, the well known Lagrange planetary equations  are
adopted.
At first order, they are
\begin{eqnarray}\dert{a}{t}&=& \cu \ \dfm ,\lb{dos}\\
\dert{e}{t}&=& \cd \ \dfm - \ctr \ \dfog ,\lb{ee}\\
\dert{i}{t}&=& \ci \cq \ \dfog - \cq \ \dfo ,\\
\dert{\O}{t}&=& \cq \ \dfi ,\lb{omeg}\\
\dert{\og}{t}&=&- \ci \cq \ \dfi + \ctr \ \dfe ,\lb{omeghin}\\
\dert{\mtc{M}}{t}&=&n- \cd \ \dfe - \cu \ \dfa
.\lb{tres}\end{eqnarray} The idea consists of using
${\mtc{L}}_{gm}$ to obtain a suitable non-central term
${\mtc{R}}_{gm}$ to be employed in these equations. This can be
done considering the Hamiltonian for the test particle \eqi
{\mtc{H}}={\bf p}\cdot{\bf v}-{\mtc{L}}.\lb{ham}\eqf Inserting
\rfr{lag2} in \rfr{ham}  one has \eqi
{\mtc{H}}={\mtc{H}}^{(0)}+{\mtc{H}}_{gm},\eqf with
${\mtc{H}}_{gm}=-\rp{m}{c}({\bf A}_{g}\cdot{\bf v})$. So it can be
posed\eqi {\mtc{R}}_{gm}=-\rp{{\mtc{H}}_{gm}}{m}=\frac{({\bf
A}_{g}\cdot{\bf v})}{c}\lb{r}.\eqf

Now it is useful to  express \rfr{r} in terms of the Keplerian
elements. Referring to eq. (\ref{acchina}), \rfr{xx}, and
recalling that in the frame $K\{x,\ y,\ z\}$ ${\bf J}=(0,\ 0,\ J)$
and that for an unperturbed Keplerian motion \eqi
\rp{1}{r}=\rp{(1+e\cos{f})}{a(1-e^2)}, \eqf it is possible to
obtain, for a perfectly spherical central body \eqi
{\mtc{R}}_{gm}= -\rp{2G}{c^2}\rp{J\ci}{r} \dot{u}=
-\rp{2GJ\ci}{c^2}\rp{(1+e\cos{f})}{a(1-e^2)}\dot{u}.\lb{ru}\eqf In
\rfr{ru}  $\dot u\simeq \dot f$ is assumed due to
 the fact that the osculating
element $\og$ may be retained almost constant on the temporal
scale of variation of the true anomaly of the test
body.\subsection{Secular gravitomagnetic effects on the Keplerian
orbital elements} The secular effects can be worked out by
adopting the same strategy followed, e.g., in \ct{ref:vin} for a
similar kind of perturbing functions. By using averaging we
implicitly make several assumptions about the system that may
restrict the validity and applicability of our result. Averaging
is generally applied to systems where the perturbing force is
sufficiently small so that, over one orbital period, the
deviations of the true trajectory from the Keplerian trajectory
are relatively small. This perturbation can generally be related
to the magnitude of the perturbing acceleration as compared to the
central-body attraction. In this case, clearly, we can view the
application of the averaging approach as valid. When eq.
(\ref{ru}) is mediated over one orbital period of the test body,
$a,\ e ,\ i,\ \O$ and $\og$ are to be considered constant
\eqi\med{{\mtc{R}}}=-\rp{1}{P}\int_{0}^{P}
\rp{G}{c^2}\rp{2J}{r}\ci du
 =-n\rp{G}{c^2}\rp{2J\ci}{ 2\p}\int_{0}^{2\p}
\rp{df}{r}=-n\rp{G}{c^2}\rp{2J\cos{i}}{ a(1-e^2)}\lb{fine}.\label{po}\eqf
 The relation $du=d\og+df=df$ has been used
 in \rfr{po}.
Eq. (\ref{po}) can now be used in determining the
 secular changes of the Keplerian orbital
elements of the test body. From it one gets
\begin{eqnarray}
\derp{\med{\mtc{R}}}{{\mtc{M}}}&=&0,\lb{soluz}\\
\derp{\med{\mtc{R}}}{\og}&=&0,\\
\derp{\med{\mtc{R}}}{\O}&=&0,\\
\derp{\med{\mtc{R}}}{i}&=&
\rp{G}{c^2}\rp{2nJ\sin{i}}{a(1-e^2)},\\
\derp{\med{\mtc{R}}}{e}&=&-
\rp{G}{c^2}\rp{4nJe\cos{i}}{a(1-e^2)^2}\lb{re}.
\end{eqnarray}
A particular care is needed for the treatment of $n$ when the
derivative of $\med{\mtc{R}}$ with respect to $a$ is taken;
indeed, it must be posed as
\eqi\derp{\med{\mtc{R}}}{a}=\left.\derp{\med{\mtc{R}}}{a}\right|_{n}+
\left.\derp{\med{\mtc{R}}}{n}\right|_{a}\derp{n}{a}.\lb{palle}\eqf
In \rfr{palle} \eqi\left.\derp{\med{\mtc{R}}}{a}\right|_{n}=
\rp{G}{c^2}\rp{2nJ\cos{i}}{a^2(1-e^2)},\lb{uffa}\eqf and
\eqi\left.\derp{\med{\mtc{R}}}{n}\right|_{a}\derp{n}{a}=
\rp{G}{c^2}\rp{3nJ\cos{i}}{a^2(1-e^2)}.\eqf From \rfr{dos} and
\rfr{soluz} it appears that there are no secular changes in the
semimajor axis\footnote{Here only the effect of the off-diagonal
gravitomagnetic components of the metric tensor is considered,
neglecting any other physical influence of gravitational and non
gravitational origin.}, and so the orbital period of the test
body, related to the mean motion by $P=2\p/n$, can be considered
constant. So in \rfr{palle} only \rfr{uffa} must be retained.
Using \rfrs{soluz}{palle} in \rfrs{dos}{tres} one obtains for the
averaged rates
\begin{eqnarray}
\left.{\dert{{a}}{t}}\right|_{{\rm LT}} &=& 0,\lb{alt}\\
\left.{\dert{{e}}{t}}\right|_{{\rm LT}}&=& 0,\\
\left.{\dert{{i}}{t}}\right|_{{\rm LT}}&=& 0,\\
\left.{\dert{{\O}}{t}}\right|_{{\rm LT}}&=&\rp{G}{c^2}\rp{2J}{a^{3}(1-e^2)^{3/2}},\\
\left.{\dert{{\og}}{t}}\right|_{{\rm LT}}&=&-\rp{G}{c^2}\rp{6J\ci}{a^{3}(1-e^2)^{3/2}},\\
\left.{\dert{{{\mtc{M}}}}{t}}\right|_{{\rm
LT}}&=&0\lb{moo}.\end{eqnarray} They are the well known
Lense-Thirring equations \ct{ref:len,ref:ciuwhe,ref:iorcim}. In
deriving them it has been assumed that the spatial average over
$f$ yields the same results for the time average
\ct{ref:arn83,ref:arn89,ref:mil}.

The effects of the quadrupole mass moments of the source are
treated in \ct{ref:iorcim}: it turns out that they are negligible
in the context of the LAGEOS Lense--Thirring experiment.

\section{Orbital perturbations induced by solid Earth and ocean tides}
\subsection{Introduction}
In this section the problem of calculating the  gravitational
time-dependent  perturbations induced by solid Earth and ocean
tides on the \kep\ of \lg\ and \lgg\ is addressed. The focus of
the analysis is on the nodes of \lg\ and \lgg\ and the \pg\ of
\lgg\ in view of their close connection to the measurement of the
\lt\ effect. The perturbative scheme adopted is that based on the
Lagrange' s planetary equations
\ct{ref:kau,ref:dow,ref:chretal,ref:cas}.

An evaluation from first principles of the amplitudes, the periods
and the initial phases of the tidal perturbations on the nodes of
LAGEOS and LAGEOS\ II and the perigee of LAGEOS\ II is useful, in
the present context, for the following reasons.

$\bullet$ It allows one to point out which tidal constituents the
Lense-Thirring shift is mainly sensitive to, so that people can
focus the researches on them

$\bullet$ It allows one to get insight in how to update the orbit
determination softwares. In this way the impact on the time series
of those tidal constituents which will turn out to be more
effective could be reduced along with the number of the parameters
to be included in the least-squares solution

$\bullet$ For a given observational time span  $T_{{\rm obs}}$ it
allows one to find those constituents whose periods  are longer
than it and consequently may act like superimposed bias (e.g. the
18.6-year and 9.3-year zonal tides and the $K_1\ l=3\ p=1\ q=-1$
ocean diurnal tide on the perigee of LAGEOS\ II) corrupting the
detection of the gravitomagnetic shift. In these cases, if we know
their amplitudes, periods and initial phases we could fit and
remove them from the time series\footnote{If the period of the
disturbing harmonic, assumed to be known, is shorter than the time
series length the perturbation can be viewed as an empirically fit
quantity. But if its period is longer than the time series length
it is not possible to fit and remove the harmonic from the data
without removing also the true trend, unless we know the initial
phase.} or, at least, it should be possible to assess the
mismodelling level induced by them on the detection of the
Lense-Thirring trend. So, also these estimates can be included in
the final error budget. (See section 4)

$\bullet$ It can be used as starting point for numerical
simulations of the combined residual data in order  to check, e.
g., the impact of the diurnal ($m$=1) and semidiurnal ($m$=2)
tidal perturbations (not cancelled out by the Ciufolini' s
observable), as done in section 4

The section is organized as follows. In subsect. 2 the essential
features of the tide generating potential, from which the tidal
forces are traditionally derived, are reviewed.

The solid Earth tidal perturbations are worked out in subsect. 3.
An evaluation of their mismodelling  and a comparison with the
\lt\ perturbations on the nodes of \lg\ and \lgg\ and the \pg\ of
\lgg\ over 4 years are also presented.

Subsect. 4-5 are devoted to the orbital perturbations due to the
ocean tides and their mismodelling in connection with the \lt\
drag. They are treated in detail because the ocean tides are known
less accurately than the solid ones, so that the impact of their
mismodelling on the \grc\ trend is more relevant.

Subsect. 6 is devoted to the conclusions.
\subsection{The tide generating potential}
On the scale of the Earth's dimensions the gravitational field of any
not too far astronomical body $B$
 cannot be considered
uniform so that the various points of our planet are acted upon by
a differential gravitational pull in the external field of $B$.
This is the origin of the Earth's solid, ocean and atmospheric
tides, mainly due to the Moon and the Sun.  It is customary to
derive the tidal forces acting on a point on the Earth' s surface
from the so called tide generating potential which, in its most
general form, for a perturbing body $B$ can be written as
\ct{ref:dro,ref:lam,ref:mel} $$ \Phi_B =
 \rp{GM_B}{d_B}\
 \sum_{l=2}^{\infty}\left(\rp{r}{d_B}\right)^{l}P_{l}(\cos{z})=$$
\eqi=\rp{3}{4}GM_B\rp{R^2}{c_B^3}\left(\rp{r}{R}\right)^2\left(\rp{c_B}{d_B}\right)^3
\left[\rp{4}{3}P_2(\cos{z})+\rp{4}{3}\rp{r}{d_B}P_3(\cos{z})+...\right]\lb{tgp}.
\eqf In \rfr{tgp} the following quantities appear

$\bullet\ R$\ \ \ mean Earth' s equatorial radius

$\bullet\ M_B$\ \ \ mass of the perturbing body

$\bullet\ d_B$\ \ \ instantaneous distance between the Earth and
the perturbing body $B$

$\bullet\ r$\ \ \ distance between  the center of mass of the
Earth and a point on its surface

$\bullet\ P_{l}(x)$\ \ \ Legendre polynomials of degree $l$
 \eqi P_l(x)=\rp{(-1)^l}{l!2^l}\rp{d^l}{dx^l}(1-x^2)^l\eqf

$\bullet\ z$\ \ \ geocentric zenithal distance of the perturbing
body

$\bullet\ c_B$\ \ \ mean distance between the Earth and the
perturbing body

The factor \eqi
\rp{3}{4}GM_B\rp{R^2}{c_B^3}\left(\rp{r}{R}\right)^2\eqf is often
rewritten as\eqi D_B(r)=\rp{M_B}{M_M}\rp{c_M^3}{c_B^3}D(r),\eqf
with \eqi D(r)=
\rp{3}{4}GM_M\rp{R^2}{c_M^3}\left(\rp{r}{R}\right)^2=D_1\left(\rp{r}{R}\right)^2\eqf
in which \eqi D_1=\rp{3}{4}\ GM_{M}\rp{R^2}{c_M^3}\eqf is the
Doodson constant for the Moon, having the dimension of an energy
per unit mass, and $M_M$ is the Moon's mass. Thus the tide
generating potential outside of the Earth $\Phi_B^l$ is a harmonic
function of degree $l$ and can be written in the form $\Phi_B^l$
$=r^l Y_l (\f,\l)$ where $Y_l (\f,\l)$ is a surface harmonic, with
$\f$ and $\l$ the terrestrial latitude and East longitude. At the
Earth's surface it becomes $\Phi_B^l ({\bf R})$ $=(R/r)^l \Phi_B^l
({\bf r})$.

It is useful to express  \rfr{tgp} in terms of the geocentric
equatorial coordinates $\{\f,\ \d_B,\ H_B\}$ of an Earth-fixed
frame where

$\bullet\ \d_B$\ \ \ declination of the perturbing body

$\bullet\ H_B$\ \ \ hour angle of the perturbing body

This can be done by means of the following spherical trigonometric
formula \eqi
\cos{z}=\sin{\f}\sin{\d_B}+\cos{\f}\cos{\d_B}\cos{H_B}.\eqf The
result is the Laplace's tidal development: it  consists of a sum
of terms each of which is the product of a factor depending only
on the latitude of a given place on the Earth (the geodetic
coefficient) and a time-dependent factor which depends on the
astronomical coordinates $\d_B$ and $H_B$ of the perturbing body.
E.g., the term of degree $l=2$ becomes
$$\Phi_B^2=D_{B}(r)\left(\rp{c_B}{d_B}\right)^3\
 \left[\left(\rp{1-3\sin^2\f}{2}\right)\left(
\rp{1}{3}-\sin^2\d_B\right)+\right.$$\eqi+\left.\sin{2\f}\sin{2\d_B}\cos{H_B}+\cos^2\f\cos^2\d_B\cos{2H_B}\
\right] \lb{tigp}. \eqf In this sum the quantities $c_B/d_B,\
\d_B,\ H_B$ exhibit a very complex behavior in time due to the
astronomical motions of the perturbing body $B$. They must be
carefully expanded in periodic components in order to obtain an
expression for the tidal potential as a sum of harmonic terms.
This can be done by using the ephemerides tables for the
perturbing bodies. Obviously, the most relevant are the Moon and
the Sun. Since the ephemerides refer to  the ecliptic and not to
the celestial equator, it is necessary to express $\d_B$ and $H_B$
in terms of ecliptical coordinates. Spherical trigonometry gives
\eqi\sin{\d_B}=\sin{\ve}\cos{\b_B}\sin{\l_B}+
\cos{\ve}\sin{\l_B},\lb{st1}\eqf
\eqi\cos{\d_B}\cos{H_B}=\cos{\b_B}\cos{\l_B}\cos{\th}+(\cos{\ve}\cos{\b_B}\sin{\l_B}
 -\sin{\ve}\sin{\b_B})\sin{\th}\lb{st2},\eqf
where

$\bullet\ \ve$\ \ \ inclination of the ecliptic  to the celestial
equator

$\bullet\ \l_B,\ \b_B$\ \ \ ecliptical longitude and latitude of
the perturbing body, respectively

$\bullet\ \th$\ \ \  sidereal time

The lunar and solar ephemerides express $d_B,\ \l_B,\ \b_B$ in
terms of time-dependent series of sines and cosines whose
arguments are linear combinations of the mean Lunar time $\t$ and
other astronomical quantities of the Moon and the Sun
\footnote{The most recent developments take in account also the
influence of other planets like Venus and Jupiter \ct{ref:roo}.}.
Among them the most useful are  the mean ecliptical lunisolar
variables $s,\ h ,\ p,\ N^{'},\  p_s$ \ct{ref:dro}. They are

$\bullet\ s$\ \ \ mean ecliptical  longitude of the Moon with
period $P_s=27.32$ days

$\bullet\ h$\ \ \ mean ecliptical  longitude of the Sun with
period $P_h=365.2422$ days

$\bullet\ p$\ \ \ mean ecliptical  longitude of the Moon's perigee
with period
 $P_p=3,232$ days (8.84 years)

$\bullet\ N^{'}=-N$\ \ \ mean ecliptical  longitude of the Moon's
node with the sign reversed,
 with period $P_N=6,798.38$ days (18.61 years). The lunar node, moving in the
Earth-fixed frame from East to West counterclockwise, is
retrograde with respect to the other lunisolar longitudes

$\bullet\ p_s$\ \ \ mean ecliptical  longitude of the perihelion
with period
 $P_{p_s}=7.65\times 10^{6}$ days (20,953 years)

Putting \rfrs{st1}{st2} in the Laplace' s tidal development, with
the ephemerides expansions for $d_B,\ \l_B,\ \b_B$, allows one to
obtain the harmonic expansion of the tide generating potential
$\Phi_B$
\ct{ref:doo,ref:cartay,ref:caredd,ref:bue,ref:tam,ref:xi,ref:harwen,ref:roo}.
It consists of a sum of terms, the constituents, with sinusoidal
time dependence where the sines and cosines have arguments
involving linear combinations $\s t$ of
 the previously defined orbital
longitudes of the Sun and the Moon \eqi \s t=j_1 \t+j_2 s+ j_3
h+j_4  p+ j_5 N^{'}+ j_6
 p_s.\eqf
The circular frequency of the tidal bulge generated by the corresponding
 constituent, viewed in the terrestrial frame, is then given by
\eqi \s\equiv 2\p f= j_1\dot \t+j_2\dot s+ j_3 \dot h+j_4 \dot p+
j_5\dot N^{'}+ j_6 \dot p_s ;\lb{freq}\eqf by considering that
$\t=\th-s$, \rfr{freq} becomes \eqi \s= j_1\dot \th+(j_2-j_1)\dot
s+ j_3 \dot h+j_4 \dot p+ j_5\dot
 N^{'}+ j_6
\dot p_s.\lb{freq2}\eqf The coefficients $j_k,\ k=1,...,6$ are
small integers which can assume negative, positive or null values.
The advantage of using such lunisolar ecliptical variables relies
in the fact that they, over an interval of a century or so, are
practically linear increasing with time. This feature will reveal
itself very useful in integrating the equations for the orbital
perturbations. Each tidal constituent is identified by the set of
the six integers $j_k,\ k=1,. ..,6$ arranged in the so called
Doodson number \eqi j_1(j_2+5)(j_3+5).(j_4+5)(j_5+5)(j_6+5).\eqf
In it the integer $j_1$ classifies the tides in long period or
zonal ($j_1=0$),
 diurnal or tesseral ($j_1=1$) and
semidiurnal or sectorial ($j_1=2$).

The tide generating potential is the cornerstone for every
calculation of tidal perturbations on the satellites' Keplerian
orbital elements. Indeed, if the Earth, globally modeled as a
nonrigid body, is acted upon by a tidal potential $\Phi_{l}$
harmonic in degree, it deforms itself giving rise, among other
things, to a periodic redistribution of the masses
 within its
volume. This deformation acts upon a point in the external space
surrounding the Earth by means of an additional potential
\ct{ref:lov}\eqi
 \phi_l({\bf r})=k_l\left(\rp{R}{r}\right)^{l+1}\Phi_{l}(R)=k_l\left(\rp{R}{r}\right)^{2l+1}
\Phi_{l}(r)\lb{cdd}\eqf where $k_l$ is one of the so called Love
numbers (they will be defined  precisely in the following
sections). A close orbiting satellite is perturbed by such an
additional potential and senses the global effect of the solid and
fluid mass redistribution in the Earth. This means that if one
wish to employ a pool of satellites in order to recover the tidal
parameters, like the Love numbers, entering  the tidal
perturbations and wants to use the so obtained values to predict
the perturbations on some particular satellite, he or she must be
very careful. Indeed, such values are aliased by the whole of the
effects sensed by the measuring satellites; it is as if they would
be ``tailored'' for the satellites used in their recovery. So,
these ``effective'' tidal parameters neither can be directly
compared to those measured on the Earth's surface or predicted by
any theoretical model and related to the physical properties of
the Earth, nor can be used in calculating {\it a priori} the tidal
perturbations on the satellites of interest for some particular
application, as is the case for the detection of the
Lense-Thirring effect. Differently stated, when the tidal
perturbations are to be predicted for a given satellite, it is
incorrect to use for it the ``effective'' values of the
geophysical parameters recovered by other satellites. It is so
customary to split the analytical expressions of the tidal
perturbations into a part due to the solid Earth tides and another
one which accounts for the effects of the oceans \ct{ref:lametal}.
After this step, their effects on the satellite of interest must
be analytically predicted using for the various parameters the
values theoretically calculated or measured in such a way that
they reflect the effective Earth's properties.
\subsection{Solid Earth tidal perturbations}
Concerning the solid Earth tides \ct{ref:wan97}, as in
\ct{ref:dow,ref:bercar}, the starting point is the
frequency-dependent
 model of \ct{ref:wah81b}. It is based on the assumption of a perfectly
elastic, hydrostatically prestressed,  ellipsoidal rotating  Earth
acted upon by the lunisolar tidal potential. The interior of our
planet is thought to be made of a solid inner core, a fluid outer
core, and a solid mantle capped by a thin continental crust,
without oceans and atmosphere. Substantially, the Earth is thought
as a set of coupled harmonic oscillators showing a variety of
resonant frequencies, the normal modes, which can be excited by
the external forcing constituents of the tide generating potential
\ct{ref:wah81a}.

The expansion adopted for it is that of Cartwrigt and Tayler in
\ct{ref:cartay} in terms of harmonic constituents of degree $l$,
order $m$ and circular frequency $\s$\eqi\Phi= \
g\sum_{l=2}^{\infty}\sum_{m=0}^{l}\left(\rp{r}{R}\right)
^{l}Y_{l}^{m}(\f,\l) \sum_{f}H_{l}^{m}(f)e^{i( \s
t+c_{lm})}\lb{ct}.\eqf Only the real part of \rfr{ct} has to be
retained. The quantities entering \rfr{ct} are

$\bullet\ Y_{l}^{m}$\ \ \ spherical harmonics\eqi
Y_{l}^{m}(\f,\l)=\sqrt{\rp{2l+1}{4\p}\rp{(l-m)!}{(l+m)!}}P_{l}^{m}(
\sin{\f})e^{im\l},\eqf in which the Condon-Shortley phase factor
$(-1)^{m}$ has been neglected in order to compare our results to
those of  \ct{ref:bercar}

$\bullet\ H_{l}^{m}(f)$\ \ \ coefficients of the harmonic
expansion.
 They contain the lunisolar ephemerides
information and define the modulus of the vertical shift $V/g$ in
the equipotential surface at the Earth's surface for
$r=R_{\oplus}$ with respect to the case in which only proper
Earth's gravity is considered. The values used in the present
calculation are those recently released in \ct{ref:roo} using the
ELP2000-85 series for the ecliptical coordinates of the Moon
\ct{ref:ctz}
 and the VSOP87 series for those of the Sun
\ct{ref:brf}. Their accuracy is of the order of $10^{-7}$ m. In
order
 to use them in place of the coefficients of \ct{ref:caredd}, which have a different
  normalization, a multiplication for a suitable
conversion factor \ct{ref:iers} has been performed. Due to the
extreme smallness of the $H_{l}^{m}(f)$ for $l \geq 3$, in the
following summations only the term of degree $l=2$ are to be
considered

$\bullet\ c_{lm}$\ \ \ additive constant. It is equal to $-\p/2$
in order to generate sines for the ${l-m}$ odd constituents, while
it is equal to $0$ in order to yield cosines for the ${l-m}$ even
constituents.

The Earth free space potential under the action of the tidal
constituent of degree $l$, order $m$ and circular frequency $\s$,
dropping the time dependence due to $e^{i\s t}$, can be written as
\eqi \phi_{l}^{m}=Re\ gH_{l}^{m}\left[
k_{lm}^{(0)}\left(\rp{R}{r}\right)^{l+1}
 Y_{l}^{m} +k_{lm}^{+} \left(\rp{R}{r}\right)^{l+3}  Y_{l+2}^{m}\right]\lb{ela},\eqf
in which

$\bullet\ r$\ \ \ distance from the Earth' s center of mass and a
point in the space outside of the Earth

$\bullet$ The parameter \eqi
k^{(0)}_{lm}(f)=\sqrt{{k_{lm}^{R}(f)}^2+{k_{lm}^{I}(f)}^2},\lb{loven}\eqf
is the modulus of the Love number $k$ \ct{ref:lov,ref:mat} for a
static, spherical Earth. In general, it is defined as the ratio of
the free space gravitational potential $\phi({\bf r})$, evaluated
at the Earth' s equator, and the tide generating potential
$\Phi({\bf r})$ calculated at the mean equatorial radius.
 If the rotation and non-sphericity of the Earth are accounted for, there
are different values of $k$ for any $l,m$ and $f$. In general,
they are evaluated by convolving, in the frequency domain, the
tide generating potential with the transfer function for a non
rigid Earth \ct{ref:dah,ref:smi}. There are several theoretical
calculations for the Love numbers $k$
\ct{ref:wah81b,ref:wan94,ref:mat,ref:iers,ref:deh}. They differ in
the choice of the Earth's interior models \ct{ref:gil,ref:dziand}
adopted for the calculation of the transfer function and the
departure from elasticity of the mantle' s behavior. Particularly
important is also if they account for, or not, the normal modes in
the diurnal band. In \rfr{loven} the IERS conventions
\ct{ref:iers} have been used for $k^{(0)}_{lm}(f)$
 \eqi k_{lm}^{I}(f)=Im\ k_{lm} + \d k_f^{anel},\eqf
\eqi k_{lm}^{R}(f)=Re\ k_{lm} + \d k_f^{el},\eqf
where  $Im\ k_{lm}$ and
 $Re\ k_{lm}$ are the frequency-independent parts of Love numbers,
and $\d k_f^{anel}$, $\d k_f^{el}$ are the frequency-dependent
corrections. They are important in the diurnal band, through $\d
k_f^{el}$, and in the zonal band with $\d k_f^{anel}$ due to the
anelasticity of the mantle. The $k_{lm}^{+}(f)$ account for the
rotation and ellipticity of our planet. In the present analysis
for the latitude dependent Love numbers $k_{lm}^+(f)$ the values
quoted in \ct{ref:deh} have been adopted.


Neglecting, in this first step, their tiny contribution,
 the free space potential
by means of which the Earth responds to the forcing lunisolar
tidal potential can be written, in the time domain, as\eqi
\phi=\rp{GM}{R}\sum_{l=2}^{\infty}\sum_{m=0}^{l}
\sqrt{\rp{2l+1}{4\p} \rp{(l-m)!}{(l+m)!}}
\rp{1}{r}\left(\rp{R}{r}\right)^{l}\times
\sum_{f}k_{lm}^{(0)}H_{l}^{m}P_{l}^{m}(\sin{\f}) {\mc{\cos{(\s t +
m\l)}}{\sin{(\s t+m\l)}}}\lb{duro}.\eqf Here and in the following,
the upper expressions refer to $l-m$ even while the lower ones
refer to $l-m$ odd. \Rfr{duro} can be put into the form\eqi
\phi=\rp{GM}{R}\sum_{l=2}^{\infty}\sum_{m=0}^{l}
\rp{1}{r}\left({\rp{R}{r}}\right)^{l}(C_{lm}\cos{m\l} +S_{lm}
\sin{m\l})P_{l}^{m}(\sin{\f}),\lb{solido}\eqf in which, assuming
temporarily for the sake of simplicity that the Love numbers are
not frequency-dependent, the coefficients $C_{lm}$ and $S_{lm}$
are
\begin{eqnarray} C_{lm} & = & A_{lm}k_{lm}^{(0)} \sum_{f} H_{l}^{m}
{\mc{\cos{\s t}}{\sin{\s t}}},\lb{clm}\\
 S_{lm} & = & A_{lm}k_{lm}^{(0)} \sum_{f} H_{l}^{m}
{\mc{-\sin{\s t}}{\cos{\s t}}}\lb{slm},\end{eqnarray} with\eqi
A_{lm}=\sqrt{\rp{2l+1}{4\p}\rp{(l-m)!}{(l+m)!}}.\eqf The $C_{lm}$
and $S_{lm}$ have the dimensions of lengths. \Rfr{solido} is
formally equal to the well known expression of the static
geopotential worked out by Kaula in ref.\ct{ref:kau}\eqi
\phi=\rp{GM}{r}\left\{1+\sum_{l=2}^{\infty}\sum_{m=0}^{l}
\left(\rp{R}{r}\right)^{l}P_{lm}(\sin{\f})
[C_{lm}\cos{m\l}+S_{lm}\sin{m\l}]\right\}.\lb{prima}\eqf  By means
of the transformation
\eqi\rp{1}{r}\left(\rp{R}{r}\right)^{l}P_{l}^{m}(\sin{\f})e^{im\l}
= \rp{1}{a}\left(\rp{R}{a}\right)^{l} \sum_{p=0}^{l}F_{lmp}(i)
\sum_{q=-\infty}^{+\infty}G_{lpq}(e)
{\mc{e^{i\ps_{lmpq}}}{-ie^{i\ps_{lmpq}}}},\lb{trans}\eqf
with\eqi\ps_{lmpq}=
(l-2p)\omega+(l-2p+q){\mtc{M}}+m(\O-\th)\lb{psi},\eqf it can be
cast into the familiar form\eqi \phi=
\rp{GM}{R}\sum_{l=2}^{\infty}\sum_{m=0}^{l}
\rp{1}{a}\left({\rp{R}{a}}\right)^{l}\sum_{p=0}^{l}
\sum_{q=-\infty}^{+\infty}F_{lmp}(i)\ G_{lpq}(e)S_{lmpq}(\omega,\
\O, {\mtc{M}},\ \th),\lb{kaula}\eqf where   $F_{lmp}(i)$ and
$G_{lpq}(e)$ are the inclination and eccentricity functions
respectively \ct{ref:kau}. The $S_{lmpq}$ is given by \eqi
S_{lmpq}=
{\mc{C_{lm}}{-S_{lm}}}\cos{\ps_{lmpq}}+{\mc{S_{lm}}{C_{lm}}}\sin{\ps_{lmpq}}.
\lb{kaulino}\eqf \Rfr{kaulino} in our case becomes\eqi
S_{lmpq}=A_{lm}k_{lm}^{(0)}\sum_{f}H_{l}^{m}\cos{( \s
t+\ps_{lmpq})}.\eqf

It is worthwhile noting that \rfr{trans} expresses a transformation of
coordinates from the  geocentric equatorial frame rotating
with the Earth to the geocentric inertial
frame  $K\{x,
\ y,\ z\}$. This fact will reveal itself of paramount importance
in evaluating the frequencies of the tidal perturbations on the Keplerian
elements.

Up to now the response of the Earth to the forcing tidal potential has been
considered as perfectly elastic; if it had been so, there would be
 no delay between
 the Earth's tidal
bulge and the Moon in the sense that when the latter passes at the
observer's meridian, say,  at $A$ the tidal bulge reaches $A$ at
exactly the same time. The reality is quite different since
complicated mechanisms of energy dissipation in the interior of
the Earth \ct{ref:var} makes its response to depart from the
perfectly elastic behavior previously sketched. A phase lag with
respect to the external lunisolar potential must be introduced in
the sense that the tidal bulge reaches the observer's meridian at
$A$ after a certain time $\D t$ with respect to the passage at $A$
of Moon. It is generally assumed that it is the Earth's mantle to
exhibit an anelastic behavior at the various frequencies of the
tide generating potential; this topic is not yet well understood,
but if one wants to account for the phase lag introduced by it, he
or she has to adopt complex frequency-dependent Love numbers
\ct{ref:mat}.
 In order to give a quantitative estimation also of
the lag angle   it has been decided to adopt the values of
ref.\ct{ref:mat} quoted in ref.\ct{ref:iers}; indeed, the authors
of ref.\ct{ref:deh} do not quote the imaginary part of
$k_{lm}^{(0)}(f)$ and
 the discrepancies with their values for $\vass{k_{lm}^{(0)}(f)}$   amount
to only $0.1\ \%$. Moreover,  concerning the effects
 of mantle's anelasticity, the
results of ref.\ct{ref:deh} are not less uncertain than the other
ones.

An equivalent form of \rfr{clm} and
\rfr{slm}, for a single constituent of frequency $\s$, which reveals
 useful in handling with complex quantities is
\eqi C_{lmf}-iS_{lmf}=\et A_{lm}k_{lm}^{(0)}H_{l}^{m}(f)e^{i\s
t},\lb{comp}\eqf with $\et=-i$ if $l-m$ is  odd  and $\et=1$ if
$l-m$ is even. If the anelasticity of the Earth's mantle is to be
accounted for writing
$k^{(0)}_{lm}(f)=k^{R}_{lm}(f)+ik^{I}_{lm}(f)$, \rfr{comp} becomes
$$ C_{lmf}-iS_{lmf}=\et
A_{lm}H_{l}^{m}(f)\times\left[(k^{R}_{lm}\cos{\s
t}-k^{I}_{lm}\sin{\s t})\right.+$$ \eqi \left.+i(k^{R}_{lm}\sin{\s
t}+k^{I}_{lm}\cos{\s t})\right],\lb{comp2}\eqf or
\begin{eqnarray} C_{lm} &=&A_{lm}\sum_{f}H_{l}^{m}
{\mc{k^{R}_{lm}\cos{\s t}-k^{I}_{lm}\sin{\s t}}{
k^{R}_{lm}\sin{\s t}+k^{I}_{lm}\cos{\s t}}},\lb{clmnew}\\
 S_{lm} & = & A_{lm} \sum_{f} H_{l}^{m}
{\mc{-k^{R}_{lm}\sin{\s t}-k^{I}_{lm}\cos{\s t}}
{k^{R}_{lm}\cos{\s t}-k^{I}_{lm}\sin{\s
t}}},\lb{slmnew}\end{eqnarray} If \rfr{clmnew} and \rfr{slmnew}
are put into \rfr{kaulino}, it is straightforward to obtain for it
\eqi S_{lmpq}=A_{lm}\sum_{f}H_{l}^{m} k^{(0)}_{lm}(f)\cos{(\s
t+\ps_{lmpq}-\d_{lmf})},\lb{lellona}\eqf with \eqi
k^{(0)}_{lm}(f)=\sqrt{{k_{lm}^{R}(f)}^2+{k_{lm}^{I}(f)}^2},\eqf
\eqi \tan{\d_{lmf}}=\rp{k^{I}_{lm}(f)}{k^{R}_{lm}(f)}.\eqf The
factor $\d_{lmf}$ is the phase lag of the response of the solid
Earth with respect to the tidal constituent of degree $l$, order
$m$ and circular frequency $\s$ induced by the anelasticity in the
mantle: notice that if $k_{lm}^{I}(f)=0$, also $\tan{\d_{lmf}}=0$.
An inspection of the values quoted in \ct{ref:iers} shows that, at
the low frequencies of the zonal constituents of the tide
generating potential, the role played by the mantle's anelasticity
is more relevant than in the diurnal and semidiurnal bands.
Indeed, while for the tesseral and sectorial tides one finds
$k_{21}^I=-0.00144$ and $k_{22}^I=-0.00130$ for all the
frequencies, in the zonal band $k_{20}^I(f)$ varies from
$-0.00541$ to $-0.00192$.

\Rfr{kaula}, with the $S_{lmpq}$ given by  \rfr{lellona},
 is the dynamical, non-central part of the geopotential due to the
response of the solid Earth to the forcing lunisolar tidal
action.  In the linear
 Lagrange equations of perturbation theory for the rates of change of the
Keplerian elements it can play
 the role of the perturbative term $\mtc{R}$.
It is an observed fact that the secular motions are the dominant perturbation
in the elements $\og,\ \O,\ {\mtc{M}}$ of
geodetically useful satellites. So, taking as constants $a,\ e,\
i$ and consider as linearly variable in time
 $\og, \O,\ {\mtc{M}}$ and $\th$,
 apart from $\s t$, the following expressions, at first order, may be
worked out
$$\D\O_f=\rp{g}{na^2\sqrt{1-e^2}\sin{i}}
\sum_{l=0}^{\infty}\sum_{m=0}^{l}
\left({\rp{R}{r}}\right)^{l+1}\times$$ \eqi\times A_{lm}
\sum_{p=0}^{l}
\sum_{q=-\infty}^{+\infty}\dert{F_{lmp}}{i}G_{lpq}\rp{1}{f_p}
k^{(0)}_{lm}H_{l}^{m}\sin{\ga_{flmpq}},\lb{juun}\eqf
$$\D\og_f=\rp{g}{na^2\sqrt{1-e^2}}
\sum_{l=0}^{\infty}\sum_{m=0}^{l}
\left({\rp{R}{r}}\right)^{l+1}\times$$\eqi\times
A_{lm}\sum_{p=0}^{l} \sum_{q=-\infty}^{+\infty}\left[
\rp{1-e^2}{e}F_{lmp}\dert{G_{lpq}}{e}- \rp{\cos{i}}
{\sin{i}}\dert{F_{lmp}}{i}G_{lpq}\right]\rp{1}{f_p}
k^{(0)}_{lm}H_{l}^{m}\sin{\ga_{flmpq}},\lb{du}\eqf
with \eqi\ga_{flmpq}=(l-2p)\omega+(l-2p+q){\mtc{M}}+m(\O-\th)+\s
t-\d_{lmf},\eqf \eqi f_p=(l-2p)\dot\omega+(l-2p+q)\dot{\mtc{M}}
+m(\dot\O-\dot\th)+\s.\lb{pert}\eqf The indirect, second order
effects due to the oblateness of the Earth \ct{ref:bal} have been
neglected in this section. Due to the secular trend of the
Lense-Thirring effect, only the perturbations whose periods are
much longer than those of the orbital satellites motions,
 which, typically,
amount to a few hours, are to be considered. This implies that in
such terms the rate of the mean anomaly does not appear and the
condition \eqi l-2p+q=0\lb{boul}\eqf must hold. Moreover, if  the
effect of Earth's diurnal rotation, which could introduce
periodicities of the order of 24 hours, is to be neglected, one
must retain only those terms in which  the non-negative
 multiplicative coefficient $j_1$ of the Greenwich sidereal time in $\s t$
coincides to  the order $m$ of the tidal constituent considered:
 in this way in $f_p$ the contributions of
$\dot \th$ are equal and opposite, and cancel out.
 With these bounds on $l,\ m,\ p$ and $q$ the circular
frequencies of the perturbations of interest  become \eqi f_p=\dot
\G_f+(l-2p)\dot \og+m\dot \O\lb{frf}\eqf in which \eqi \dot
\G_f=(j_2-m) \dot s+j_3 \dot h+j_4 \dot p+j_5 \dot N^{'}+j_6 \dot
p_s. \eqf In the performed calculation only the degree $l=2$
constituents have been considered due to the smallness of the
$k_{3m}^{(0)}$ and $H_{3}^{m}(f)$. For $l=2$, $p$ runs from $0$ to
$2$, and so, in virtue of the condition
 $l-2p+q=0$,
$q$ assumes the values $-2,\ 0,\ 2$. From an inspection of the
table of the eccentricity functions $G_{lpq}(e)$ in \ct{ref:kau}
it turns out that $G_{20-2}=G_{222}=0$, while
$G_{210}=(1-e^2)^{-3/2}$. For this combination of $l,\ p$ and $q$
the condition $l-2p=0$ holds:  the frequencies of the
perturbations are, in this case, given by\eqi f_p =\dot
 \G_f +m\dot \O\lb{frt}.\eqf
Passing from $\s$ to $\dot \G_f +m\dot \O$, as previously noticed,
is equivalent to leaving the
 Earth-fixed frame of reference
for the inertial one,
in which the relevant physical feature is the
 orientation of the tidal bulges  with respect to the orbital plane of
satellite. Indeed, $\dot \G_{f}+m \dot \O$ could be
considered as the relative frequency of motions of the
tidal bulge and the orbital plane, both viewed in the inertial frame.
$\dot\G_f$ is the opposite of
 the inertial rate of the tidal bulge obtained from the
Earth-fixed one
subtracting the Earth's rotation rate $\dot \th$ times $m$ in order to account
for the symmetry of the bulges.
The
periodicities  introduced in the perturbations
 by \rfr{frt} are, in general, much longer
that the ones present in the tidal potential because, while the latter
are dominated by $j_1\dot \th$ due to the rotation of the Earth-fixed frame,
 the former are
mainly determined by the precessional  rates $m\dot\O$
 of the orbital frames of the different satellites in the inertial frame.
This explain the major variety of periodicities in the orbital
elements' perturbations of the near-Earth's satellites with
respect to the tides sensed by an observer on the Earth surface
which are mainly concentrated in the diurnal and semidiurnal
bands. In  \rfrs{juun}{du} it is worthwhile noting that the
amplitudes of the perturbations  are inversely proportional to the
frequency of the perturbation $f_p$. This implies that some tidal
constituent which in the  Earth-fixed frame  has an high
frequency, in the inertial frame may induce relevant perturbations
because its frequency  greatly decreases; this fact could
compensate an eventual small value of its coefficient in the
harmonic expansion of the tide generating potential (calculated in
the terrestrial frame). In other words, constituents which on the
Earth's surface would produce small tides may become important in
perturbing the orbits of near-Earth' s satellites due to the
changes in their frequencies when viewed in the inertial frame.
\subsection{Discussion of the numerical results}
In view of their direct implication in \rfr{unolt}, in the
following we shall focus our attention to the nodes of \lg\  and
\lgg\ and the perigee of \lgg. In Table \ref{unod}, Table
\ref{dued}, and Table \ref{tred}
 the  results for them are shown;
  since the
observable quantity for $\og$ is
 $ea\dot \og$ \ct{ref:ciu96}, the calculation for the perigee of
LAGEOS, due to the notable smallness of the  eccentricity of its orbit, have
 not been performed.
 \begin{table}[ht!]
\caption[Solid Earth tidal perturbations on the node $\O$
 of LAGEOS.]{\footnotesize{Perturbative amplitudes on the node $\O$ of LAGEOS
due
 to solid Earth
tides for $l=2,\ p=1,\ q=0$. In the first column the Doodson
number of each constituent is quoted followed by the Darwin's
name, when it is present. The tidal lines are listed in order of
decreasing periods. The coefficients $H_l^m(f)$ are those recently
calculated by Roosbeek and multiplied by suitable normalization
factors (IERS standards) in order to make possible a comparison
with those of Cartwright and Edden. $\tan{\d_{lmf}}$ expresses the
phase lag of the solid Earth response with respect to the tidal
potential due to the anelasticity in the mantle. The periods are
in days, the amplitudes in mas and the $H_{l}^{m}(f)$ in meters.
}} \label{unod}
\begin{center}
\begin{tabular}{llllll}
\noalign{\hrule height 1.5pt} Tide & Period &  Amplitude & $k_2$
Love number & $H_{l}^{m}(f)$ &
$\tan{\d_{lmf}}$\\
\hline 055.565 & 6,798.38 & -1,079.38 & 0.315 & 0.02792 & -0.01715
\\  055.575 & 3,399.19 & 5.23 & 0.313 & 0.000272 & -0.015584
\\  056.554 $S_a$ & 365.27 & 9.96 & 0.307 & -0.00492 &
-0.01135 \\ 057.555 $S_{sa}$ & 182.62 & 31.21 & 0.305 & -0.03099 &
-0.01029 \\  065.455 $M_m$ & 27.55 & 5.28 & 0.302 & -0.03518 &
-0.00782 \\
075.555 $M_{f}$ & 13.66 & 4.94 & 0.301 & -0.06659 & -0.007059 \\
\\ 165.545 & 1232.94 & -41.15 & 0.259 & -0.007295 & -0.00554\\
165.555 $K_1$& 1,043.67 & 1,744.38 & 0.257 & 0.3687012 & -0.0055933 \\
165.565 & 904.77 & 203.02 & 0.254 & 0.050028 & -0.005653\\
 163.555 $P_1$ & -221.35 & 136.44 & 0.286 & -0.12198 &
-0.005017\\ 145.555 $O_1$& -13.84& 19 & 0.297 & -0.26214 &
-0.00484 \\  135.655 $Q_1$& -9.21& 2.42& 0.297 & -0.05019 &
-0.00483 \\ \\ 274.556 & -1,217.55 & 1.68 & 0.301 & 0.000625 &
-0.00431 \\  274.554 & -1,216.73 & -6.63 & 0.301 & -0.00246 &
-0.00431\\  275.555 $K_2$ & 521.835 & -92.37 & 0.301 & 0.0799155 &
-0.00431 \\  273.555 $S_2$& -280.93 & 182.96 & 0.301 & 0.2940 &
-0.00431 \\  272.556 $T_2$& -158.80 & 6.04 & 0.301 & 0.0171884 &
-0.00431 \\  255.555 $M_2$& -14.02 & 19.63 & 0.301 & 0.6319 &
-0.00431\\  245.655 $N_2$& -9.29 & 2.49 & 0.301 & 0.12099 &
-0.00431\\
\noalign{\hrule height 1.5pt}
\end{tabular}
\end{center}
\end{table}
\begin{table}[ht!]
\caption[Solid Earth  tidal perturbations on the node $\O$
 of LAGEOS\ II.]{\footnotesize{Perturbative
 amplitudes on the node $\O$ of LAGEOS\ II due to solid Earth
tides for $l=2,\ p=1,\ q=0$. In the first column the Doodson
number of each constituent is quoted followed by the Darwin's
name, when it is present. The tidal lines are listed in order of
decreasing periods. The coefficients $H_l^m(f)$ are those recently
calculated by Roosbeek and multiplied by suitable normalization
factors (IERS standards) in order to make possible a comparison
with those of Cartwright and Edden. $\tan{\d_{lmf}}$ expresses the
phase lag of the solid Earth response with respect to the tidal
potential due to the anelasticity in the mantle. The periods are
in days, the amplitudes in mas and the $H_{l}^{m}(f)$ in meters.}}
\label{dued}
\begin{center}
\begin{tabular}{llllll}
\noalign{\hrule height 1.5pt}
 Tide & Period & Amplitude & $k_2$
Love number & $H_{l}^{m}(f)$ &
$\tan{\d_{lmf}}$\\
\hline 055.565  & 6,798.38  & 1,982.16  & 0.315 & 0.02792  &
-0.01715 \\  055.575  & 3,399.19 & -9.61 & 0.313 & -0.000272
& -0.015584 \\
056.554 $S_a$  & 365.27  &  -18.28  & 0.307 & -0.00492  & -0.01135 \\
057.555 $S_{sa}$ & 182.62   & -57.31  & 0.305 & -0.03099  & -0.01029 \\
 065.455 $M_m$  & 27.55  & -9.71 &  0.302 & -0.03518 &
-0.00782 \\
075.555 $M_f$  &  13.66 & -9,08   & 0.301 & -0.06659   & -0.007059 \\
\\ 165.565  & -621.22   & -58.31  & 0.254 & 0.050028  & -0.005653
\\  165.555  $K_1$& -569.21  &-398  & 0.257 &
0.3687012  & -0.0055933 \\
 165.545  & -525.23  & 7.33  & 0.259 &
-0.007295  & -0.005541
\\  163.555  $P_1$& -138.26  & 35.65  & 0.286 & -0.1219 &
-0.005017 \\  145.555  $O_1$&-13.33& 7.66 & 0.297
& -0.26214  & -0.00484 \\
135.655  $Q_1$& -8.98& 0.98 & 0.297 & -0.05019  & -0.00483  \\
\\ 275.555  $K_2$& -284.6 &-92.51  & 0.301 & 0.079915  &
-0.004318 \\  274.556  &-159.96  &-0.40  & 0.301 & 0.000625 &
-0.00431 \\  274.554  & -159.95 & 1.6  & 0.301 & -0.00246 &
-0.0043 \\  273.555  $S_2$&-111.24  & -133.04  & 0.301 & 0.29402 &
-0.00431 \\  272.556 $T_2$& -85.27 & -5.96 & 0.301 & 0.017188 &
-0.00431\\  255.555  $M_2$&-13.03 & -33.05 & 0.301 & 0.6319  &
-0.004318 \\  245.655 $N_2$&-8.84
& -4.35 & 0.301 & 0.12099  & -0.00431  \\
\noalign{\hrule height 1.5pt}
\end{tabular}
\end{center}
\end{table}
\clearpage
\begin{table}[ht!]
\caption[Solid Earth  tidal perturbations on the perigee $\og$
 of LAGEOS\ II.]{\footnotesize{Perturbative
 amplitudes on the perigee $\og$ of LAGEOS\ II due to solid Earth
tides for $l=2,\ p=1,\ q=0$. In the first column the Doodson
number of each constituent is quoted followed by the Darwin's
name, when it is present. The tidal lines are listed in order of
decreasing periods. The coefficients $H_l^m(f)$ are those recently
calculated by Roosbeek and multiplied by suitable normalization
factors (IERS standards) in order to make possible a comparison
with those of Cartwright and Edden. $\tan{\d_{lmf}}$ expresses the
phase lag of the solid Earth response with respect to the tidal
potential due to the anelasticity in the mantle. The periods are
in days, the amplitudes in mas and the $H_{l}^{m}(f)$ in meters.}}
\label{tred}
\begin{center}
\begin{tabular}{llllll}
\noalign{\hrule height 1.5pt} Tide & Period  & Amplitude  & $k_2$
Love number & $H_{l}^{m}(f)$ & $\tan{\d_{lmf}}$\\ \hline
 055.565  & 6,798.38  & -1,375.58  & 0.315 & 0.02792  & -0.01715
\\  055.575  & 3,399.19 & 6.66 & 0.313 & -0.000272 &
-0.015584 \\
056.554 $S_a$  & 365.27  &  12.69  & 0.307 & -0.00492  & -0.01135 \\
057.555 $S_{sa}$ & 182.62   & 39.77  & 0.305 & -0.03099  &
-0.01029 \\ 065.455 $M_m$  & 27.55  & 6.74  &  0.302 & -0.03518 &
-0.00782
\\
075.555 $M_f$  &  13.66 & 6.30  & 0.301 & -0.06659   & -0.007059 \\
\\ 165.565  & -621.22   & 290.43  & 0.254 & 0.050028  &
-0.005653 \\
165.555  $K_1$& -569.21  & 1,982.14 & 0.257 & 0.3687012  &
-0.0055933 \\ 165.545  & -525.23  & -36.52  & 0.259 & -0.007295  &
-0.005541
\\
163.555  $P_1$& -138.26  & -177.56  & 0.286 & -0.1219  & -0.005017
\\ 145.555  $O_1$&-13.33 & -38.16 & 0.297 & -0.26214  & -0.00484
\\
135.655  $Q_1$& -8.98& -4.92 & 0.297 & -0.05019  & -0.00483  \\
\\ 275.555  $K_2$& -284.6 & -88.19 & 0.301 & 0.079915  &
-0.004318 \\  274.556  &-159.96  &-0.38  & 0.301 & 0.000625 &
-0.00431 \\  274.554  & -159.95 & 1.52  & 0.301 & -0.00246 &
-0.0043 \\  273.555  $S_2$&-111.24  &-126.83   & 0.301 & 0.29402 &
-0.00431 \\  272.556 $T_2$& -85.27 & -5.68 & 0.301 & 0.017188 &
-0.00431\\  255.555  $M_2$&-13.03& -31.9 & 0.301 & 0.6319 &
-0.004318 \\  245.655  $N_2$&-8.84&-4.15
& 0.301 & 0.12099  & -0.00431  \\
\noalign{\hrule height 1.5pt}
\end{tabular}
\end{center}
\end{table}
The tidal lines for which the analysis was performed have
 been chosen  also in order to make a comparison with the perturbative
 amplitudes of the ocean tides
based on the results of EGM96 gravity model \ct{ref:lem} which
will be shown in the next sections. It is interesting to notice
that the periods of many of these tidal perturbations
 are almost equal to or
longer than an year, a time interval in which their effect may alias the
Lense-Thirring effect, if the observations are taken on a such temporal
scale; indeed, only if the data were acquired on  time intervals of the order
of many years it should be
possible to clearly resolve the action of these tidal lines with respect to
 the secular
Lense-Thirring trend. And also in this case, it could remain the
aliasing effect of the 18.6-year and 9.3-year  tides. Up to now,
in refs.\ct{ref:ciu97,ref:ciu98} the authors have analyzed the
data acquired on a 4 years interval for a suitable combination of
residuals showing that many tidal signals with periods shorter
than it can be resolved by means of suitable fits.

The results presented in Table \ref{unod}, Table \ref{dued} and
Table \ref{tred} can be compared to those of
\ct{ref:dow,ref:bercar}. See also \ct{ref:chr}. In doing so it
must be kept in mind that both these authors have neglected not
only the contribution of $k_{lm}^{+}$ but also the anelasticity
and the frequency dependence of the Love numbers $k_{lm}^{(0)}$.
Moreover, Dow includes in his analysis also the indirect
influences of the oblateness of the Earth  for $O_1$, $Q_1$, $M_2$
and $N_2$. Obviously, in their analyses the LAGEOS\ II is not
present since it was launched only in 1992. Another important
factor to be considered is the actual sensitivity in measurements
of $\O$ and $\og$, in the sense that the eventual discrepancies
between the present results and the other ones must be not smaller
than the experimental error in the Keplerian elements if one wish
to check the theoretical assumptions behind the different models
adopted.
 Carpino has analyzed
 the inclination and the node of LAGEOS only.  His value
for the important zonal 18.6-year tide is $-1087.24$ mas, while
Table \ref{unod} gives $-1,079.38$ mas; the difference amounts to
$7.86$ mas, the $0.72$ $\%$ of the ``elastic'',
frequency-independent Carpino's value. Considering that for the
node $\O$ of LAGEOS the present accuracy is of the order of the
mas, $7.86$ mas could be in principle detected, allowing for a
discrimination between the different models adopted in the
calculation. For the $K_1$ tide, one of the most powerful
constituent in perturbing the satellites' orbits,  Table
\ref{unod} quotes  $1,744.38$ mas against $2144.46$ mas of
Carpino's result; the gap is $400.08$ mas, the $18.6$ $\%$ of
Carpino. In the sectorial band, the present analysis quotes  for
the $K_2$ $-92.37$ mas and Carpino $-97.54$ mas; there is a
difference of $5.17$ mas, the $5.3$ $\%$ of the Carpino's value.
It must pointed out that there are other tidal lines for which the
difference falls below the mas level, as is the case for the
9.3-year tide. As it could be expected, the major differences
between the present ``anelastic'',
 frequency-dependent
calculations and
 the other ones based on a single, real value for the Love number
$k_2$ lie in the diurnal band: in it the contribution of
anelasticity is not particularly relevant, but, as already pointed
out, the elastic part of $k_{21}^{(0)}$ is strongly dependent on
frequencies of the tidal spectrum. May be interesting to notice
that when the calculation  have been  repeated with the same value
$k_2=0.317$ adopted by Carpino,  his results have been obtained
again.

Another feature which characterizes this study is the calculation of the phase
lag of the solid Earth' s  response with respect
 to the tide generating potential.
In   Table \ref{unod}, Table \ref{dued} and
 Table \ref{tred} $\tan{\d_{lmf}}$ is always negative and does not exceed
$10^{-2}$ in absolute value showing that the response of the
Earth is slightly retarded with respect to the forcing lunisolar tidal
potential. In absolute value, $\tan{\d_{20f}}$ is greater than
 $\tan{\d_{21f}}$ and $\tan{\d_{22f}}$ of one order of magnitude, confirming that
the behavior of the solid Earth's response is more influenced by the
anelasticity in the zonal band that in the diurnal and semidiurnal ones.

Up to now the effects induced by the Earth's flattening and the
Earth's rotation have been neglected. If they have to be analyzed
it is necessary to take in \rfr{ela} the part\eqi Re\
g\sum_{l=2}^{\infty}\sum_{m=0}^{l}\sum_f
H_{l}^{m}(f)\left(\rp{R}{r}\right)^{l+3} k_{lm}^{+}(f)
Y_{l+2}^{m}(\f, \l)\lb{kappa+}\eqf and work out it in the same
manner as done for the spherical non-rotating Earth's
contribution. In applying the transformation to the orbital
elements given by \rfr{trans} one has to substitute everywhere $l$
with $l+2$. So the equations for the perturbations on the node and
the perigee become
$$\D\O^{+}_f=\rp{g}{na^2\sqrt{1-e^2}\sin{i}}
\sum_{l=0}^{\infty}\sum_{m=0}^{l}
\left({\rp{R}{r}}\right)^{l+3}A_{l+2 \ m}H_{l}^{m}\times $$ \eqi
\times k^{+}_{lm}\rp{1}{f_p}\sum_{p=0}^{l}
\sum_{q=-\infty}^{+\infty}\dert{F_{l+2\ mp}}{i}G_{l+2\ pq}
\sin{\ga_{fl+2\ mpq}},\lb{juun+}\eqf
$$ \D\og^{+}_f=\rp{g}{na^2\sqrt{1-e^2}}
\sum_{l=0}^{\infty}\sum_{m=0}^{l}
\left({\rp{R}{r}}\right)^{l+3}A_{l+2\ m}
k^{+}_{lm}H_{l}^{m}\rp{1}{f_p}\times $$ \eqi\times\sum_{p=0}^{l}
\sum_{q=-\infty}^{+\infty}\left[ \rp{1-e^2}{e}F_{l+2\
mp}\dert{G_{l+2\ pq}}{e}-\rp{\cos{i}} {\sin{i}}\dert{F_{l+2\
mp}}{i}G_{l+2\ pq}\right]\sin{\ga_{fl+2\ mpq}}.\lb{beh}\eqf The
corrections  induced by the Earth's flattening and rotation, due
to the smallness of $k^{+}_{lm}$, have been calculated only for
those tidal lines which turned out to be the most effective in
perturbing the node and the perigee of LAGEOS and LAGEOS\ II, i.e.
the zonal 18.6-year tide and the $K_1$. Adopting the values quoted
in ref.\ct{ref:deh} for $k^{+}_{lm}$ and \rfrs{juun+}{beh} it is
possible to obtain the results summarized in Table \ref{k+}.
\begin{table}[h!]\caption[Corrections due to Earth's flattening and rotation to
solid tidal perturbations.]{\footnotesize{Corrections, in mas, to
the Earth solid tidal perturbations on the nodes of LAGEOS and the
perigee of LAGEOS\ II due to the Earth's flattening and Earth's
rotation for $l=2,\ p=2,\ q=0$. The values adopted for
$k_{2m}^{+}$ are those quoted in ref.\ct{ref:deh}.  }} \label{k+}
\begin{center}
\begin{tabular}{lllll}
\noalign{\hrule height 1.5pt} Tide & $\D\O_{{\rm LAGEOS}}$
 & $\D\O_{{\rm LAGEOS\ II}}$  & $\D\og_{{\rm LAGEOS\ II}}$ &
$k_{2m}^{+}$\\
\hline 055.565  & 0.43 & 1.28 & 0.19 & -0.00094 \\  165.555
$K_1$ & -0.97 & -9.49 & 0.85 & -0.00074 \\
\noalign{\hrule height 1.5pt}
\end{tabular}
\end{center}
\end{table}
The quoted values, with the exception of the node of
 LAGEOS\ II, fall below the mas level turning out, at the
present, undetectable.
\subsection{The mismodelling in the nodes and the perigee}
Concerning the  errors in the values released in Table \ref{unod},
Table \ref{dued}, and Table \ref{tred}, the major source of
uncertainty in the perturbative amplitudes lies in the Love
numbers $k_{2m}^{(0)}(f)$. and the orbital injection errors $\d i$
affecting the inclination $i$.

About the latter, by assuming  $\d i=0.5$ mas \ct{ref:ciu89}, we
have calculated $\vass{\derp{A(\O)}{i}}\d i$ and
$\vass{\derp{A(\og)}{i}}\d i$ for $055.565,\ K_1,\ $and $S_2$
which are the most powerful tidal constituents in perturbing
LAGEOS and LAGEOS\ II orbits. The results are of the order of
$10^{-6}$ mas, so that we can neglect the effect of uncertainties
in the inclination determination.

Concerning the Love numbers $k^{(0)}_{2m}(f)$, from a preliminary
point of view, theoreticians claim that they are able to compute
them with an accuracy of $10^{-4}$ using the specified Earth
model. But the Earth models may differ. Probably uncertainty in
Earth models may bring errors in computation of the Love numbers
of the second order about $10^{-3}$ ({ L. Petrov,} private
communication, 1999). First of all, this means that the level of
accuracy of 1 mas, i.e. $10^{-9}$ rad, on $\O$ and $\og$ given by
\rfr{juun} and \rfr{du} can be well reached, since the uncertainty
in $H_{l}^{m}(f)$ amounts  to $10^{-7}$ m. Second, assuming a mean
value of $0.3$ for $k_2$, it can be stated that the uncertainty in
the solid Earth tidal perturbation due to a single tide generating
potential's constituent amount to almost $0.3\ \%$ for $\O$ and
$\og$.

More specifically, we have assessed the uncertainties in them by
calculating for certain tidal constituents the factor $\d
k_2/{k_2}$; ${k_2}$ is the average on the values released by the
most reliable models and $\d k_2$ is its standard deviation.
According to the recommendations of the Working Group of
Theoretical Tidal Model of the Earth Tide Commission ({\bf
http://www.astro.oma.be/D1/EARTH$\_$TIDES/wgtide.html}), in the
diurnal band we have chosen the values released in
refs.\ct{ref:mat,ref:iers} and the two sets of ref.\ct{ref:deh}.
For the zonal and sectorial bands we have included also the
results of ref.\ct{ref:wan94}. The uncertainties calculated in the
Love numbers $k_2$ span from 0.5$\%$ to 1.5$\%$ for the tides of
interest. However, it must be noticed that the worst known Love
numbers are those related to the zonal band of the tidal spectrum
due to the uncertainties in the anelasticity of the Earth' s
mantle. These results have been obtained in order to calculate the
mismodelled amplitudes of the solid tidal perturbations
$\d\O^{{\rm I}},\ \d\O^{{\rm II}}$ and $\d\og^{{\rm II}}$ for the
nodes of \lg\ and \lgg\ and the perigee of \lgg; they have been
subsequently compared with the gravitomagnetic precessions over 4
years $\D\O^{{\rm I}}_{{\rm LT}}=124$ mas, $\D\O^{{\rm II}}_{{\rm
LT}}=126$ mas and $\D\og^{{\rm II}}_{{\rm LT}}=-228$ mas. In Table
\ref{mismo1} we have quoted only those tidal lines whose
mismodelled perturbative amplitudes are greater than $1\%$ of the
gravitomagnetic perturbations. It turns out that only 055.565
18.6-year and $K_1$ exceed this cutoff.
\begin{table}[ht!]\caption[Mismodeled solid tidal perturbations on
nodes of LAGEOS and LAGEOS\ II and perigee of LAGEOS
II.]{\footnotesize{Mismodeled solid tidal perturbations on the
nodes $\O$ of LAGEOS  and LAGEOS\ II and the perigee $\og$ of
LAGEOS\ II compared to their gravitomagnetic precessions over 4
years $\D\O^{{\rm I}}_{{\rm LT}}=124$ mas \ $\D\O^{{\rm II}}_{{\rm
LT}}=126$ mas \ $\D\og^{{\rm II}}_{{\rm LT}}=-228$ mas. Only those
constituents which exceed the 1 $\%$ cutoff have been quoted. The
ratios are in percent and the $\delta X$ in mas.}} \label{mismo1}
\begin{center}
\begin{tabular}{llllllll}
\noalign{\hrule height 1.5pt}
 Tide
      & $\frac{\d k_2}{k_2}$  & $\d\O^{{\rm I}}$  & $\frac{\d\O^{{\rm I}}}{\D\O^{{\rm I}}_{{\rm LT}}}$  &
$\d\O^{{\rm II}}$  & $\frac{\d\O^{\rm II}}{\D\O^{\rm II}_{\rm
LT}}$  & $\d\og^{{\rm II}}$  & $\frac{\d\og^{{\rm
II}}}{\D\og^{{\rm II}}_{{\rm LT}}}$ \\ \hline
 055.565 & 1.5 & -16.5 & 13.3 & 30.3 & 24 & -21 & 9.2 \\
165.555 $K_1$& 0.5 & 9 & 7.2 & -2 & 1.6 & 10.2 & 4.4 \\
\noalign{\hrule height 1.5pt}
\end{tabular}
\end{center}
\end{table}

\subsection{Gravitational potential of the ocean Earth tides}
One of the early  attempts to explain the phenomenon of fluid
tides \ct{ref:zah} in its globality is the equilibrium theory
\ct{ref:def,ref:mel} by Newton. It
 is based on the assumption that at every time
the free water surface of the oceans coincides exactly with  the  spheroidal
equipotential surface at $r=R$ due to the combined action of the Earth's proper
gravity
and the lunisolar tidal potential. Differently stated,
 the oceanic tidal bulge at every
instant coincides exactly with the envelope of the forces produced
by the tide generating potential at the Earth's surface. At every
tidal constituents $\Phi(f)$ corresponds an equilibrium partial
tide $\et_f$. Referring to \rfr{ct} it can be write
\eqi\et_{f}=\rp{\Phi_{f}(R)} {g}=A_{lm}H_{l}^{m}
P_{l}^{m}(\sin{\f})\cos(\s t+m\l+c_{lm})\lb{etide}.\eqf For $l=2$
and $m=0$ \rfr{etide} represents a standing wave which crests at
$t=0$, while for $l=2$ and $m=1,2$ it refers to running waves
around the oceans from East to West. If the equilibrium theory had
been valid, it would mean that the water masses  acted upon by the
tidal potential are  deprived almost entirely of their inertia in
the sense that they would adapt instantaneously to their
equilibrium positions on the  equipotential surface which, in
fact,  changes in time at the frequencies of the lunisolar tidal
 potential.
The reality is quite different, because of the complex hydrodynamical behavior
 of the oceans; at every equilibrium partial tide $\et_f$ corresponds an effective
partial tide \ct{ref:henmun} \eqi \z_f=\x_f(\vth, \l) \cos{[ \s
t-\d_f(\vth, \l)]}, \eqf where
 $\x_f(\vth, \l)$
and $\d_f(\vth, \l)$ \ct{ref:sch} are defined as harmonic
 constants: $\x_f(\vth, \l)$ is
half the difference between high water and low water for a given place while
$\d_f(\vth, \l)$ is the phase lag of the ocean tide with respect to the
equilibrium one  when the phase of the latter is calculated
at Greenwich meridian $\l=0$: the delay at any other meridian is obtained
simply adding $m\l$ to $\d_f(\vth, \l)$.
Moreover, $\z_f$ at every time $\ol{t}$ crests on the
``cotidal'' line $\s \ol{t}= \d_f(\vth, \l)$; in general, referring to a given point
$(\vth,\ \l)$, e.g., on the shoreline, the cotidal lines related
to two consecutive instants may depart from the shore into the sea or vice versa,
giving rise to
outcoming and incoming waves with respect to that place.

Such departures from the equilibrium theory are due to several
reasons. First of all, the fluid  elements can freely flow and
once they had been put in motion, their notable inertia prevent
them to change instantaneously their state of motion making them
stop at the equipotential surface. Second, dissipative phenomena
and non-linear interactions among the various partial tides and
other ocean currents  must be taken in account. This implies that
the response of the water masses to the forcing lunisolar
potential presents a phase lag with respect to it: the sea surface
is not an equipotential one so that tangential forces acts on the
water particles generating tidal currents.
 The equilibrium theory may be considered an
almost good
approximation to the reality only at very low frequencies of the tidal
potential, like as the Moon and the Sun would remain fix in the space with
respect to the Earth.

The equations of motion governing the complex hydrodynamics of the
tidal currents are the so called Lagrange Tidal Equations (LTE)
\ct{ref:neu,ref:pek,ref:hen72,ref:hen73,ref:gil}, which, in
general, are not linear. The simplest form for them is obtained
considering a spherical, rotating Earth entirely covered by an
ocean made of a perfect and incompressible fluid; if only the
tangential motion is considered, neglecting the squares and the
 cross products of the velocity field's components, it is possible to obtain
the following linearized LTE
\begin{eqnarray}
\derp{v_{\vth}}{t}&=&+2\orot v_{\l}\cos{\vth}-\rp{g}{R}\derp{(
\z-{\et})}{\vth},\\
\derp{v_{\l}}{t}&=&-2\orot v_{\vth}\cos{\vth}-\rp{g}{R\sin{\vth}}\derp{(
\z-{\et})}{\l},\\
\derp{\z}{t}  &=&  -\rp{1}{R\sin{\vth}}[\derp{(hv_{\l})}{\l}+
\derp{(h\sin{\vth}\ v_{\vth})}{\vth}],\lb{lte}
\end{eqnarray}
 in which\\
$\vth$\ \ \ terrestrial colatitude. \\
$h$\ \ \ depth of the ocean, $h=h(\vth, \l)$, m.\\
$\z$\ \ \ non-equilibrium, ocean partial tide, m.\\
$v_{\vth}$ and $v_{\l}$\ \ \ tangential components of the water's
velocity field.\\
The LTE are linear, and so each ocean partial tide $\z_f$  can be
treated independently of the other ones. The fundamental problem
of the ocean tides consists of the determination of the harmonic
constants for every place on the Earth: this can be obtained, in
principle, resolving the LTE for a given constituent, but this
task is in many cases prohibitive due to the great calculational
complexity
 needed to obtain
realistic results. Another strategy consists of  employing
geodetic
 artificial satellites, as is done in the realization of the various Earth
Gravity Model of the Goddard Space Flight Center among which EGM96
\ct{ref:lem} is the most recent, or in  altimetric measurements
\ct{ref:shu}.

In order to calculate the gravitational effect of the water masses raised
 by the
tidal forces of a given constituent $\Phi_l^m(f)$ it  can be
considered a spherical layer
 with
radius $R$ endowed with a bidimensional surface mass
 density $\m_f(\vth, \l)=\r \z_f(\vth,
\l)$, where $\r$ is the volumetric ocean water density assumed constant,
 and calculate its potential as
\eqi U_f=G  \int{\rp{\m_f(\vth^{'}, \l^{'})}{\vass{{\bf r}-{\bf
r}^{'}}}d \S^{'}}= G  \int{\rp{\r \z_f(\vth^{'},
\l^{'})}{\vass{{\bf r}-{\bf r}^{'}}}d \S^{'}}\lb{pot}.\eqf In
\rfr{pot} the geodetic convention for the potential is used
\ct{ref:kau}: in it $U$ is the potential usually defined in
physics with the sign reversed. Let us rewrite $\z_f$ as
 \eqi\z_f  =  \x_f\cos(\s t-\d_f)=\x_f\cos{\d_f} \cos{\s t}+
\x_f\sin{\d_f} \sin{\s t}.\lb{eta}\eqf The following step consists
of  expanding
 \rfr{eta}
in spherical harmonics \ct{ref:dow,ref:chretal}
\begin{eqnarray}\x_f\cos\d_f &=& \sum_{l=0}^{\infty}\sum_{m=0}^{l}(
a_{lmf}\cos{m\l}+b_{lmf}\sin{m\l})P_{l}^{m}(\sin{\f}),\lb{uuno}\\
\x_f\sin\d_f &=& \sum_{l=0}^{\infty}\sum_{m=0}^{l}(
c_{lmf}\cos{m\l}+d_{lmf}\sin{m\l})P_{l}^{m}(\sin{\f}).\lb{dddue}
\end{eqnarray}
The coefficients of the harmonic expansion, which have the
dimension of lengths, can be calculated if the harmonic constants
are known by means of the following formulas
\begin{eqnarray} a_{lmf}&=&\rp{(2l+1)(2-\d_{0m})}{4\p}\rp{(l-m)!}
{(l+m)!}\times\int\x_f\cos{\d_f}\cos{m\l}P_l^m(\sin{\f})d\S,\lb{sc1}\\
b_{lmf}&=&\rp{(2l+1)(2-\d_{0m})}{4\p}\rp{(l-m)!}
{(l+m)!}\times\int\x_f\cos{\d_f}\sin{m\l}P_l^m(\sin{\f})d\S,\lb{scu}\\
c_{lmf}&=&\rp{(2l+1)(2-\d_{0m})}{4\p}\rp{(l-m)!}
{(l+m)!}\times\int\x_f\sin{\d_f}\cos{m\l}P_l^m(\sin{\f})d\S,\lb{sc3}\\
d_{lmf}&=&\rp{(2l+1)(2-\d_{0m})}{4\p}\rp{(l-m)!}
{(l+m)!}\times\int\x_f\sin{\d_f}\sin{m\l}P_l^m(\sin{\f})d\S.\lb{sc}
\end{eqnarray}\Rfrs{sc1}{sc} are the connection between the
harmonic constants obtained through oceanographic models and the
data acquired by geodetic satellites, in the sense that if the
coefficients $a_{lmf},\ b_{lmf},\ c_{lmf},\ d_{lmf} $, for a given
tidal line, are known from orbital analysis, by means of
\rfr{sc1}-\rfr{sc}, the essential features of that partial tide
may be recovered. With \rfrs{uuno}{dddue}  \rfr{eta} becomes \eqi
\z_f=\sum_{l=0}^{\infty}\sum_{m=0}^{l}(C_{lmf}\cos{m\l}+S_{lmf}\sin{m\l})
P_{l}^{m}(\sin{\f})\lb{etadue}\eqf in which
\begin{eqnarray}C_{lmf} &=& a_{lmf}\cos{\s t}+c_{lmf}\sin{\s t},\lb{betty}\\
S_{lmf} &=& b_{lmf}\cos{\s t}+d_{lmf}\sin{\s
t}\lb{july}.\end{eqnarray} By inserting \rfr{etadue} in \rfr{pot}
and expanding in spherical harmonics also the term $1/\vass{{\bf
r}-{\bf r}^{'}}$, it can be obtained\eqi U_f=4\p G R
\r\sum_{l=0}^{\infty}\sum_{m=0}^{l}\rp{1}{2l+1}
\left({\rp{R}{r}}\right)^{l+1} (C_{lmf}\cos{m\l}+S_{lmf}\sin{m\l})
P_{l}^{m}(\sin{\f}).\lb{step}\eqf In \rfr{step}, using
\rfr{betty}-\rfr{july}, one  can further expand
 $(C_{lmf}\cos{m\l}+S_{lmf}\sin{m\l})$
by introducing the  prograde (westwards) and retrograde waves
\ct{ref:sch}\eqi
C_{lmf}\cos{m\l}+S_{lmf}\sin{m\l}=\sum_{+}^{-}K_{lmf}^{\pm}\cos{(
m \l \pm \s t)} \pm S_{lmf}^{\pm}\sin{(m \l \pm \s
t)}\lb{pollo}.\eqf In \rfr{pollo} and in the following,
expressions like
 $\sum_{+}^{-}A^{\pm}\cos{(a\pm b)}$ must be interpreted as $A^{+}\cos{(a+b)
}+A^{-}\cos{(a-b)}$. The sign $+$ refers to the progressive (westwards) waves
while the
sign $-$ is
for the waves
moving from West to
East. From \rfrs{betty}{july} and \rfr{pollo}
one has\begin{eqnarray} a_{flm}&=&K^+_{flm}+K^-_{flm},\lb{ola}\\
b_{flm}&=&S^+_{flm}-S^-_{flm},\\
c_{flm}&=&S^+_{flm}+S^-_{flm},\\
d_{flm}&=&-K^+_{flm}+K^-_{flm}.\lb{ole}
\end{eqnarray}
From these equations, by defining
\begin{eqnarray} K_{lmf}^{\pm} & =&
 C_{lmf}^{\pm}\cos{\ve_{flm}^{\pm}},\lb{kv}\\
S_{lmf}^{\pm}  &=& C_{lmf}^{\pm}\sin{\ve_{flm}^{\pm}},
\lb{v}\end{eqnarray} it is possible to express the coefficients of
the  expansion of the harmonic constants $a_{lmf},\ b_{lmf},\
c_{lmf},\ d_{lmf}$ in terms of $C_{lmf}^{\pm},\ \ve_{lmf}^{\pm}$
which can be recovered by geodetic satellites
\begin{eqnarray}C_{lmf}^{+}\cos{\ve_{lmf}^{+}}
 &=&  \rp{a_{lmf}-d_{lmf}}{2}\lb{y},\\
C_{lmf}^{-}\cos{\ve_{lmf}^{-}}
 &=&  \rp{a_{lmf}+d_{lmf}}{2}\lb{ii},\\
C_{lmf}^{+}\sin{\ve_{lmf}^{+}}
 &=&  \rp{c_{lmf}+b_{lmf}}{2}\lb{iii},\\
C_{lmf}^{-}\sin{\ve_{lmf}^{-}}
 &=&  \rp{-b_{lmf}+c_{lmf}}{2}\lb{iiii}.\end{eqnarray}
\Rfr{scu} and \rfr{sc} show that, for $m=0$, it turns out
$b_{l0f}=d_{l0f}=0$ identically; this implies through
\rfrs{y}{iiii} that \eqi C_{l0f}^{+}=C_{l0f}^{-}, \lb{zeroo}\eqf
\eqi \ve_{l0f}^{+}=\ve_{l0f}^{-}.\lb{zero}\eqf \Rfrs{zeroo}{zero}
will be used in the next section. It is interesting to notice that
\rfrs{y}{iiii}, or also \rfrs{ola}{ole}, show that if it were
$c_{lmf}=d_{lmf}=0$, the distinction between the prograde and
retrograde waves would disappear and only one kind of waves would
remain, as in the equilibrium theory. Indeed, \rfrs{sc3}{sc} point
out that $c_{lmf}=d_{lmf}=0$ if and only if $\d_f=0$, where $\d_f$
is the phase lag of the ocean response with respect to the
lunisolar tidal constituent due to the ocean hydrodynamics.

\Rfr{step}, with \rfr{pollo} and \rfrs{kv}{v}, becomes\eqi U_f=
4\p G R \r\sum_{l=0}^{\infty}\sum_{m=0}^{l}\sum_{+}^{-}
\left({\rp{R}{r}}\right)^{l+1}\rp{C_{lmf}^{\pm}}{2l+1} \cos{( \s t
\pm m\l - \ve_{lmf}^{\pm} )}P_{l}^{m}(\sin{\f}).\lb{giusi}\eqf
\Rfr{giusi} expresses the gravitational potential of the Earth's
 ocean waters, thought as a spherical layer of mass $\r \z_f$ raised by the
action of a given lunisolar tidal constituent of circular frequency $\s$.

But these enormous water masses act upon the sea floor and the
whole of the solid Earth which are deformed and attracted by them.
This effect has also to be considered. It can be done according to
\rfr{cdd} and introducing the so called load Love numbers
$k^{'}_{l}$ \ct{ref:far,ref:dow,ref:pag}. The global response of
the Earth oceans to the forcing tidal constituent of frequency
$\s$ can be written as $$ U_f=U_{\rm oc}+U_{\rm load}= 4\p G R
\r\sum_{l=0}^{\infty}\sum_{m=0}^{l}\sum_{+}^{-}
\left({\rp{R}{r}}\right)^{l+1}\rp{(1+k^{'}_{l})}{2l+1}\times$$\eqi
\times C_{lmf}^{\pm} \cos{( \s t \pm m\l-
\ve_{lmf}^{\pm})}P_{l}^{m}(\sin{\f}),\lb{pezzo}\eqf with $U_f <
U_{\rm oc}$, i.e. $k^{'}<0$. This fact can be intuitively guessed
by thinking about $U$ in terms of shift of the equipotential
surfaces $h=U/g$: if the sea floor is raised by the tidally driven
water masses the total shift is smaller than it would be if we
considered the solid Earth as completely rigid.
\subsection{Ocean tidal orbital perturbations}
\Rfr{pezzo} can be fruitfully rewritten in a form more suitable
for calculating the perturbations induced on the Keplerian orbital
elements of artificial satellites \ct{ref:fel,ref:goa}. To this
aim the following quantities are
defined\begin{eqnarray}A_{lmf}^{\pm} & =&
 4\p G R \r\rp{(1+k^{'}_{l})}{2l+1}C_{lmf}^{\pm},\\
{\mtc{C}}_{lmf}^{\pm} &=&  A_{lmf}^{\pm}\cos{(\mp \s t \pm \ve_{lmf}^{\pm})},\\
{\mtc{S}}_{lmf}^{\pm} &=&  A_{lmf}^{\pm}\sin{(\mp \s t \pm
 \ve_{lmf}^{\pm})}.\end{eqnarray} $U_f$ becomes
\eqi U_f= \sum_{l=0}^{\infty}\sum_{m=0}^{l}\sum_{+}^{-}
\left({\rp{R}{r}}\right)^{l+1}({\mtc{C}}_{lmf}^{\pm}\cos{m\l}
+{\mtc{S}}_{lmf}^{\pm}
\sin{m\l})P_{l}^{m}(\sin{\f}).\lb{marilena}\eqf \Rfr{marilena} is
formally equal to the expression of the static gravitational
potential of the Earth worked out by Kaula. So, with the same
mathematical menagerie used for the geopotential in
ref.\ct{ref:kau}, it is possible to write \rfr{marilena} in terms
of the Keplerian orbital elements of a test body in the field of
the Earth as \eqi U_f=
\sum_{l=0}^{\infty}\sum_{m=0}^{l}\sum_{+}^{-}
\left({\rp{R}{r}}\right)^{l+1}A_{lmf}^{\pm}\sum_{p=0}^{l}
\sum_{q=-\infty}^{+\infty}F_{lmp}(i)G_{lpq}(e)
{\mc{\cos{\ga_{flmpq}^{\pm}}}{\sin{\ga_{flmpq}^{\pm}}}},\lb{traf}\eqf
in which \eqi\ga_{flmpq}^{\pm}=(l-2p)\omega+(l-2p+q){\mtc{M}}+
m(\O-\th)\pm( \s t-\ve_{lmf}^{\pm})\lb{gamma}.\eqf The equations
for the tidal ocean perturbations may be worked out as already
done for the solid Earth tides in subsect. 4. At first order, one
obtains
$$ \D\O_f=\rp{1}{na^2\sqrt{1-e^2}\sin{i}}
\sum_{l=0}^{\infty}\sum_{m=0}^{l}\sum_{+}^{-}
\left({\rp{R}{r}}\right)^{l+1}A_{lmf}^{\pm}\times$$ \eqi
\times\sum_{p=0}^{l}
\sum_{q=-\infty}^{+\infty}\dert{F_{lmp}}{i}G_{lpq}\rp{1}{f_p}
\mc{\sin{\ga_{flmpq}^{\pm}}}{-\cos{\ga_{flmpq}^{\pm}}}^{l-m \
\textrm{even}}_{l-m\  \textrm{odd}},\lb{juuno}\eqf
$$ \D\og_f=\rp{1}{na^2\sqrt{1-e^2}}
\sum_{l=0}^{\infty}\sum_{m=0}^{l}\sum_{+}^{-}
\left({\rp{R}{r}}\right)^{l+1}A_{lmf}^{\pm}\times$$
\eqi\times\sum_{p=0}^{l}\sum_{q=-\infty}^{+\infty}\left[
\rp{1-e^2}{e}F_{lmp}\dert{G_{lpq}}{e}-\rp{\cos{i}}
{\sin{i}}\dert{F_{lmp}}{i}G_{lpq}\right]\rp{1}{f_p}
\mc{\sin{\ga_{flmpq}^{\pm}}}{-\cos{\ga_{flmpq}^{\pm}}}^{l-m \
\textrm{even}}_{l-m\ \textrm{odd}},\lb{duec}\eqf
in which \eqi f_p=(l-2p)\dot \og+(l-2p+q)\dot {\mtc{M}}+ m(\dot
\O-\dot \th)\pm
 {\s}.\lb{fr}\eqf
It should be noticed that the frequencies of the perturbations
given by \rfr{fr} are different, in general, from the frequencies
of the solid Earth tidal perturbations given by \rfr{pert}.
 While for the solid tides the diurnal modulation due to
$\dot \th$ cancels out automatically if one considers those terms
 in which $j_1=m$,
 for the ocean tides, in general, this does not happen because
of the presence of the Eastwards waves due to the non-equilibrium
pattern of the ocean tidal bulge.

Long periodicities can be obtained considering those combinations of $l,\ p$
and $q$ for
which $l-2p+q=0$ holds and retaining  the Westward prograde terms with
$j_1=m$. Only
in this way in \rfr{fr} the contributions of
$\dot \th$ are equal and opposite, and cancel out.
 With these bounds on $l,\ m,\ p$ and $q$ the
frequencies of the perturbations of interest  become \eqi f_p=\dot
\G_f+(l-2p)\dot \og+m\dot \O\lb{ff}.\eqf It is worthwhile noting
that the frequencies $f_p$ are identical to those of solid tidal
perturbations \ct{ref:dow,ref:bercar}, so that the satellites
cannot distinguish one effect from the other, a feature which will
turn out to be important in the recovery of the coefficients
$C^{\pm}_{lmf}$.

For $l=2$, as for the solid tides, $l-2p=0$ holds. The frequencies
of the perturbations are, in this case, given by\eqi f_p=\dot
 \G_f +m\dot \O\lb{ft}.\eqf
If $l$ is odd the
situation is different because now all the terms with $p$ running from
0 to $l$ must be considered and  $q$ is
different from zero: $q=2p-l$. So, the rates of the perturbations include the
contribute of $\dot \og$.

Up to now   the prograde wave terms only have been considered in
order to deal with long period perturbations. Yet, there is a case
in which also the retrograde terms fit into this scheme; it is
possible to show that for $l$ even and $m=0$ the frequencies of
prograde and retrograde terms coincide, in absolute values, and
the perturbation amplitudes are twice the amplitudes of the
prograde terms only. Indeed, according to \rfrs{juuno}{duec}, $\D
\O _f,\ \D \og_f$
 are proportional to \eqi\sum_{+}^{-}A_{flm}^{\pm}\rp{1}{f_p}
{\mc{\sin{\ga_{flmpq}^{\pm}}}{-\cos{\ga_{flmpq}^{\pm}}}};\lb{pippo}\eqf
in this case $l-m$ is even and retaining only those terms for
which $l-2p=0$ one has
 \eqi\ga_{f l 0 \rp{l}{2} 0}^{\pm}=\pm(\s t-\ve_{fl0}^{\pm}),\eqf
 \eqi f_p=\pm \s.\eqf Expanding \rfr{pippo} it can be
obtained $$ A_{fl0}^{+}\rp{1}{\s}\sin{(\s t-\ve_{fl0}^{+})}+
A_{fl0}^{-}\rp{1}{-\s}\sin{[-(\s t-\ve_{fl0}^{-})]}=$$ \eqi
=A_{fl0}^{+}\rp{1}{\s}\sin{(\s t-\ve_{fl0}^{+})}+
A_{fl0}^{-}\rp{1}{\s}\sin{(\s t-\ve_{fl0}^{-})}.\lb{oiu}\eqf Since
$A_{fl0}^{+}=A_{fl0}^{-}$ and $\ve_{fl0}^{+}=\ve_{fl0}^{-}$, as
shown by \rfr{zeroo} and \rfr{zero}, \rfr{oiu} becomes\eqi\rp{2
A_{fl0}^{+}}{\s}\sin{(\s t-\ve_{fl0}^{+})}.\eqf

\Rfrs{juuno}{duec} have been adopted in order to compute the
amplitudes of the ocean tidal  perturbations on the Keplerian
elements $\O$ for LAGEOS and LAGEOS\ II and $\og$ for LAGEOS\ II
which pertain directly the \lt\ experiment. Also in  this section
the perturbations of second order \ct{ref:bal} due to the
oblateness of the Earth have not been considered, as already done
for the solid tides. The
 inclination and
eccentricity functions used are those quoted in \ct{ref:kau}.
 For the numerical values of the various geophysical parameters which figure in
\rfrs{juuno}{duec} the IERS standards \ct{ref:iers} have been
used, while the EGM96 gravity model \ct{ref:lem} has been adopted
for the choice of the tidal constituents and their
 coefficients $C_{lm}^{\pm}$ in the ocean
tidal potential $U_f$.
The calculation have been performed, considering
 only the progressive waves,  for the following tidal lines\\
$\bullet$ $M_m\ (065.455),\ S_a\ (056.554),\ M_f\ (075.555),\ S_{sa}\ (057.555)$;\\
$\bullet$ $K_1\ (165.555),\ O_1\ (145.555),\ P_1\ (163.555),\ Q_1\ (135.655);$\\
$\bullet$ $K_2\ (275.555)$, $M_2\ (255.555)$, $S_2\ (273.555)$,
$N_2\ (245.655)$,
$T_2\ (272.556)$.\\
For each of these tidal constituents the following terms have been calculated\\
$ l=2,\ p=1,\ q=0$ because the eccentricity
functions $G_{lpq}(e)$ for $p=0,\ q=-2$ and $p=2,\ q=2$ vanish.\\
$l=3,\ p=1,\ q=-1$ and $l=3,\ p=2,\ q=1$ because $G_{30-3}$ and
 $G_{333}$ are not quoted in \ct{ref:kau} due to their smallness: indeed, the
$G_{lpq}(e)$ are proportional to $e^{|q|}$.\\
$ l=4,\ p=2,\ q=0$ because the other admissible combinations of $l,\ p$
and $q$ give rise to negligible eccentricity functions. So, also for $l=4$ the
condition
 $l-2p=0$,
 in practice, holds and the constituents of degree $l=2$ and $l=4$ generate
detectable perturbations with identical periods.

Similar analysis can be found in \ct{ref:chr,ref:dow}. It should
be noticed that when Christodoulidis performed his study, which is
relative to only five constituents, neither
 the LAGEOS
nor the LAGEOS\ II were in orbit, while Dow has sampled the tidal
spectrum for LAGEOS in a poorer manner with respect to this study
 in the sense that, for each constituent,
 only the terms of degree $l=2$ have been
considered with the exception of the  $K_1$ whose $l=4$
contribution has been also analyzed.  Moreover, when these works
have been realized there were a few coefficients $C_{lmf}^+$
available with relevant associated errors.

Some explanations are needed about the determination of the
coefficients $C^{\pm}_{lmf}$. In EGM96 the geopotential is recovered through both
altimeter and surface gravity information,  and satellites data. The pool of
near-Earth  satellites employed,
in general,
are not perceptibly perturbed by the entire tidal spectrum, but they are
sensitive only to some certain tidal lines, depending on the features of their
orbits. Moreover, on a large enough temporal scale
they cannot distinguish between solid and prograde ocean tidal perturbations
because their frequencies are the same. Finally, there are also other periodic
physical phenomena different from the tides that affect the orbits of the geodetic
satellites; in many cases their periodicities are similar to that of the tides,
particularly in the zonal band. These facts imply that in the
analytical expressions of the perturbations
it is
necessary to assume
as known {\it a priori}
from various reliable models some solid or ocean tidal terms and consider variable
the other ones in
order to adjust them by means of the experimentally determined values of the
perturbations;  all the constituents considered, both those
 held fixed and the other recovered, must be
 capable to influence  perceptibly the satellites employed.
 The strategy followed in EGM96 has consisted in adopting the
frequency-dependent Wahr model for the solid tides \ct{ref:wah81b}
with its values for the Love numbers and the $H_{l}^{m}(f)$ of
Cartwright and Edden, calculating a certain number of ocean tidal
terms by means of oceanographic models, when it has been possible
to solve the LTE, or by means of other algorithms \ct{ref:cas},
considering also the retrograde waves, and choosing the 13 ocean
tidal terms listed above to be adjusted. The values recovered for
the coefficients $C_{20f}^+$ account for also the retrograde terms
because, as previously noticed, their periods are equal to those
of prograde ones. The terms whose value has been considered given
{\it a priori} constitute the so called background; it necessarily
contains terms up to sixth degree because the attenuation due to
altitude makes the near-Earth satellites almost insensitive to
larger degree terms:
 indeed, in \rfrs{juuno}{duec} the $A^{\pm}_{lmf}$ are
proportional to
\eqi\left({\rp{R}{r}}\right)^{l+1}\rp{1}{2l+1}.\eqf The values
obtained for the coefficients $C_{lmf}^+$ are, in general, biased
by  the effects of the anelasticity of the solid Earth's mantle
and by all other phenomena which have not been explicitly modeled
in the background. For example, in the values recovered for the
$S_a$ are included climatological effects which have not
gravitational origin; in the $S_2$ coefficients are also included
the variations of the atmospheric pressure due to the atmospheric
tides.
\subsection{Discussion of the numerical results}
In Table \ref{otto}, Table \ref{nove} and Table \ref{dieci} the
present results for the nodes $\O$ of LAGEOS and LAGEOS\ II, and
the argument of perigee $\og$ for LAGEOS\ II are quoted.

\begin{table}[ht!]
\caption[Ocean tidal perturbations on the node $\O$ of
LAGEOS.]{\footnotesize{Perturbative
 amplitudes on the node $\O$ of LAGEOS due to ocean
tides. $P$ indicates the periods in days, $A$ the amplitudes in
mas and $E$ the percent error in the $C_{lmf}^{+}$. The values
employed for them and the related errors are those quoted in EGM96
model.}} \label{otto}
\begin{center}
\tiny{\begin{tabular}{lllllllllllll} \noalign{\hrule height 1.5pt}
  Tide & \multicolumn{3}{l} { $l=2,\ p=1,\ q=0$} &
\multicolumn{3}{l}
 { $l=3,\ p=1,\ q=-1$} & \multicolumn{3}{l} { $l=3,\ p=2,\ q=1$} &
\multicolumn{3}{l} { $l=4,\ p=2,\ q=0$}\\ \cline{2-13}
      & $P$  & $A$  & $E$ & $P$  & $A$ & $E$
      & $P$  & $A$ & $E$ & $P$  & $A$ & $E$\\
\hline 065.455 $M_{m}$ & 27.55   & -0.54 & 14.4  & 28 & $-10^{-4}$
& 66.6  & 27.11 &
 $10^{-4}$ &
 66.6  & - & - & - \\ 056.554 $S_{a}$ & 365.27 & -20.55 & 6.7 & 464.67 & $-10^{-2}$ & 10 &
300.91 &
 $10^{-2}$ & 10
& - & - & -\\  075.555 $M_{f}$ & 13.66 & -0.62 & 7.8 & 13.77 &
$-10^{-4}$ & 112 & 13.55 & $10^{-4}$
  & 112 & -
& - & -\\  057.555 $S_{sa}$ & 182.62 & -5.98 & 9.4 & 204.5 &
$-10^{-3}$ & 27.2 & 164.9 &
 $10^{-3}$  & 27.2
& - & - & -\\ \\ 165.555 $K_1$ & 1,043.67 & 156.55 & 3.8 & 2,684.2
&
-0.36 & 5.2 & 647.76 & $10^{-2}$  & 5.2 & 1,043.67 & 4.63 & 3.9\\
163.555 $P_1$ & -221.35 & -11.49 & 8.1 & -195.95 & $10^{-3}$ &
18.5 & -254.3 &
 $-10^{-3}$ &
18.5 & -221.35 & -0.32 & 8\\ 145.555 $O_1$ & -13.84 & -2 & 2.9 &
-13.72 & $10^{-3}$ & 3.2 & -13.95 & $10^{-4}$ & 3.2 & -13.84 &
$10^{-2}$ & 5.7 \\ 135.655 $Q1$ & -9.21 & -0.28 & 13.5 & -9.16 &
$10^{-4}$ & 25
 & -9.26 & $10^{-5}$ & 25 & -9.21 &
$-10^{-3}$ & 20\\ \\ 275.555 $K_2$ & 521.83 & -6.24 & 11.1 & 751.5
& $-10^{-2}$ & 5.5 & 399.7 &
 $10^{-2}$  & 5.5 &
521.83 & -9.58 & 15.4\\  273.555 $S_2$ & -280.93 & 9.45 & 3.9 &
-241.24 & $10^{-3}$  & 7.1 & -336.25 &
 $-10^{-2}$  & 7.1 &
-280.93 & 15.08 & 5.2\\  272.556 $T_2$ & -158.8 & 0.28 & 75 &
-145.3 & $10^{-4}$  & 50 & -175 & $-10^{-3}$  & 50 & -158.8 & 0.44
& 100
\\ 255.555 $M_2$ & -14.02 & 2.03 & 0.9 & -13.9 & $-10^{-4}$ &
7.4 & -14.14 &
 $10^{-3}$  & 7.4 &
-14.02 & 2.08 & 2.8\\  245.655 $N_2$ & -9.29 & 0.3 & 4.6 & -9.2 &
$10^{-5}$ & 12.5 & -9.3 & $10^{-4}$  & 12.5 & -9.29 & 0.3 & 8.3\\
\noalign{\hrule height 1.5pt}
\end{tabular}}
\end{center}
\end{table}
\begin{table}[ht!]
\caption[Ocean tidal perturbations on the node $\Omega$
 of LAGEOS\ II.]{\footnotesize{Perturbative amplitudes
 on the node $\Omega$ of LAGEOS\ II due to
 ocean
tides. $P$ indicates the periods in days, $A$ the amplitudes in
mas and $E$ the percent error in the $C_{lmf}^{+}$. The values
employed for them and the related errors are those quoted in EGM96
model.}} \label{nove}
\begin{center}
\tiny{\begin{tabular}{lllllllllllll} \noalign{\hrule height 1.5pt}
Tide & \multicolumn{3}{l} { $l=2,\ p=1,\ q=0$} &
\multicolumn{3}{l}
 { $l=3,\ p=1,\ q=-1$} & \multicolumn{3}{l} { $l=3,\ p=2,\ q=1$} &
\multicolumn{3}{l} { $l=4,\ p=2,\ q=0$}\\ \cline{2-13}
      & $P$  & $A$  & $E$ & $P$  & $A$ & $E$
      & $P$  & $A$ & $E$ & $P$  & $A$ & $E$\\
\hline

065.455 $M_{m}$  & 27.55  &  1  &  14.4  &  26.65  &  $10^{-3}$  &
66.6  &
 28.5  &  -$10^{-3}$ & 66.6  &  - & - & - \\
 056.554 $S_{a}$ & 365.27 & 37.71 & 6.7 & 252.8 & 0.13 & 10 & 657.5
& -0.35 & 10 & - & - & -\\  075.555 $M_{f}$ & 13.6 & 1.13 & 7.8 &
13.43 & $10^{-4}$ & 112 & 13.89 & $-10^{-4}$
 & 112 & - &
- & - \\  057.555 $S_{sa}$ & 182.62 & 10.98 & 9.4 & 149.41 &
$10^{-2}$ & 27.2 & 234.8 &
 $-10^{-2}$ & 27.2 & - & - & -\\ \\
165.555 $K_1$ & -569.21 & -35.69 & 3.8 & -1,851.9 & -1.02 & 5.2 &
-336.3 & $-10 ^{-3}$ & 5.2 & -569.21 & 41.58 & 3.9 \\  163.555
$P_1$ & -138.26 & -3 & 8.1 & -166.23 & $-10^{-2}$ & 18.5 & -118.35
& $-10^{-4}$ & 18.5 & -138.26 & 3.29 & 8 \\ 145.555 $O_1$ & -13.3
& -0.8 &  2.9 & -13.5 & -$10^{-2}$ & 3.2 &  -13.12 & $-10^{-4}$ &
3.2 & -13.3 & 0.7 & 5.7 \\  135.655 $Q_1$ & -8.98 & -0.11 & 13.5 &
-9.08 & -$10^{-3}$ & 25 & -8.89  & $10^{-5}$ & 25 & -8.98 & 0.11 &
20 \\ \\ 275.555 $K_2$ & -284.6 & -6.24 & 11.1 & -435.38 & -0.13 &
5.5 & -211.4 & $10^{-2}$ & 5.5 & -284.6 & -5.95 & 15.4
\\ 273.555 $S2$ & -111.2 & -6.87 & 3.9 & -128.6 & $-10^{-2}$
& 7.1 & -97.9 &
 $10^{-2}$ &
7.1 & -111.2 & -6.79 & 5.2 \\  272.556 $T_2$ & -85.27 & -0.277 &
75 & -95.14 & $-10^{-3}$ &  50 &
 -77.25 & $10^{-3}$ & 50 & -85.27 & -0.274 & 100 \\
255.555 $M_2$ & -13.03 & -3.46 & 0.9 & -13.2 & -$10^{-3}$ & 7.4 &
-12.83
 & $10^{-3}$
& 7.4 & -13.03 & -2.2 & 2.8 \\  245.655 $N_2$ & -8.8 & -0.46 & 4.6
& -8.9 & $-10^{-3}$ & 12.5 & -8.7 & $10^{-4}$ & 12.5 & -8.8 &
-0.34 & 8.3 \\
\noalign{\hrule height 1.5pt}
\end{tabular}}
\end{center}
\end{table}
\begin{table}[ht!]
\caption[Ocean tidal perturbations on the perigee $\omega$
 of LAGEOS\ II.]{\footnotesize{Perturbative amplitudes
 on the perigee $\omega$ of LAGEOS\ II due to
 ocean
tides. $P$ indicates the periods in days, $A$ the amplitudes in
mas and $E$ the percent error in the $C_{lmf}^{+}$. The values
employed for them and the related errors are those quoted in EGM96
model.}} \label{dieci}
\begin{center}
\tiny{\begin{tabular}{lllllllllllll} \noalign{\hrule height 1.5pt}
Tide & \multicolumn{3}{l} { $l=2,\ p=1,\ q=0$} &
\multicolumn{3}{l}
 { $l=3,\ p=1,\ q=-1$} & \multicolumn{3}{l} { $l=3,\ p=2,\ q=1$} &
\multicolumn{3}{l} { $l=4,\ p=2,\ q=0$}\\ \cline{2-13}
      & $P$  & $A$  & $E$ & $P$  & $A$ & $E$
      & $P$  & $A$ & $E$ & $P$  & $A$ & $E$\\
\hline 065.455 $M_{m}$ & 27.55 & -0.69 & 14.4 & 26.65 & -1.53 &
66.6 & 28.5 & 1.64  & 66.6
 & - & - &-
\\
056.554 $S_{a}$ & 365.27 & -26.17 & 6.7 & 252.8 & -114.35 & 10 &
657.55 & 297.34
 & 10 &
- & - & - \\  075.555 $M_{f}$ & 13.66 & -0.78 & 7.8 & 13.43 &
-0.58 & 112 & 13.89 & 0.60 & 112 &- & - & - \\  057.555 $S_{sa}$ &
182.6 & -7.62 & 9.4 & 149.41 & -22.95  & 27.2  & 234.8 & 36.07 &
27.2 & - & - & - \\ \\ 165.555 $K_1$ & -569.21 & 177.76 & 3.8 &
-1,851.9 & -1,136 & 5.2 & -336.28 & 346.6
 & 5.2 & -569.21 & -3.95 & 3.9 \\
163.555 $P_1$ & -138.26 & 14.95 & 8.1 & -166.2 & -28.97 & 18.5 &
-118.35 & 34.67
 & 18.5 & -138.2 & -0.31 & 8 \\
145.555 $O_1$ & -13.3 & 4 & 2.9 & -13.55 & -13.7 & 3.2 &
 -13.12 & 22.3
& 3.2 & -13.3 & -$10^{-2}$ & 5.7 \\  135.655 $Q_1$ & -8.98 & 0.58
& 13.5 & -9.08 & -1.17 & 25 & -8.89 & 1.92
 & 25
& -8.98 & -$10^{-2}$ &  20 \\ \\ 275.555 $K_2$ & -284.6 & -5.95 &
11.1 & -435.3 & 214.23 & 5.5 & -211.4 & 87.3
 & 5.5 &
-284.6 & -2.49 & 15.3 \\  273.555 $S_2$ & -111.2 & -6.55 & 3.9 &
-128.6 &  98.47 & 7.1 & -97.9 & 62.9
 & 7.1 &
-111.2 & -2.85 & 5.2 \\  272.556 $T_2$ & -85.2 & -0.26 & 75 &
-95.1 & 5.2 & 50 & -77.2 & 3.54 & 50 & -85.2
 &
-0.11 & 100 \\  255.555 $M_2$ & -13.03 & -3.3 & 0.9 & -13.2 & 9.7
& 7.4 & -12.83 & 7.9
 & 7.4 &
-13.03 & -0.92 & 2.8 \\  245.655 $N_2$ & -8.48 & -0.44 & 4.6 &
-8.94 & 1.95 & 12.5 & -8.75 & 1.6
 & 12.5
& -8.84 & -0.14 & 8.3 \\
\noalign{\hrule height 1.5pt}
\end{tabular}}
\end{center}
\end{table}
From an accurate inspection of Table \ref{otto}, Table \ref{nove}
 it is possible to notice that, for the nodes $\O$, only the even
degree terms give an appreciable contribute; the $l=3$ terms are totally
negligible. This so because $\D \O_f$ is proportional to the
 $G_{lpq}(e)$ functions which, in turn, are, in general,  proportional to
$e^{|q|}$: in this case  the eccentricity functions  are
$G(e)_{31-1}=e(1-e^2)^{-5/2}=G(e)_{321}$ and the eccentricities of
\lg satellites are of the order of $10^{-2}$ or less.

Among the long period zonal tides, the Solar annual tide $S_a$
(056.554) exerts the most relevant action on the nodes, with an
associated percent error in the amplitudes of $6.7$ $\%$. It is
interesting to compare for this tidal line the ocean tidal
perturbations of degree $l=2$ $A^{{\rm ocean}}_{{\rm
LAGEOS}}(\O)=-20.55$ mas,
 $A^{{\rm ocean}}_{{\rm LAGEOS}\ {\rm II}}(\O)=37.7$ mas, with those due to the solid Earth tides
$A^{{{\rm solid}}}_{{\rm LAGEOS}}(\O)=9.96$ mas, $A^{{{\rm
solid}}}_{{\rm LAGEOS}\ {\rm II}}(\O)=-18.28$ mas. The ocean
amplitudes amount to $204$  $\%$ of the solid ones, while for the
other zonal constituents they vary from $9.9$ $\%$ for the $M_m$
to the $18.5$ $\%$ of the $S_{sa}$. This seems to point toward
that the recovered value  of $C_{20f}^{+}$ for the $S_a$ is biased
by other climatological effects than the tides; indeed, for all
the other tidal lines, zonal or not, the ocean tidal perturbations
of degree $l=2$ amount to almost $10$ $\%$ of the solid Earth
tides perturbations.

The terms of degree
$l=2$ of the tesseral tides $K_1$ (165.555) and $O_1$ (145.555)
induce very large
perturbations on the node of LAGEOS
 and, to a lesser extent,  of
LAGEOS\ II; Table \ref{otto} and Table \ref{nove} quote $156.55$
mas, $151.02$ mas for the former and $-35.69$ mas, $-34.43$ mas
for the latter. The associated percent errors are $3.8$ $\%$ and
$2.9$ $\%$, respectively. For the terms of degree $l=4$, which
have the same periods of those of degree $l=2$, the situation is
reversed: Table \ref{otto} and Table \ref{nove} quote $4.63$ mas
and $3.54$ mas for LAGEOS and $41.58$ and $31.81$ mas for LAGEOS\
II. The percent errors associated with $K_1$ and $O_1$, for $l=4$,
are $3.9$ $\%$ and $5.7$ $\%$ respectively. For all the tesseral
lines investigated the ocean tidal perturbations of degree $l=2$
are, in general, the $10$ $\%$ of the solid tidal perturbations.

Among the sectorial tides, the most relevant in perturbing the
nodes of LAGEOS satellites, on large temporal scales, is the $M_2$
(255.555): Table \ref{otto} and Table \ref{nove} quote $-75.59$
mas ($l=2$) and $-77.40$ mas ($l=4$) for LAGEOS,  and $-75.65$ mas
($l=2$) and $-48.07$ mas ($l=4$)
 for LAGEOS\ II. The associated percent errors are $0.9$ $\%$ for
$l=2$ and $42.8$ $\%$ for $l=4$.
The $M_2$ ocean perturbation of degree $l=2$ amounts to  $10.3$ $\%$ of the
corresponding solid tide. The
same  holds for the other sectorial tides of degree $l=2$.

About the periodicities of these perturbations in relation to the detection of
Lense-Thirring effect, the same considerations already exposed for the solid
tides hold also in this case. Moreover, it
must be pointed out the aliasing role played by the  zonal 18.6-year and
 9.3-year tides in the extracting any secular effect, like the gravitomagnetic
precession, from a record whose duration is shorter than their
very long periods. Recently both Starlette and LAGEOS SLR
satellites passed their 19th year in orbit and this span of time
is now adequate to get reliable information about these tides
\ct{ref:che95,ref:ean} due to their slow frequencies they can be
correctly modelled in terms of the equilibrium theory through the
$H_l^m$ coefficients and a complex Love number accounting for the
anelasticity of the mantle. So, concerning them the results quoted
for the solid tides can be considered adequately representative.
In Table \ref{dieci}
 the  amplitudes of the perturbations
 on the
argument of perigee $\og$ for LAGEOS\ II
 are quoted.
For this orbital element the factor \eqi \rp{1-e^2}{e}F_{lmp}
\dert{G_{lpq}}{e}-\rp{\cos{i}}{\sin{i}}\dert{F_{lmp}}{i}G_{lpq}\eqf
to which $\D \og_f$ is proportional makes the contributions of the
$l=3$ terms not negligible. For the even degree terms the
situation is quite similar to that of $\O$ in the
sense that the most influent tidal lines are the $S_a,\ K_1,\ O_1$ and $M_2$.\\
Once again, among the long period tides the $S_a$ exhibits a
characteristic behavior. Indeed, its $l=3$ contributions are much
stronger than those of the other zonal tides. This fact could be
connected to the large values obtained in its $C_{30f}^{+}$
coefficient and people believe that it partially represents north
to south hemisphere mass transport effects with an annual cycle
nontidal in origin. The $l=3$ terms present, in general, for the
perigee of LAGEOS\ II a very interesting spectrum also for the
tesseral and sectorial bands:  for LAGEOS\ II
there are lots of tidal lines which induce, on large temporal
scales, very relevant perturbations on $\og$, with periods of the
order of an year or more. In particular, it  must be quoted the
effect of $K_1$ line for $p=1,\ q=-1$:  the perturbation induced
amount to -1,136 mas with period of -1,851.9 days. These values
are comparable to the effects induced by the solid Earth tidal
constituent of degree $l=2$ on the node $\O$ of LAGEOS. Comparing
the degree $l=2$ terms with those of the solid tides, it can be
noticed that also for the perigee the proportions are the same
already seen for the nodes.
\subsection{The mismodelling in the nodes and the perigee}
Concerning the mismodelling on the ocean tidal perturbations, the
main source of uncertainties in them is represented by the load
Love numbers $k^{'}_{l}$ and  the coefficients $C^{+}_{lmf}$. In
regard to the ocean loading, the first calculations of the load
Love numbers $k^{'}_l$ can be found in refs.\ct{ref:far}.
Pagiatakis, in ref.\ct{ref:pag}, in a first step has recalculated
$k^{'}_{l}$ for an elastic, isotropic and non-rotating Earth: for
$l<800$ he claims that his estimates differ from those by Farrell
in ref.\ct{ref:far}, calculated with the same hypotheses, of less
than $1\%$. Subsequently, he added to the equations, one at a
time, the effects of anisotropy, rotation and dissipation; for low
values of $l$ their effects on the results of the calculations
amount to less than 1$\%$. It has been decided to calculate
$\vass{\derp{A(\og)}{k^{'}_{l}}}\d k^{'}_{l}$ of the perigee of
LAGEOS\ II for $K_1\ l=3\ p=1$ which turns out to be the most
powerful ocean tidal constituent acting upon this orbital element.
First, we have calculated mean and standard deviation of the
values for $k^{'}_{3}$ by Farrell and Pagiatakis obtaining $\d
k^{'}_{3}/k^{'}_{3}=0.9\%$, in according to the estimates by
Pagiatakis. Then, by assuming in a pessimistic way that the global
effect of the departures from these symmetric models yield to a
total $\d k^{'}_{3}/k^{'}_{3}=2\%$, we have obtained $\d\og^{{\rm
II}}=5.5$ mas which corresponds to $2\%$ of $\D\og^{{\rm
II}}_{{\rm LT}}$ over 4 years. Subsequently, for this constituent
and for all other tidal lines we have calculated the effect of the
mismodelling of $C^{+}_{lmf}$ as quoted in EGM96 \ct{ref:lem}.  In
Table \ref{mismo2} we compare the so obtained mismodelled ocean
tidal perturbations to those generated over 4 years by the
Lense-Thirring effect.
\begin{table}[ht!]
\caption[Mismodeled ocean tidal perturbations on the nodes of
LAGEOS and LAGEOS\ II and the perigee of LAGEOS\
II.]{\footnotesize{Mismodeled ocean
 tidal perturbations on the nodes $\O$ of LAGEOS  and LAGEOS\ II and the perigee $\og$ of
LAGEOS\ II compared to their gravitomagnetic precessions over 4
years $\D\O^{{\rm I}}_{{\rm LT}}=124$ mas $\D\O^{{\rm II}}_{{\rm
LT}}=126$ mas $\D\og^{{\rm II}}_{{\rm LT}}=-228$ mas. The effect
of the ocean loading has been neglected.  When the $1\%$ cutoff
has not been reached a - has been inserted. The values quoted for
$K_1\ l=3\ p=1$ includes also the mismodelling in the ocean
loading coefficient $k^{'}_{3}$ assumed equal to $2\%$. The ratios
are in percent and the mismodelled amplitudes $\delta X$ in mas.
}} \label{mismo2}
\begin{center}
\begin{tabular}{llllllll}
\noalign{\hrule height 1.5pt} { Tide } & {$\frac{\d
C^{+}}{C^{+}}$} & {$\d\O^{{\rm I}}$ } & {$\frac{\d\O^{{\rm
I}}}{\D\O^{{\rm I}}_{{\rm LT}}}$}
  & {$\d\O^{{\rm II}}$ } &
{$\frac{\d\O^{{\rm II}}}{\D\O^{{\rm II}}_{{\rm LT}}}$ } &
{$\d\og^{{\rm II}}$ }& {$\frac{\d\og^{{\rm II}}}{\D\og^{{\rm
II}}_{{\rm LT}}}$}\\  \hline $S_{a}$ $l$=2 $p$=1 $q$=0& 6.7 & 1.37
& 1.1 & 2.5 & 1.9 & - & - \\  $S_{a}$ $l$=3 $p$=1 $q$=-1 & 10 & -
& - & - & - & 11.4 & 5
\\  $S_{a}$ $l$=3 $p$=2 $q$=1 & 10 & - & - & - & - & 29.7 & 13
\\  $S_{sa}$ $l$=3 $p$=1 $q$=-1 & 27.2 & - & - & - & - & 6.2 &
2.7 \\ \\ $K_1$ $l$=2 $p$=1 $q$=0& 3.8 & 5.9 & 4.7 & 1.3 & 1 &
6.75& 2.9\\  $K_1$ $l$=3 $p$=1 $q$=-1& 5.2 & - & - & - & - & 64.5
& 28.3\\  $K_1$ $l$=3 $p$=2 $q$=1& 5.2 & - & - & - & - & 18 &
7.9\\  $K_1$ $l$=4 $p$=2 $q$=0& 3.9 & - & - & 1.6 & 1.2 & - & -\\
\\ $P_1$ $l$=3 $p$=1 $q$=-1& 18.5 & - & - & - & - & 5.3 &
2.3\\ $P_1$ $l$=3 $p$=2 $q$=1& 18.5 & - & - & - & - & 6.4 & 2.8\\
\\ $K_2$ $l$=3 $p$=1 $q$=-1& 5.5 & - & - & - &
- & 11.7 & 5\\
$K_2$ $l$=3 $p$=2 $q$=1& 5.5 & - & - & - & - & 4.8 & 2\\
\\ $S_2$ $l$=3 $p$=1 $q$=-1& 7.1 & - & - & - & - & 6.9 & 3\\
 $S_2$ $l$=3 $p$=2 $q$=1& 7.1 & - & - & - & - & 4.4 & 1.9\\
\\ $T_2$ $l$=3 $p$=1 $q$=-1& 50 & - & - & - & - & 2.6 & 1.1\\
\noalign{\hrule height 1.5pt}
\end{tabular}
\end{center}
\end{table}
It turns out that the perigee of LAGEOS\ II is more sensitive to
the mismodelling of the ocean part of the Earth response to the
tide generating potential. In particular, the effect of $K_1\ l=3\
p=1\ q=-1$ is relevant with a total $\d\og=
\vass{\derp{A(\og)}{C^{+}_{lmf}}}\d
C^{+}_{lmf}+\vass{\derp{A(\og)}{k^{'}_{l}}}\d k^{'}_{l}$ of 64.5
mas amounting to 28.3 $\%$ of $\D\og^{{\rm II}}_{{\rm LT}}$ over 4
years.
\subsection{Conclusions}
Concerning the orbital tidal perturbations on LAGEOS and LAGEOS\
II the following improvements with respect to the previous works
have been reached

$\bullet$ The node and the perigee have been considered

$\bullet$ The analysis has been extended also to LAGEOS\ II,
launched in 1992

$\bullet$ Concerning the solid tides, the most recent available
frequency-dependent Love numbers $k_2$ have been used instead of a
single-valued Love number $k_2=0.317$. Moreover, the
latitude-dependence of the Love number $k^{+}$ has been considered
for some selected tidal lines

$\bullet$ About the ocean tides, their orbital perturbations have
been extensively calculated for $l=2,3,4$ by using the most recent
available model EGM96. Interesting are the results obtained for
the $K_1,\ l=3\ m=1\ p=1\ q=-1$ oceanic constituent. Indeed it
induces on the perigee of LAGEOS\ II  perturbations whose
amplitudes are of the order of thousands of mas and the periods
amounts to some years as the solid tidal perturbations

$\bullet$ An evaluation of the mismodelling on such orbital
perturbations, for certain tidal constituents, have been performed

The calculations performed here have a general validity because
they can be extended to any artificial satellite. The results
presented here for the LAGEOS can be used, from one hand, in
improving the modelling of their orbital perturbations and, from
another hand, for the correct evaluation of the error budget in
any space-based experiment devoted to the measurement of some
particular feature of the Earth' s gravitational field.

In the context of the general relativistic Lense-Thirring
experiment, the calculations performed here have been used, in
view of a refinement of the error budget, in order to check
preliminarily which tidal constituents are really important in
perturbing the combined residuals so to fit and remove them from
the data, if possible, or, at least, to evaluate the systematic
error induced by them. Table  \rfr{mismo1} and Table \rfr{mismo2}
show that, over a 4 years time span the nodes of the two LAGEOS
are sensitive to the even components of the $18.6$-year line, the
$K_1$ and the $S_a$ at a 1$\%$ level at least. Moreover, the
perigee of LAGEOS\ II turns out to be very sensitive to the $l=3$
part of the ocean tidal spectrum.

The results obtained in this section for the nodes of \lg\ and
\lgg\ and the perigee of \lgg\  will be the starting point for the
numerical simulations with MATLAB, described  in section 4 and
section 6, performed to assess quantitatively the  direct impact
of the tidal perturbations on the \grc\ \lt\ experiment, currently
implemented, and on the proposed gravitoelectric experiment.
\section{The impact of the Earth tides on the determination of the \lt\ effect}
\subsection{Introduction}
According to \ct{ref:ciu96}, the \lt\  effect  could be detected,
in the field of the Earth, by analyzing the orbits of the two
laser-ranged LAGEOS and LAGEOS\ II satellites. The observable
adopted is the combination of orbital residuals of \rfr{unolt}:
the determination of the scaling parameter $\m_{{\rm LT}}$, 1 in
 General Relativity and 0 in
Newtonian mechanics, is influenced by a great number of
gravitational and non-gravitational perturbations acting upon
LAGEOS and LAGEOS\ II.  Among the perturbations of gravitational
origin a primary role is played by the solid Earth and ocean
tides. Their effects on the orbital elements of LAGEOS and LAGEOS
II has been extensively analyzed in section 3.

In \ct{ref:ciu96} it is claimed the combined residuals
$y\equiv\d\O^{{\rm I}}_{\rm exp}+c_1\d\O^{{\rm II}}_{\rm
exp}+c_2\d\og^{{\rm II}}_{\rm exp}$ allow one to cancel out the
static and dynamical perturbations of degree $l=2,4$ and order
$m=0$ of the terrestrial gravitational field; however, this is not
so for the tesseral ($m=1$) and sectorial ($m=2$) tides. This
section aims to assess quantitatively how the solid and ocean
Earth tides of order $m=0,1,2$ affect the recovery of $\m_{\rm
LT}$ in view of a refinement of the error budget of the
gravitomagnetic LAGEOS experiment.

Concerning the zonal tides, the results obtained in section 3 for
the amplitudes of their perturbations have been used directly in
\rfr{unolt} in order to test in a preliminary way if the $l=2,4$
$m=0$ tidal perturbations really cancel out. In regard to the
tesseral and sectorial tides the analysis of their impact on the
\lt\ measurement has been done by simulating the real residual
curve and analyzing it. The $T_{{\rm obs}}$ chosen span ranges
from 4 years to 8 years. This is so because 4 years is the length
of the latest time series actually analyzed in \ct{ref:ciu98} and
8 years is the maximum length obtainable today (Summer 2000)
because LAGEOS\ II has been launched in 1992. The analysis
includes also the long-period signals due to solar radiation
pressure and the $J_3$ geopotential zonal harmonic acting on the
perigee of LAGEOS\ II. In our case we have a signal built up with
the secular \lt\ trend\footnote{In fact, the combined residuals
are affected also by the secular contribution due to $l=6,8,..$
zonal terms of the geopotential. Their effect amounts to almost
13$\%$ of the \lt trend \ct{ref:ciu98}. See also Appendix A.} and
a certain number of long-period harmonics, i.e. the tesseral and
sectorial tidal perturbations and the other signals with known
periodicities.

The part of interest for us is the secular trend while the
harmonic part represents the noise. We address the problem of how
the harmonics affect the recovery of the secular trend on given
time spans $T_{{\rm obs}}$ and various samplings $\D t$.

Among the long-period perturbations we distinguish between those
signals whose periods are shorter than $T_{{\rm obs}}$ and those
signals with periods longer than $T_{{\rm obs}}$. While the former
average out if $T_{{\rm obs}}$ is an integer multiple of their
periods, the action of the latter is more subtle since they could
resemble a trend over temporal intervals too short with respect to
their periods. They must be considered as biases on the
Lense-Thirring determination affecting its recovery by means of
their mismodelling. Thus, it is of the  utmost importance that we
reliably assess  their effect on the determination of the trend of
the gravitomagnetic effect. It would be useful to direct the
efforts of the community  (geophysicists, astronomers and space
geodesists) towards the improvement of our knowledge of those
tidal constituents to which $\m_{\rm LT}$ turns out to be
particularly sensitive (for the LAGEOS orbits).

This investigation will quantify unambiguously what one means with
statements like: $\D\m_{{\rm tides}}\leq X\%\m_{{\rm LT}}$. In
this section we shall try to put forward a simple and meaningful
approach. It must be pointed out though that it is not a
straightforward application of any exact or rigorously proven
method; on the contrary, it is, at a certain level, heuristic and
intuitive, but it has the merit of yielding reasonable and simple
answers and allowing for their critical discussion.


The section is organized as follows. In subsect. 2  the effects of
the $l=2,4$ $m=0$ tidal constituents on the combined residuals is
examined; it turns out that, not only they affect it at a level
$<<1\%$, but this feature also extends, within certain limits, to
the $l=3,\ m=0$ part of the tidal spectrum. In subsect. 3 the
features of the simulation procedure of the observable curve  are
outlined. In subsect. 4 the effects of the harmonics of the order
$m=1,2$ with period shorter than 4 years are examined by comparing
the least squares fitted values of $\m_{\rm LT}$ in two different
scenarios: in the first one the simulated curve is complete and
the fitting model  contains all the most relevant signals  plus a
straight line, while in the second one some selected harmonics are
removed from the simulated curve which is fitted by means of the
straight line only.  Subsect. 5 addresses the topic of the
harmonics with periods longer than 4 years.  Subsect. 6 is devoted
to the conclusions.
\subsection{Systematic error due to the  zonal tides}
The combined residuals by Ciufolini  are useful since they should
vanish if calculated for the even zonal contributions $C_{20}$ and
$C_{40}$ of the geopotential \ct{ref:ciu96}. More precisely, the
right side of \rfr{unolt} should become equal to zero if the left
side were calculated for any of these two even zonal
contributions, both of static and dynamical origin; the nearer to
zero is the right side, the smaller is the systematic uncertainty
in $\m_{\rm LT}$ due to the contribution considered.

In order to test this important feature for the case of tides, in
a very conservative way the results obtained in section 3, Table
\ref{unod}, Table \ref{dued}, Table \ref{tred} for the solid Earth
tides and Table \ref{otto}, Table \ref{nove}, Table \ref{dieci}
for the ocean tides, have been used in \rfr{unolt} by assuming,
for the sake of clarity and in order to make easier the comparison
with ref.\ct{ref:ciu97}, an observational period of 1 year and the
nominal values of the calculated tidal perturbative amplitudes, as
if the zonal solid and ocean tides were not at all included in the
GEODYN II dynamical models so that the residuals should account
entirely for them.

The results are released in Table \ref{ciuflt1} and Table
\ref{ciuflt2}.
\begin{table}[ht!]
 \caption[Solid Earth even
zonal tidal contributions to $\D\m_{{\rm
LT}}$.]{\footnotesize{Contribution of the even zonal solid tidal
constituents to $\D\m_{\rm LT}$ by means of the formula
$\d\dot\O^{{\rm I}}+\d\dot\O^{{\rm II}}\times 0.295-
\d\dot\og^{{\rm II}}\times 0.35=60.2\times \m_{\rm LT}$ for $l=2,\
m=0,\ p=1,\ q=0$.}} \label{ciuflt1}
\begin{center}
\begin{tabular}{lllll}
\noalign{\hrule height 1.5pt}
 Tide & $A(\O_{{\rm I}})$
 (mas)& $A(\O_{{\rm II}})$ (mas) & $A(\og_{{\rm II}})$ (mas)&
$\D\m_{\rm LT}$\\ \hline
055.565 & -1,079.38 & 1,982.16 & -1,375.58 & -0.219 \\
055.575 & 5.23 & -9.61 & 6.66 & $1.06\times 10^{-3}$\\
056.554 $S_a$ & 9.95 & -18.28 & 12.69 & $2.02\times 10^{-3}$ \\
 057.555 $S_{sa}$ & 31.21 & -57.31 & 39.77 & $6.33\times 10^{-3}$
\\  065.455 $M_m$ & 5.28 & -9.71 & 6.74 & $1.07\times 10^{-3}$ \\
075.555 $M_{f}$ & 4.94 & -9.08 &
6.3 & $1\times 10^{-3}$ \\
\noalign{\hrule height 1.5pt}
\end{tabular}
\end{center}
\end{table}

\begin{table}[ht!]
\caption[Ocean even zonal tidal contributions to $\D\m_{{\rm
LT}}$.]{\footnotesize{Contribution of the even zonal ocean tidal
 constituents to
$\D\m_{\rm LT}$ by means of the formula $\d\dot\O^{{\rm
I}}+\d\dot\O^{{\rm II}}\times 0.295- \d\dot\og^{{\rm II}}\times
0.35=60.2\times \m_{\rm LT}$ for $l=2,\ m=0,\ p=1,\ q=0$.}}
\label{ciuflt2}
\begin{center}
\begin{tabular}{lllll}
\noalign{\hrule height 1.5pt}
 Tide & $A(\O_{{\rm I}})$
 (mas)& $A(\O_{{\rm II}})$ (mas) & $A(\og_{{\rm II}})$ (mas)&
$\D\m_{\rm LT}$\\
\hline 056.554 $S_{a}$ & -20.55 & 37.71 & -26.17 & $-3.68\times
10^{-3}$
\\  057.555 $S_{sa}$ & -5.98 & 10.98 & -7.62 & $-1.28\times
10^{-3}$ \\  065.455 $M_{m}$ & -0.54 & 1 & -0.69 & $-8.78\times
10^{-5}$ \\  075.555 $M_{f}$ & -0.62 & 1.13 & -0.78 &
$-1.73\times 10^{-4}$ \\
\noalign{\hrule height 1.5pt}
\end{tabular}
\end{center}
\end{table}
 The values of $\D\m_{\rm LT}$ quoted there for the various
zonal tidal lines may be considered as the systematic error in
$\m_{\rm LT}$ due to the chosen constituents, if considered one at
a time by neglecting any possible reciprocal correlation among the
other tidal lines. Table \ref{ciuflt1} and Table \ref{ciuflt2}
show that the percent error in the general relativistic value of
$\m_{\rm LT}$ due to the 18.6-year tide, the most insidious one in
recovering the LT since it superimposes to the gravitomagnetic
trend over time spans of a few year, amounts to 21.9 $\%$, while
for all the other zonal tides it decreases to 0.1 $\%$ or less.
This means that, even if neglected in the satellite orbit
determination models, the $l=2\ m=0$ tides, with the exception of
the 18.6-year tide, do not affect the recovery of $\m_{{\rm LT}}$
by means of the combined residuals.

It is interesting to compare the present results to those released
in ref.\ct{ref:ciu97} for the 18.6-year tide. The value -0.219 due
to the solid component for $\m_{\rm LT}$ quoted in Table
\ref{ciuflt1} must be compared to -0.361 in ref.\ct{ref:ciu97},
with an improvement of $39.3$ $\%$. In the cited work there is no
reference to any estimate of the mismodelling of the 18.6-year
tide, so that we have used the nominal tidal perturbative
amplitudes released in it: $A(\O^{{\rm I}})=-997$ mas, $A(\O^{{\rm
II}})=1,805$ mas and $A(\og^{{\rm II}})=-1,265$ mas. These figures
for the perturbative amplitudes due to the solid Earth tide of
18.6-year are notably different from those quoted in the present
study.  In ref.\ct{ref:ciu97} the theoretical framework in which
they have been calculated (F. Vespe, private communication, 1999)
 is based on the assumption
of a spherical, static, elastic Earth with a single
nominal value of $k_2=0.317$ used  for  the
entire tidal spectrum. The inclusion of the tiny corrections due to the
 Earth's flattening and rotation on the
perturbative amplitudes of $\O$ and $\og$ for the 18.6-year tide
could allow one to slightly improve the related uncertainty in
$\m_{\rm LT}$; it would amount to $20.6\ \%$. But since the
present-day accuracy in laser ranging measurements could hardly
allow one to detect these small effects, their utilization in
\rfr{unolt} is debatable.

Remember that the result quoted for the 18.6-year tide is obtained
in the worst possible case, i.e. a time span of only 1 year and
the assumption that the residuals have been built up by neglecting
completely the zonal tides in the dynamical models used. If, more
realistically, we calculate \rfr{unolt} with the mismodelled
amplitudes quoted in the first row of Table \ref{mismo1} for the
18.6-year tide we obtain, over 1 year, $\D\m=-3.51\times10^{-3}$.
This strongly highlights that many efforts, either theoretical or
experimental, must be done in order to modelling as accurately as
possible such a constituent, so that it can be included in the
nominal background of the orbital determination softwares like
GEODYN II at a satisfactory level of accuracy. The calculations
performed here point toward this goal in the sense that, if we put
our values for the perturbative amplitudes due to the 18.6-year
tide in the models of, e.g., GEODYN II and subsequently build up
the orbital residuals, we expect that the contribution of such
semisecular constituent to $\D\m_{\rm LT}$ amounts to the value
quoted here.

Even though a cancellation is not expected as for the first two
even zonal constituents, by calculating the left hand side of
\rfr{unolt} for the other tides yields, at least, an order of
magnitude of their effect on $\m_{\rm LT}$. An interesting,
unpredicted feature stands out for the odd zonal ocean tides. The
contribution of $l=3$ zonal ocean tidal nominal perturbations over
1 year  to $\D\m_{\rm LT}$ can be found in Table \ref{ciuflt3} and
Table \ref{ciuflt4}.
\begin{table}[ht!]
\caption[Ocean odd zonal tidal contributions to $\D\m_{{\rm LT}}$
for  $p=1,\ q=-1$.]{\footnotesize{Contribution of the odd
 zonal ocean tidal constituents to
$\D\m_{\rm LT}$ by means of the formula $\d\dot\O_{{\rm
I}}+\d\dot\O^{{\rm II}}\times 0.295- \d\dot\og^{{\rm II}}\times
0.35=60.2\times \m_{\rm LT}$ for  $l=3,\ m=0,\ p=1,\ q=-1$.}}
\label{ciuflt3}
\begin{center}
\begin{tabular}{lllll}
\noalign{\hrule height 1.5pt}
 Tide & $A(\O_{{\rm I}})$
 (mas)& $A(\O_{{\rm II}})$ (mas) & $A(\og_{{\rm II}})$ (mas)&
$\D\m_{\rm LT}$\\ \hline
056.554 $S_{a}$ & -0.063 & 0.13 & -114.35 & $0.66$ \\
057.555 $S_{sa}$ & $-9\times 10^{-3}$ & 0.028 & -22.95 & $0.133$ \\
065.455 $M_{m}$ & $-4\times 10^{-4}$ & $1\times 10^{-3}$ & -1.53
& $-8.93\times 10^{-3}$ \\
 075.555 $M_{f}$ & $-1\times 10^{-4}$ &
$7\times 10^{-4}$ & -0.58 &
 $3.41\times 10^{-3}$ \\
\noalign{\hrule height 1.5pt}
\end{tabular}
\end{center}
\end{table}

\begin{table}[ht!]
\caption[Ocean odd zonal tidal contributions  to $\D\m_{{\rm LT}}$
for  $l=3,\ m=0,\ p=2,\ q=1$.]{\footnotesize{Contribution of the
odd
 zonal ocean tidal constituents to
$\D\m_{\rm LT}$ by means of the formula $\d\dot\O^{{\rm
I}}+\d\dot\O^{{\rm II}}\times 0.295- \d\dot\og^{{\rm II}}\times
0.35=60.2\times \m_{\rm LT}$ for  $l=3,\ m=0,\ p=2,\ q=1$.}}
\label{ciuflt4}
\begin{center}
\begin{tabular}{lllll}
\noalign{\hrule height 1.5pt} Tide & $A(\O_{{\rm I}})$
 (mas)& $A(\O_{{\rm II}})$ (mas) & $A(\og_{{\rm II}})$ (mas)&
$\D\m_{\rm LT}$\\
\hline 056.554 $S_{a}$ & 0.047 & -0.36 & 297.34 & $-1.72$ \\
057.555 $S_{sa}$ & $7\times 10^{-3}$ & -0.044 & 36.07 & $-0.209$ \\
065.455 $M_{m}$ & $4\times 10^{-4}$ & $-2\times 10^{-3}$ & 1.642 &
$-9.55\times 10^{-3}$ \\
075.555 $M_{f}$ & $1.79\times 10^{-4}$ & $-7\times 10^{-4}$ &
-0.60 &
$-3.52\times 10^{-3}$ \\
\noalign{\hrule height 1.5pt}
\end{tabular}
\end{center}
\end{table}
By inspecting them it is clear that the sensitivity of perigee of
LAGEOS\ II to the $l=3$ part of the ocean tidal spectrum may
affect the recovery of the Lense-Thirring parameter $\m_{\rm LT}$
by means of the proposed combined residuals, especially as far as
$S_a$ and $S_{sa}$ are concerned. This fact agrees with the
results of Table \ref{mismo2} which tells us that the mismodelled
parts of $S_{a}$ and $S_{sa}$ are not negligible fractions of
$\D\og^{{\rm II}}_{{\rm LT}}$. However, if the mismodelled
amplitudes are employed in \rfr{unolt} it can be seen that, over 1
year, a cancellation of the order of $10^{-1}$ ($S_{a}\ l=3\ p=2$)
and $10^{-2}$ ($S_{a}\ l=3\ p=1;\ S_{sa}\ l=3\ p=1,2$) takes
place. The contributions of the mismodelling on $M_{m}$ and
$M_{f}$ are completely negligible. So, we can conclude that also
the $l=3$ part of the zonal ocean tidal spectrum may not affect
the combined residuals in a notable manner if the $l=3$ part of
$S_a$ and $S_{sa}$ is properly accounted for.
\subsection{The simulated residual
signal} The first step of our strategy  was to generate with
MATLAB a time series which simulates, at the same length and
resolution,  the real residual curve obtainable from $y=\d\O_{{\rm
exp}}^{{\rm I}}+c_1\d\O_{{\rm exp}}^{{\rm II}}+c_2\d\og_{{\rm
exp}}^{{\rm II}}$ through, e.g., GEODYN. This simulated curve
(``Input Model" - IM in the following) was constructed with

$\bullet$ The secular Lense-Thirring trend as predicted by the
General Relativity\footnote{Remember that in the dynamical models
of GEODYN II it was set purposely equal to 0 so that the residuals
absorbed (contained) entirely the relativistic effect
\ct{ref:ciu97}.}

$\bullet$ A certain number of sinusoids of the form $\d
A_k\cos{(\rp{2\p}{P_k}t+\f_k)}$ with known periods $P_k,\ k=1,..N$
simulating the mismodelled tides and other long-period signals
which, to the level of assumed mismodelling, affect the combined
residuals

$\bullet$ A  noise of given amplitude which simulates the
experimental errors in the laser-ranged measurements and,
depending on the characteristics chosen for it, some other
physical forces.

In a nutshell \eqi {\rm IM}={\rm LT}+[{\rm mismodelled\
tides}]+[{\rm other\ mismodelled\ long\ period\ signals}]+[{\rm
noise}].\eqf The harmonics included in the IM
are the following\\
$\bullet\ K_1,\ l=2$ solid and ocean; node of LAGEOS (P=1,043.67 days)\\
$\bullet\ K_1,\ l=2$ solid and ocean; node and perigee of LAGEOS\
II
(P=-569.21 days)\\
$\bullet\ K_1,\ l=3,\ p=1$  ocean; perigee of LAGEOS\ II (P=-1,851.9 days)\\
$\bullet\ K_1,\ l=3,\ p=2$ ocean; perigee of LAGEOS\ II (P=-336.28 days)\\
$\bullet\ K_2,\ l=3,\ p=1$ ocean; perigee of LAGEOS\ II (P=-435.3 days)\\
$\bullet\ K_2,\ l=3,\ p=2$  ocean; perigee of LAGEOS\ II (P=-211.4 days)\\
$\bullet\ 165.565,\ l=2$ solid; node of LAGEOS (P=904.77 days)\\
$\bullet\ 165.565,\ l=2$ solid; node and perigee of LAGEOS\ II
(P=-621.22 days)\\
$\bullet\ S_2,\ l=2$ solid and ocean; node of LAGEOS (P=-280.93 days)\\
$\bullet\ S_2,\ l=2$ solid and ocean; node and perigee of LAGEOS\
II
(P=-111.24 days)\\
$\bullet\ S_2,\ l=3,\ p=1$ ocean; perigee of LAGEOS\ II (P=-128.6 days)\\
$\bullet\ S_2,\ l=3,\ p=2$ ocean; perigee of LAGEOS\ II (P=-97.9 days)\\
$\bullet\ P_1,\ l=2$ solid and ocean; node of LAGEOS (P=-221.35 days)\\
$\bullet\ P_1,\ l=2$ solid and ocean; node and perigee of LAGEOS\
II
(P=-138.26 days)\\
$\bullet\ P_1,\ l=3,\ p=1$ ocean; perigee of LAGEOS\ II (P=-166.2 days)\\
$\bullet\ P_1,\ l=3,\ p=2$ ocean; perigee of LAGEOS\ II (P=-118.35 days)\\
$\bullet\ \textrm{Solar\ Radiation\ Pressure},$  perigee of LAGEOS\ II (P=-4,241 days)\\
$\bullet\ \textrm{Solar\ Radiation\ Pressure},$ perigee of LAGEOS\ II (P=657 days)\\
$\bullet\ \textrm{perigee\ odd\ zonal}\ C_{30}$, perigee of LAGEOS\ II (P=821.79) \\
In the following the signals due to solar radiation pressure will
be denoted as {\rm SRP}(P) where P is their period; the effects of
the eclipses and Earth penumbra have not been accounted for. Many
of the periodicities listed above have been actually found in the
Fourier spectrum of the real residual curve \ct{ref:ciu97}.
Concerning $K_1\ l=3\ p=1$ and {\rm SRP}(4,241), see subsect. 5.

When the real data are collected they refer to a unique,
unrepeatable situation characterized by  certain starting and
ending dates for $T_{{\rm obs}}$. This means that each analysis
which could be carried out in the real world necessarily refers to
a given set of initial phases and noise for the residual curve
corresponding to the chosen observational period; if the data are
collected over the same $T_{{\rm obs}}$ shifted backward or
forward in time, in general, such features of the curve will
change. We neither know a priori when the next real experiment
will be performed, nor which will be the set of initial phases and
the level of experimental errors accounted for by the noise.
Moreover, maybe the dynamical models of the orbit determination
software employed are out of date in regard to the perturbations
acting upon satellites' orbits or do not include some of them at
all. Consequently, it would be incorrect to work with a single
simulated curve, fixed by an arbitrary set of $\d A_k,\ \f_k$ and
noise, because it could refer to a situation different from that
in which, in the real world, the residuals will actually
correspond.

The need for great flexibility in generating the IM becomes
apparent: to account for the entire spectrum of possibilities
occurring when the real analysis will be carried out. We therefore
decided to build into the MATLAB routine the capability to  vary
randomly the initial phases $\f_k$, the noise and the amplitude
errors $\d A_k$. Concerning the error amplitudes of the harmonics,
they can be  randomly varied so that $\d A_k$ spans $[0,\ \d
A_k^{\rm nom}]$ where $\d A^{\rm nom}_k$ is the nominal
mismodelled amplitude calculated taking into account Table
\ref{mismo1} and Table \ref{mismo2}; it means that we assume  they
are reliable estimates of the differences between the real data
and the dynamical models of the orbital determination softwares,
i.e. of the residuals. The MATLAB routine allows also the user to
decide which harmonic is to be included in the IM; it is also
possible to choose the length of the time series $T_{{\rm obs}}$,
the sampling step $\D t$ and  the amplitude of the noise.

In order to assess quantitatively this feature we proceeded as
follows. First, over a time span of 3.1 years, we fitted the IM
with a straight line only, finding for a choice of random phases
and noise which qualitatively reproduces the curve shown in
ref.\ct{ref:ciu97}, the value of 38.25 mas for the root mean
square of the post fit IM. The value quoted in
ref.\ct{ref:ciu97}is 43 mas. Second, as done in the cited paper,
we fitted the complete IM with the LT plus a set of long-period
signals (see subsect. 4) and then we subtracted the so adjusted
harmonics from the original IM obtaining a  ``residual" simulated
signal curve. The latter was subsequently fitted with a straight
line only, finding a rms post fit of 12.3 mas (it is nearly
independent of the random phases and the noise) versus 13 mas
quoted in \ct{ref:ciu97}. A uniformly distributed noise with
nonzero average and amplitude of 50 mas was used (see subsect. 4).
These considerations suggest that the simulation procedure adopted
is reliable, replicates the real world satisfactorily and yields a
good starting point for conducting our sensitivity analysis.

\subsection{Sensitivity analysis}
A preliminary analysis was carried out in order to evaluate the
importance of the mismodelled solid tides on the recovery of
$\m_{\rm LT}$. We calculated the average\eqi \rp{1}{T_{{\rm
obs}}}\int_0^{T_{{\rm obs}}}[{\rm solid\ tides}]dt,\eqf where
$[\textrm{solid\ tides}]$ denotes the analytical expressions of
the mismodelled solid tidal perturbations as given by
\rfrs{juun}{du}. Subsequently, we compared it to the value of the
gravitomagnetic trend for the same $T_{{\rm obs}}$.   The tests
were repeated by varying $\D t$, $T_{{\rm obs}}$ and the initial
phases. They have shown that the mismodelled part of the solid
Earth tidal spectrum is entirely negligible with respect to the LT
signal, falling always below $1\%$ of the gravitomagnetic shift of
the combined residuals accumulated over the examined $T_{{\rm
obs}}$.

For the ocean tides and the other long-period signals the problem
was addressed in a different way. First, for a given $\D t$ and
different time series lengths, we included in the IM the effects
of LT, the noise and the solid tides only: subsequently we fitted
it simply by means of a straight line. In a second step we have
simultaneously added, both to the IM and the fitting model (FM
hereafter), all the ocean tides, the solar radiation pressure {\rm
SRP}(657) signal and the odd zonal geopotential harmonic. We then
compared the fitted values of $\m_{\rm LT}$ and $\d\m_{\rm LT}$
recovered from both cases in order to evaluate $\D\m_{\rm LT}$ and
$\D\d\m_{\rm LT}$. The least squares fits \ct{ref:bar,ref:dra}
were performed by means of the MATLAB routine ``nlleasqr" (see,
e.g., {\bf http://www.ill.fr/tas/matlab/doc/mfit.html}); as
$\d\m_{\rm LT}$ we have assumed the square root of the diagonal
covariance matrix element relative to the slope parameter. It
simulates the formal experimental error which, in the context of
the \grc\ \lg\ experiment, is far smaller than the systematic
errors which, consequently, must be carefully assessed.
  Notice that the {\rm SRP}(4,241) has never been included in the FM (See
section 5), while $K_1\ l=3\ p=1$ has been included for $T_{{\rm
obs}}>5$ years. For both of the described scenarios we have taken
the average for $\m_{\rm LT}$ and $\d\m_{\rm LT}$ over 1,500 runs
performed by varying randomly the initial phases, the noise and
the amplitudes of the mismodelled signals in order to account for
all possible situations occurring in the real world, as pointed
out in the previous section. The large number of repetitions was
chosen in order to avoid that statistical fluctuations in the
results could ``leak" into $\D\ol{\m_{\rm LT}}$ and
$\D\ol{\d\m_{\rm LT}}$ and corrupt them at the level of $1\%$.
With 1,500 runs the standard deviations on $\ol{\m_{\rm LT}}$ and
$\ol{\d\m_{\rm LT}}$ are of the order of $10^{-3}$ or less, so
that we can reliably use the results of such averages for our
comparisons of $\ol{\m_{\rm LT}}$(no tides) vs $\ol{\m_{\rm
LT}}$(all tides).

Before implementing such strategy for different $T_{{\rm obs}},\
\D t$ and noise we carefully analyzed the $\D t=15$ days, $T_{{\rm
obs}}$=4 years experiment considered in \ct{ref:ciu98} by trying
to obtain the quantitative features outlined there, so that we
start from a firm and reliable basis. This goal was achieved  by
proceeding as outlined in the previous section and adopting a
uniformly distributed random noise with an amplitude of 35
mas. 

\begin{figure}[ht!]
\begin{center}
\includegraphics*[width=10cm,height=7cm]{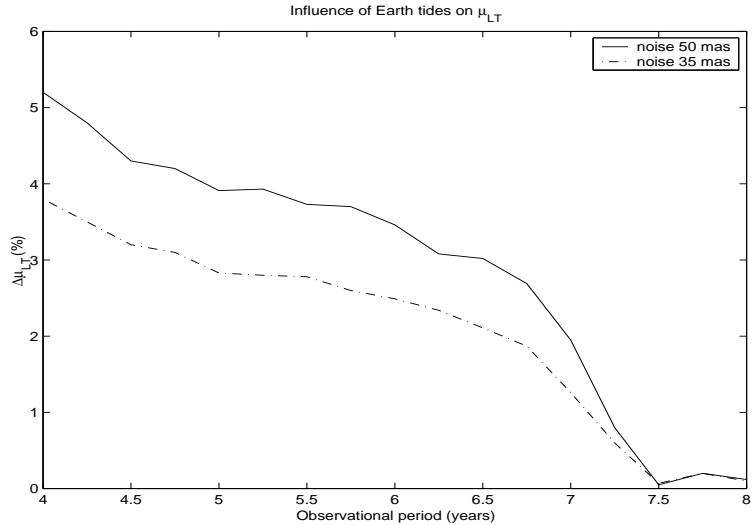}
\end{center}
\caption[Systematic error induced by Earth's solid and ocean tides
on the determination of Lense-Thirring trend over various time
spans. ]{\footnotesize Effects of the long period signals on the
recovery of $\m_{{\rm LT}}$ for $\D t$=15 days and different
choice of uniform random noise. Each point in the curves
represents an average over 1,500 runs performed by varying
randomly the initial phases and the noise' s pattern. $\D\m_{\rm
LT}$ is the difference between the least squares fitted value of
$\m_{\rm LT}$ when both the IM and the FM includes the LT plus all
the harmonics and that obtained without any harmonic both in the
IM and in the FM.} \label{dmu}
\end{figure}

Fig. \ref{dmu} shows the results for $\D\ol{\m_{\rm LT}}$ obtained
with $\D t=15$ days and a uniform random noise with amplitudes of
50 mas and 35 mas corresponding to the characteristics of the real
curves in \ct{ref:ciu97,ref:ciu98}.
 Notice that our
estimates for the case $T_{{\rm obs}}=4$ years almost coincide
with those in \ct{ref:ciu98} who claim $\D\m_{{\rm tides}}\leq
4\%$. Notice that for $T_{{\rm obs}}>7$ years the effect of tidal
perturbations errors falls around $2\%$. By choosing different $\D
t$ does not introduce appreciable modifications to the results
presented here. This fact was tested by repeating the set of runs
with $\D t=7$ and $22$ days.

Up to this point we dealt with the entire set of long-period
signals affecting the combined residuals; now we ask if it is
possible  to assess individually the influence of each tide on the
recovery of LT. We shall focus our attention on the case $\D t=15$
days, $T_{{\rm obs}}$=4 years.

In order to evaluate the influence of each tidal constituent on
the recovery of $\m_{\rm LT}$ two complementary approaches could
be followed in principle. The first one consists of starting
without any long-period signal either in the IM or in the FM, and
subsequently adding to them one harmonic at a time, while
neglecting all the others. In doing so it is implicitly assumed
that each constituent is mutually uncorrelated with any other one
present in the signal. In fact, the matter is quite different
since if for the complete model case we consider the covariance
and correlation matrices of the FM adjusted parameters it turns
out that the LT is strongly correlated, at a level of
$\vass{Corr(i,j)}> 0.9$, with certain harmonics, which happen to
be
mutually correlated too. These are\\
$\bullet\ K_1,\ l=2$ (P=1,043.67 days; 1.39 cycles completed)\\
$\bullet\ K_1,\ l=2$ (P=-569.21 days; 2.56 cycles completed)\\
$\bullet\ K_1,\ l=3,\ p=2$ (P=-336.28 days; 4.34 cycles completed)\\
$\bullet\ K_2,\ l=3,\ p=1$ (P=-435.3 days; 3.35 cycles completed)\\
$\bullet\ \textrm{Solar\ Radiation\ Pressure},$ (P=657 days; 2.22 cycles completed)\\
$\bullet\ \textrm{perigee\ odd\ zonal}\ C_{30}$, (P=821.79; 1.77 cycles completed)\\
Their FM parameters are indeterminate in the sense that the
values estimated are smaller than the relative uncertainties assumed to be
$\sqrt{Cov(i,i)}$. On the contrary, there are other signals which are
poorly correlated to the LT and whose reciprocal
correlation too is very low and that are well determined.
These are\\
$\bullet\ K_2,\ l=3,\ p=2$ (P=-211.4 days;  6.91 cycles completed)\\
$\bullet\ S_2,\ l=3,\ p=1$ (P=-128.6 days;  11.3 cycles completed)\\
$\bullet\ S_2,\ l=3,\ p=2$ (P=-97.9 days;  14.9 cycles completed)\\
$\bullet\ P_1,\ l=3,\ p=1$ (P=-166.2 days; 8.78 cycles completed)\\
$\bullet\ P_1,\ l=3,\ p=2$ (P=-118.35 days; 12.3 cycles completed)\\
It is interesting to notice that over $T_{{\rm obs}}=4$ years the
signals uncorrelated with the LT have in general described many
complete cycles, contrary to these correlated with LT. This means
that  the signals that average out over $T_{{\rm obs}}$
decorrelate with LT, allowing the gravitomagnetic trend to emerge
clearly against the background, almost not affecting the LT
recovery. This feature has been tested as follows. In a first
step, all the uncorrelated tides have been removed from both the
IM (50 mas uniform random noise) and the FM, leaving only the
correlated tides in. The runs were then repeated,  all other
things being equal,  the following values were recorded:
$\ol{\m_{\rm LT}}=1.2104,\ \ol{\d\m_{\rm LT}}=0.1006$ with a
variation with respect to the complete model case ($\ol{\m_{\rm
LT}}=1.2073$, $\ol{\d\m_{\rm LT}}=0.1295$ ) of: $\D\ol{\m_{\rm
LT}}\simeq 0.3\%,\ \D\ol{\d\m_{\rm LT}}\simeq 2.9\%$. Conversely,
if all the signals with strong correlation  are removed from the
simulated data and from the FM, leaving only the uncorrelated ones
in, we obtain: $\ol{\m_{\rm LT}}=1.1587,\ \ol{\d\m_{\rm LT}}=
0.0186$ with $\D\ol{\m_{\rm LT}}\simeq 4.8\%,\ \D\ol{\d\m_{\rm
LT}}\simeq 11\%$. Notice that the sum of both contributions for
$\D\ol{\m_{\rm LT}}$ yields exactly the overall value
$\D\ol{\m_{\rm LT}}=5.2\%$ as obtained in the previous analysis
(cf. fig. \ref{dmu} for $T_{{\rm obs}}=4$ years). This highlights
the importance of certain long period signals in affecting the
recovery of LT and justify the choice of treating them
simultaneously as it was done in obtaining fig. \ref{dmu}.
Moreover, it clearly indicates that the efforts of the scientific
community should be focused on the improvement of the knowledge of
these tidal constituents. It has been tested that for $T_{{\rm
obs}}=8$ years all such ``geometrical" correlations among the LT
and the harmonics nearly disappear, as it would be intuitively
expected.
\subsection{The effect of the very long periodic harmonics}
In this section we shall deal with those signals whose periods are
longer than 4 years which could  corrupt the recovery of LT
resembling superimposed trends if their periods are considerably
longer than the  adopted time series length.

In section 3 we have shown the existence of a very long periodic
ocean tidal perturbation acting upon the perigee of LAGEOS\ II. It
is the $K_1\ l=3\ p=1\ q=-1$ constituent with period $P=1,851.9$
days (5.07 years)  and nominal amplitudes of $-1,136$ mas. In
ref.\ct{ref:luc}, for the effect of the direct solar radiation
pressure on the perigee of LAGEOS\ II, it has been explicitly
calculated, by neglecting the effects of the eclipses, a signal
{\rm SRP}(4,241) with $P=4,241$ days (11.6 years) and nominal
amplitude of 6400 mas. The mismodelling on these harmonics, both
of the form $\d A\sin{(\rp{2\p}{P}t+\f)}$, amounts to 64.5 mas for
the tidal constituent, as shown in section 3.6, and to 32 mas for
{\rm SRP}(4,241), according to ref.\ct{ref:luc}.

About the actual presence of such semisecular harmonics in  the
spectrum of the real combined residuals, it must be pointed out
that, over $T_{{\rm obs}}=3.1$ years \ct{ref:ciu97}, it is not
possible that so low frequencies could be resolved by Fourier
analysis. Indeed, according to refs.\ct{ref:god,ref:pri}, when a
signal is sampled over a finite time interval $T_{{\rm obs}}$ it
induces a sampling also in the spectrum. The lowest frequency that
can be resolved is \eqi f_{\rm min}=\rp{1}{2T_{{\rm
obs}}},\lb{freqz}\eqf i.e. a harmonic must describe, at least,
half a cycle over $T_{{\rm obs}}$ in order to be detected in the
spectrum. $f_{\rm min}$ is called elementary frequency band and it
also represents the minimum separation between two frequencies in
order to be resolved. For our two signals we have
$f(K_1)=5.39\times 10^{-4}$ cycles per day (cpd) and $f({\rm
SRP})=2.35\times 10^{-4}$ cpd; over 3.1 years $f_{\rm
min}=4.41\times 10^{-4}$ cpd. This means that the two harmonics
neither can be resolved as distinct nor they can be found in the
spectrum at all. In order to resolve them we should wait for
$T_{{\rm obs}}=4.5$ years which corresponds to their separation
$\D f=3.04\times 10^{-4}$ cpd, according to \rfr{freqz}. In view
of the potentially large aliasing effect of these two harmonics on
the LT it was decided to include both $K_1\ l=3\ p=1$ and {\rm
SRP}(4,241) in the simulated residual curve at the level of
mismodelling claimed before.

In order to obtain an upper bound of their contribution to the
systematic uncertainty in  $\m_{{\rm LT}}$ we proceeded as
follows. First, we calculated the temporal average of the
perturbations induced on the combined residuals by the two
harmonics over different $T_{{\rm obs}}$ according to \eqi
I=\rp{1}{T_{{\rm obs}}}\int_{0}^{T_{{\rm obs}}}c_2\d A
\sin{(\rp{2\p}{P} t+\f)}dt =c_2\rp{\d
A}{\t}(\cos{\f}+\sin{\f}\sin{\t}-\cos{\f}\cos{\t}),\lb{ave}\eqf
with $\t=2\pi\rp{T_{{\rm obs}}}{P}$.
It turns out that, for given $T_{{\rm obs}}$,  the averages are
periodic functions of the initial phases $\f$ with period $2\p$.
Since the gravitomagnetic trend is  positive and  growing in time,
we shall consider only the positive values of the temporal
averages corresponding to those portions of sinusoid which are
themselves positive. Moreover, notice that, in this case, one
should consider if the perturbation' s arc is rising or falling
over the considered $T_{{\rm obs}}$: indeed, even though the
corresponding averages could be equal, it is only in the first
case that the sinusoid has an aliasing effect on the LT trend. The
values of $\f$ which maximize the averages were found numerically
and for such  values  the maxima of the averages were calculated.
Subsequently, these results in mas were compared to the amount of
the predicted gravitomagnetic shift accumulated over the chosen
$T_{{\rm obs}}$ by the combined residuals $y_{{\rm LT}}=60.2$
(mas/y)$\times T_{{\rm obs}}$ (y). The results are shown in Table
\ref{mediak1} and Table \ref{mediasrp} and, as previously
outlined, represent a pessimistic estimate.
\begin{table}[ht!]
\caption[Impact of the averaged mismodelled $K_1\ l=3\ p=1$ tide
on the Lense-Thirring trend.]{\footnotesize{Effect of the averaged
mismodelled harmonic $K_1\ l=3\ p=1$ on the Lense-Thirring trend
for different $T_{{\rm obs}}$. In order to obtain upper limits the
maximum value for the average of the tidal constituent has been
taken, while for the gravitomagnetic effect it has been simply
taken the value $\dot y_{{\rm LT}}\times T_{{\rm
obs}}$.}}\label{mediak1}
\begin{center}
\begin{tabular}{llll}
\noalign{\hrule height 1.5pt}  \footnotesize{$T_{{\rm obs}}$
(years)} & \footnotesize{max $c_2\left\langle\d\og^{{\rm
II}}\right\rangle$ (mas)} & \footnotesize{$y_{{\rm LT}}$ (mas)} &
\footnotesize{max $c_2\left\langle\d\og^{{\rm II}}\right\rangle/y_{{\rm LT}}\  (\%)$ }\\
\hline
4 & 5.6 & 240.8 & 2.3  \\
5 & 0.3 & 301 &  0.09\\
6  & 3.3 & 361.2 & 0.9 \\
7 & 4.8 &  421.4 & 1.1\\
\noalign{\hrule height 1.5pt}
\end{tabular}
\end{center}
\end{table}

\begin{table}[ht!]
\caption[Impact of the averaged mismodelled ${\rm SRP}(4,241)$
harmonic on the Lense-Thirring trend.]{\footnotesize{Effect of the
averaged mismodelled harmonic ${\rm SRP}(4,241)$ on the
Lense-Thirring trend for different $T_{{\rm obs}}$. In order to
obtain upper limits the maximum value for the average of the
radiative harmonic has been taken, while for the gravitomagnetic
effect it has been simply taken the value $\dot y_{{\rm LT}}\times
T_{{\rm obs}}$.}}\label{mediasrp}
\begin{center}
\begin{tabular}{llll}
\noalign{\hrule height 1.5pt} \footnotesize{$T_{{\rm obs}}$
(years)} & \footnotesize{max $c_2\left\langle\d\og^{{\rm
II}}\right\rangle$ (mas)} & \footnotesize{$y_{{\rm LT}}$ (mas)} &
\footnotesize{max $c_2\left\langle\d\og^{{\rm II}}\right\rangle/y_{{\rm LT}}\  (\%)$} \\
\hline
4 & 9.1 & 240.8 & 3.7  \\
5 & 8 & 301 &  2.6\\
6  & 6.8 & 361.2 & 1.8 \\
7 & 5.6 &  421.4 & 1.3\\
\noalign{\hrule height 1.5pt}
\end{tabular}
\end{center}
\end{table}
It is interesting to notice that an analysis over $T_{{\rm
obs}}=5$ years, that should not be much more demanding than the
already performed works, could cancel out the effect of the $K_1\
l=3\ p=1$ tide; in this scenario the effect of {\rm SRP}(4,241)
should amount, at most to $2.6\%$ of the LT effect. Notice also
that for $T_{{\rm obs}}=7$ years, a time span sufficient for the
LT to emerge on the background of  most of the other tidal
perturbations, the estimates of fig. \ref{dmu} are compatible with
those of Table \ref{mediak1} and Table \ref{mediasrp} which
predict an upper contribution of $1.1\%$ from the $K_1 \ l=3\ p=1$
and $1.3\%$ from {\rm SRP}(4,241) on the LT parameter $\m_{\rm
LT}$.

An approach, similar to that used in \ct{ref:ves} in order to
assess the influence of the eclipses and Earth penumbra effects on
the perigee of LAGEOS\ II has been applied also to our case. It
consists of fitting with a straight line only the mismodelled
perturbation to be considered and, subsequently, comparing the
slope of such fits to that due to gravitomagnetism which is equal
to 1 in units of 60.2 mas/y. We have applied this method to $K_1 \
l=3\ p=1$ and {\rm SRP}(4,241) with the already cited mismodelled
amplitudes and by varying randomly the initial phases $\f$ within
$[-2\p,\ 2\p]$. The mean value of the fit' s slope, in units of
60.2 mas/y, over 1,500 runs is very close to zero. Concerning the
upper limits of $\D\m_{\rm LT}$ derived from these runs, they
agree with those released in Table \ref{mediak1} and Table
\ref{mediasrp} up to 1-2 $\%$.

The method of temporal averages can be successfully applied also
for the $l=2\ m=0$ 18.6-year tide. It is a very long period zonal
tidal perturbation which could potentially reveal itself as the
most dangerous in aliasing the results for $\m_{\rm LT}$ since its
nominal amplitude is very large and its period is much longer than
the $T_{{\rm obs}}$ which could be adopted for real analysis.
Ciufolini in ref.\ct{ref:ciu96} claims that the combined residuals
have the merit of cancelling out all the static and dynamical
geopotential' s contributions of degree $l=2,4$ and order $m=0$,
so that the 18.6-year tide would not create problems. This topic
was quantitatively addressed in a preliminary way. Indeed, in
order to make comparisons with other works, we have simply
calculated for $T_{{\rm obs}}=1$ year the combined residuals with
only the mismodelled amplitudes of the 18.6-year tidal
perturbations on the nodes of LAGEOS and LAGEOS\ II and the
perigee of LAGEOS\ II. This means that the dynamical pattern over
the time span of such important tidal perturbation was not
investigated. We did this by calculating the average over
different $T_{{\rm obs}}$. The results show clearly that the
18.6-year tide does not affect at all the estimation of $\m_{\rm
LT}$ if we adopt as observable the combined residuals proposed by
Ciufolini. Indeed, for $T_{{\rm obs}}=4$ years, it turns out that
the average effect will reach, at most, $0.08\%$ of the
gravitomagnetic shift over the same time span.

This feature of the 18.6-year tide is confirmed also by fitting
with a straight line only the sinusoid of this perturbation on the
combined residual: the adjusted slope amounts, at most,  to less
than 1$\%$ of the gravitomagnetic effect for different $T_{{\rm
obs}}$.



\subsection{Conclusions}
The detection of the Lense-Thirring effect can be achieved by
means of the \rfr{unolt}. Ciufolini in ref.\ct{ref:ciu96} claims
that the static and dynamical terms of degree $l=2,4$ and order
$m=0$ of the geopotential do not affect his proposed combined
residuals.

In this section we have tested this statement in regard  not only
to the $l=2,4$ $m=0$ solid and ocean Earth tides, but also to the
$m=1,2$ ones and to the other long-period perturbations which
affect the combined residuals.

Concerning the zonal tides, in order to compare our results to
those released in \ct{ref:ciu97}, we have preliminarily used the
nominal perturbative amplitudes by calculating \rfr{unolt} on a
time span of 1 year. As far as the $l=2\ m=0$ tides are concerned,
the 18.6-year tide cancels out at a level of $10^{-1}$ only. But
if we repeat the calculations with the mismodelled amplitudes over
1 year the accuracy of the cancellation grows to $10^{-3}$. This
shows that the more accurate the dynamical models employed in
building up the orbital residuals are, the more accurate the
recovery of the Lense-Thirring becomes. The $l=4$ zonal tides do
not create problems. Also the $l=3\ m=0$ tides, and this is an
unpredicted feature, although at a lesser extent, cancel out at a
level ranging from $10^{-1}$-$10^{-2}$ if they are properly
accounted in the residuals' construction. This points towards a
better understanding, from both a theoretical and experimental
point of view, of the $l=3,\ m=0$ part of the tidal spectrum. The
results presented in this section not only confirm the usefulness
of the formula by Ciufolini for the $l=2,4\ m=0$ tides, but also
extend its validity to the $l=3\ m=0$ part of the tidal response
spectrum.

In regard to the tesseral and sectorial tides, on one hand, by
simulating the real residual curve in order to reproduce as
closely as possible the results obtained for the $\D t=15$ days $
T_{{\rm obs}}=4$ years scenario published in \ct{ref:ciu98} it has
been possible to refine and detail the estimates for it. On the
other hand,  this procedure also extends them to longer
observational periods in view of new, more sophisticated analysis
to be completed in the near future based on real data analyzed
with the orbit determination software GEODYN II in collaboration
with the teams from the Joint Center for Earth Systems Technology
at NASA Goddard Space Flight Center and at the University of Rome
La Sapienza. Since such numerical analysis is very demanding in
terms of both time employed and results analysis burden, it should
be very useful to have  a priori estimates which could better
direct the work. This could be done, e.g., by identifying which
tidal constituents $\m_{{\rm LT}}$ is  more or less sensitive to
in order to seek improved dynamical models for use in GEODYN.

As far as the perturbations generated by the solid Earth tides,
the high level of accuracy with which they are known has yielded a
contribution to the systematic errors in $\m_{{\rm LT}}$ which
falls well below 1$\%$, so that they are of no concern at present.

Concerning the ocean tidal perturbations and the other long-period
harmonics, for those  whose periods are shorter than 4 years, the
role played by $T_{{\rm obs}}$, $\D t$ and the noise has been
investigated. It turned out that $\D t$ has no discernible effect
on the adjusted value of $\m_{\rm LT}$, while $T_{{\rm obs}}$ is
very important and so is the noise. The main results for these are
summarized in fig. \ref{dmu}, which tells us that the entire set
of long-period signals, if properly accounted for in building up
the residuals, affect the recovery of the Lense-Thirring effect at
a level not worse than $4\%-5\%$ for $T_{{\rm obs}}=4 $ years; the
error contribution diminishes to about $2\%$ after 7 years of
observations.

We have also shown which tides are strongly anticorrelated and
correlated with the gravitomagnetic trend over 4 years of
observations. The experimental and theoretical efforts should
concentrate on improving these constituents in particular. This
geometrical correlation tends to diminish as $T_{{\rm obs}}$
grows. This can be intuitively recognized by noting that the
longer  $T_{{\rm obs}}$ is, the larger the number of cycles these
periodic signals are sampled over and cleaner the way in which the
secular Lense-Thirring trend emerges against the background
``noise".

The ocean tide constituents $K_1\ l=3\ p=1$ and the solar
radiation pressure harmonic {\rm SRP}(4,241) generate
perturbations on the perigee of LAGEOS\ II with periods of 5.07
years and 11.6 years respectively, so that they act on the
Lense-Thirring effect as biases and corrupt its determination with
the related mismodelling: indeed they may resemble  trends if
$T_{{\rm obs}}$ is shorter than their periods. They were included
in the simulated residual curve and their effect was evaluated in
different ways with respect to the other tides. An upper bound was
calculated for their action and it turns out that they contribute
to the systematic uncertainty in the recovery of $\m_{{\rm LT}}$
at a level of less than 4$\%$  depending on $T_{{\rm obs}}$ and
the initial phases $\f$. The results are summarized in Table
\ref{mediak1} and Table \ref{mediasrp}. An observational period of
5 years, which seems to be a reasonable choice in terms of time
scale and computational burden, allows one to average out the
effect of the $K_1\ l=3\ p=1$.

The strategy followed for the latter harmonics has been extended
also to the $l=2\ m=0$ zonal 18.6-year tide. The obtained results
confirm the claim in \ct{ref:ciu96} that it does not affect the
combined residuals.

In conclusion, the strategy presented here could be used as
follows. Starting from a simulated residual curve based on the
state of art of the real analysis performed until now, it provides
helpful indications in order to improve the force models of the
orbit determination software as far as tidal perturbations are
concerned and to perform new analysis with real residuals.
Moreover, when real data will be collected for a given scenario it
will be possible to use them in our software  in order to adapt
the simulation procedure to the new situation; e.g. it is expected
that the noise level in the near future will diminish in view of
improvements in laser ranging technology and modelling. Thus, we
shall repeat our analysis for $\D\m_{{\rm tides}}$ when these new
results become available.
\section{Measuring the gravitoelectric perigee advance with LAGEOS and LAGEOS\ II}
In order to test general relativity versus alternative metric
theories of gravitation, following a method introduced by
Eddington (1922), Robertson (1962) and Schiff (1967), one can
expand the spacetime metric (which describes the gravitational
interaction) at the order beyond Newtonian theory (post-Newtonian)
and then multiply each post-Newtonian term by a dimensionless
parameter, the PPN parameters, to be experimentally determined.
Using this method, Nordtvedt and Will have developed the PPN
(Parameterized-Post-Newtonian) formalism \ct{ref:wil93}, a
powerful and useful tool for testing general relativity and
alternative metric theories. Therefore, the PPN formalism is a
post--Newtonian parameterized expansion of the metric tensor ${\bf
g}$ and of the energy-momentum tensor ${\bf T}$ in terms of small
known classical potentials.

In the slow-motion and weak-field approximation, at the so-called
post-Newtonian limit, the structure of known metric theories of
gravity is identical apart from the numerical values of the PPN
parameters which appear in the expansion of the metric
coefficients \ct{ref:wil93,ref:ciuwhe,ref:wil01}. Especially
meaningful are the parameters $\beta$ and $\ga$, i.e the usual
Eddington-Robertson-Schiff parameters used to describe the
``classical" tests of \gr. $\ga$ measures how much space curvature
is produced by unit rest mass and in the standard parameterized
post-Newtonian gauge $\b$ accounts for the level of nonlinearity
in the superposition law for gravity \ct{ref:mis}. In general
relativity $\beta = 1$ and $\ga = 1$.

To-day a very accurate measurement of $\ga$ was performed by using
the Viking time delay \ct{ref:rea}: the results was $\ga=1.000\pm
2\times 10^{\rm -3}$. The quoted uncertainty allows for possible
systematic errors. Other recent measurements exploit the
gravitational bending of the electromagnetic waves at various
wavelength \ct{ref:fro} obtaining an uncertainty of the order of
$10^{\rm -3}$; e.g., according to the VLBI radio analysis by
\ct{ref:leb} $\ga=0.9996\pm1.7\times 10^{-3}$, while in
ref.\ct{ref:fro} $\ga=0.997\pm 3\times 10^{\rm -3}$ is quoted
based on the astrometric observations of the electromagnetic waves
deflection in the visible. An improvement in the accuracy of two
orders of magnitude \ct{ref:fro} is expected from the future GAIA
astrometric mission \ct{ref:gai}. Lunar laser ranging (LLR)
measurements of the geodetic precession yields for $\ga$ an
accuracy of $10^{\rm -2}$ \ct{ref:willetal}. However a careful
discussion of the different terms in the lunar motion yields
eventually $\ga=1.000\pm 5\times 10^{\rm -3}$ as the present LLR
result \ct{ref:fro}.

The value of the parameter $\b$, which cannot be measured
independently, can be obtained from the measured values of $\ga$
and of some combinations of $\ga$ and $\b$: $\et=\nvd$ and $\ppn$
are the most accurately determined. General relativity predicts
$\et=0$ and $\ppn=1$. For $\beta$ in ref.\ct{ref:and} by using LLR
data $\delta\beta=4.7\times 10^{\rm -4}$ is reported.

The combination $\et$ measures the possible violation of the
strong equivalence principle, i.e. the so-called Nordvedt effect
\ct{ref:nor}. $\et$ has been measured by analyzing the motion of
the Moon via LLR data analyses
\ct{ref:sha76,ref:willetal,ref:and}. The most recent determination
of $\et$ by LLR is $\et=0.0002\pm 8\times 10^{\rm -4}$
\ct{ref:and}.

The combination $\ppn$ is related to the well known pericenter
shift of a test body induced by the Schwarzschild' s
gravitoelectric part of the metric for a static, spherically
symmetric distribution of mass-energy \ct{ref:mis,ref:ciuwhe}.
This effect has been accurately measured for the Mercury
perihelion shift in the field of the Sun by means of the echo
delays of radar signals transmitted from Earth to Mercury
\ct{ref:sha72} yielding $\ppn=1.005\pm 7\times 10^{\rm -3}$.
Inclusion of the probable contribution of systematic errors raises
the uncertainty to about $2\times 10^{\rm -2}$.
In this case the major sources of systematic error are the
poorly-known variation in topography on the planet's surface and
the uncertainties in the radar scattering law
\ct{ref:sha90,ref:pit}.

It should be considered that the interpretation of the perihelion
advance of Mercury as a test of general relativity must cope with
the uncertainty in the mass quadrupole moment of the Sun which
also contributes to the perihelion advance. From the expression of
the perihelion precession and from the 1976 experimental
uncertainties in $\dot \omega$, one can easily calculate that for
Mercury orbiting the Sun, any value of $J_{\rm 2 \odot}$ larger
than about 3$\times 10^{\rm -6}$ would disagree with the general
relativistic prediction of 42.98 arcseconds/century; indeed,
according to some authors measured values of $J_{\rm 2 \odot}$ may
be as large as $ \sim 5.5\times 10^{\rm -6}$ (for a comprehensive
discussion see refs.\ct{ref:ciuwhe,ref:pir}). Laboratory
experiments have been also proposed \ct{ref:cac} to measure the
solar mode $l = 1$, to get information on the rotation of the core
of the Sun from the $l = 1$ rotational frequency splitting.
Nevertheless, according to more recent determinations of $J_{\rm 2
\odot}$ using the analysis of the Sun's pressure modes, both from
the ground-based network of observatories and the space based
SOHO, $J_{\rm 2 \odot }=(2.3\pm 0.1)\times 10^{-7}$
\ct{ref:sha99}; in ref.\ct{ref:godroz} a theoretical estimate of
$J_{\rm 2 \odot }=(2.0\pm 0.4)\times 10^{-7}$ is reported. Thus,
according to these values of $J_{\rm 2 \odot }$ the observed
perihelion advance of Mercury is well in agreement with the
general relativistic predictions.

In order to measure $J_{\rm 2 \odot}$, among other astrodynamical
and relativistic parameters, several space missions have been
proposed using a variety of techniques: the most recent and
promising are SORT, IPLR and ASTROD \ct{ref:ni}.
In \ct{ref:ciumat92} a measurement of the LAGEOS laser ranged
satellite perigee shift in the field of the Earth is quoted, but
the accuracy amounts only to $2\times 10^{\rm -1}$. More accurate
measurements of $\ppn$ might be performed in the future by means
of an ESA Mercury orbiter \ct{ref:balo,ref:miletal01}.

In this section we explore the possibility of performing a
complementary measurement of $\ppn$ in the gravitational field of
the Earth by using some suitable combinations of the orbital
residuals of the presently (or proposed) existing spherical
geodetic laser ranged satellites with particular emphasis to \lg\
and \lgg\, in order to exploit the relevant experience obtained
with the \grc\ measurements in
\ct{ref:ciufetal96,ref:ciu97,ref:ciu98}. Moreover, as a secondary
outcome of the proposed experiment, by combining the accurate
determinations of the Nordtvedt parameter $\et$ and of $\ppn$ from
SLR it would be possible to obtain independent values for $\ga$
and $\b$.

This section is organized as follows. In subsect. 1 we compare the
present experimental accuracy in satellite laser-ranging
measurements with the relativistic expressions of the perigee
shift of the two \lg\ satellites. In subsect. 2 we analyze some of
the most important sources of systematic errors: the even zonal
harmonics of the static part of the geopotential, the Earth solid
and ocean tidal perturbations, the direct solar radiation pressure
and the mismodelling of the inclination. Subsect. 3 is devoted to
the discussion of the obtained results and to the conclusions.
\subsection{The relativistic perigee precession of LAGEOS type satellites}

As known, in the slow-motion and weak-field approximation, the
Schwarzschild metric generated by a static, spherically symmetric
distribution of mass-energy induces an additional post-Newtonian
``gravitoelectric" force which acts on the orbit of a test body by
shifting its pericenter; in the PPN formalism the pericenter rate
can be written as \eqi \dot\og_{\rm GR}=\rp{3 n G M}{c^2 a
(1-e^2)}\times\ppn.\lb{peg}\eqf  In the following we define
$\n\equiv \ppn$. General Relativity predicts that the perigee
shifts for \lg\ and \lgg\ amount to  3,312.35 mas/y and 3,387.46
mas/y, respectively.

Following ref.\ct{ref:ciu96} the actual experimental precision
allows for detecting such rates for both \lg. Indeed, for the
perigee the observable quantity is $r \equiv ea\dot\og$ and at
present its measurement error amounts to about $\d r_{\rm exp}\leq
1$ cm for the two LAGEOS, over several orbits and for a given set
of force models (i.e. not including modelling errors). Since the
\lg\ eccentricity is $e_{\rm I}=4.5\times 10^{\rm -3}$ the
accuracy in detecting the perigee is $\d\og^{\rm I}_{\rm exp}= \d
r_{\rm exp}/(e_{\rm I}a_{\rm I}) \simeq 37$ mas. So, over 1 year
the relative accuracy of the measurement of the relativistic
perigee shift would be $ \sim 1 \times 10^{\rm -2}$. For \lgg\
this measurement accuracy is better than that of \lg\, indeed the
\lgg\ eccentricity is $e_{\rm II}=1.4\times 10^{\rm -2}$, so that
$\d\og^{\rm II}_{\rm exp}\simeq 12$ mas; this may yield an
accuracy of about $3\times 10^{\rm -3}$ over 1 year. These
considerations rule out the possibility of directly using the
perigee of the other existing spherical geodetic laser-ranged
satellites Etalon-1, Etalon-2, Ajisai, Stella, Westpac-1 because
their eccentricities are even smaller than that of \lg. On the
contrary, Starlette has an eccentricity of the order of $2\times
10^{\rm -2}$; however, since it orbits at a lower altitude it is
more sensitive than the \lg\ satellites to atmospheric drag and to
Earth's zonal harmonics, so that it would be difficult to process
its data at an acceptable level of accuracy. Accordingly, we will
focus on the perigee of \lg\ and especially of \lgg\ in order to
accurately detect the gravitoelectric relativistic shift in the
gravitational field of Earth.
\subsection{The systematic errors}
The most important source of systematic error in such measurements
is represented by the mismodelling induced by the even zonal
harmonics of the Earth gravitational field \ct{ref:kau} on the
classical perigee precession. By using the covariance matrix of
the EGM96 Earth gravity model \ct{ref:lem} and adding the
correlated terms in a root-sum-square fashion up to degree $l=20$
we obtain for \lg\ a systematic error $\d\n/\n_{\rm
zonals}=8.1\times 10^{\rm -3}$, whereas for \lgg\ we have
$\d\n/\n_{\rm zonals}=1.5\times 10^{\rm -2}$.

Since the major source of uncertainty lies in the first two even
zonal harmonics $\d J_2$ and $\d J_4$, following \ct{ref:ciu96}
for the \lt\ \lg\ experiment, we search for suitable combinations
of orbital residuals of the existing SLR satellites in order to
eliminate most static and dynamical even zonal terms of the
geopotential. In Table \ref{ppncombix} we report the most
promising combinations: their general form is \eqi \d\dot\og^{\rm
II}+\somma{j=1}{N}c_{\rm j}\d\dot\O^{\rm j}+c_{\rm
N+1}\d\dot\og^{\rm I}=x_{\rm GR}\ \n,\lb{zione}\eqf in which $N$
is the number of the nodes of different SLR satellites employed,
$x_{\rm GR}$ is the slope, in mas/y, of the relativistic trend to
be measured and $\d\n/\n_{\rm zonals}$ is the systematic error
induced by the even zonal harmonics up to degree $l=20$ calculated
with EGM96 covariance matrix. It is important to stress that the
use of LAGEOS, due to their altitude, makes our measurement
substantially insensitive to the errors in the zonal harmonics of
degree $l>20$, so that our estimates of $\d\n/\n_{\rm zonals}$
presented here are valid even in the case that the EGM96
covariance matrix for higher degrees $l>20$ would not be accurate
enough.
\begin{table}[ht!]
\caption{PPN combined residuals.}\label{ppncombix}
\begin{center}
\begin{tabular}{lllllll}
\noalign{\hrule height 1.5pt}
     & $\O^{{\rm II}}$ & $\O^{{\rm I}}$ & $\O^{{\rm Aj}}$ & $\og^{{\rm I}}$\\
\hline $\k$ & $c_1$ & $c_2$ & $c_3$ & $c_4$  & $x_{{\rm GR}}$
(mas/y) & $\d\n/\n_{{\rm zonals}}$
\\
\hline

{\sl 1} & $-0.868$  & $-2.855$  & 0  & 0  &  3,387.46 & $6.59\times 10^{-3}$ \\

{\sl 2} & $-2.514$ & $-4.372$ & 0  & 2.511 & 11,704.92 & $1.1\times 10^{-3}$\\

{\sl 3} & $-1.962$ & $-3.693$ & $0.0366$ & $1.370$  & 7,928.51 & $8.1\times 10^{-4}$\\
\noalign{\hrule height 1.5pt}
\end{tabular}
\end{center}
\end{table}
In addition to \lg\ and \lgg, we have only considered Ajisai since
it is well tracked, contrary to, e.g., the Etalons, and it would
be less demanding than the other satellites to reduce its laser
ranged data to a level of accuracy comparable to that of \lg\ and
\lgg. Moreover, the other SLR \sat s orbit at lower altitudes,
therefore they are more sensitive to the higher degree terms of
the geopotential. Consequently, as confirmed by numerical
calculations, the inclusion in their data in the combined
residuals would increase $\d\n/\n_{\rm zonals}$,.

Note that the systematic error induced by the mismodelled secular
rates of the even zonal harmonics of the geopotential is really
critical because it can be considered as an unavoidable part of
the total error in the experiment. Indeed, the resulting aliasing
trend cannot be removed from the data and nothing can be done
about it apart from assessing as more reliably as possible the
related error.\footnote{About combination \textsl{3}, it should be
noted that in order to obtain more reliable and accurate estimates
of the systematic error due to the even zonal harmonics of the
geopotential it should be better to extend the calculations to
higher degrees than $l=20$ due to the sensitivity of Ajisai to
such higher degree terms. }

It is important to point out that the values quoted in Table
\ref{ppncombix} for $\d\n/\n_{\rm zonals}$ will be reduced in the
near future when the data from the CHAMP mission will be released.

In regard to the evaluation of the impact of other sources of
systematic errors in this section we consider in detail only the
combination {\sl 1} of Table \ref{ppncombix} so to exploit the
background acquired with the \lt\ \lg\ experiment
\ct{ref:ciufetal96,ref:ciu97,ref:ciu98}. Moreover, reducing the
Ajisai' s data to an acceptable level of accuracy for our
measurement would neither be a trivial nor an immediate task to be
performed and the inclusion of the perigee of \lg\ would raise the
experimental error. E.g., for combination \textsl{2} of Table
\ref{ppncombix}, over 1 year, the impact of the error in measuring
the perigee rate of LAGEOS amounts to $2.5\times(3.1\times 10^{\rm
-3})=7.7\times 10^{\rm -3}$ while for combination \textsl{3} it is
$1.370\times(4.7\times 10^{\rm -3})=6.4\times 10^{\rm -3}$.
\subsection{Discussion and conclusions}
In Table \ref{ppn8anni} we summarize the results obtained for a 8
years long time span with 7 days arc lengths. The results and the
methods of sections 3 and 4 and of ref.\ct{ref:luc,ref:lucluc}
have been used.

\begin{table}[ht!]
\caption[PPN preliminar error budget: $T_{obs}=8$ years, $\D t=7$
days.]{Preliminary error budget: $T_{obs}=8$ years, $\D t=7$
days.}\label{ppn8anni}
\begin{center}
\begin{tabular}{ll}
\noalign{\hrule height 1.5pt}

Even zonal harmonics & $6.59\times 10^{-3}$  \\

$J_3$ geopotential & $3.2\times 10^{-4}$ \\

Tides & $4.4\times 10^{-4}$ \\

Non-gravitational effects & $1\times 10^{-4}$ \\


Measurement error in LAGEOS II perigee & $4\times 10^{-4}$ \\

\hline

Total systematic error & $7.3\times 10^{-3}$\\



 \noalign{\hrule height 1.5pt}
\end{tabular}
\end{center}
\end{table}
In regard to the non--gravitational errors, according to the
results of ref.\ct{ref:luc} for the direct solar radiation
pressure and the Earth's albedo and to the results of
ref.\ct{ref:lucluc} for the thermal thrust perturbations and the
asymmetric reflectivity, the corresponding uncertainty would
amount to almost $1\times 10^{-2}$ over 7 years. In obtaining this
result a very pessimistic approach has been adopted by assuming a
mismodeling of 20$\%$ for all the perturbing effects except for
the direct solar radiation pressure which has been assumed to have
an uncertainty at the 0.5$\%$ level. However, we stress that only
the Earth's thermal thrust, or Yarkovski-Rubincam effect, induces
a mismodeled linear trend whose impact would amount to about
$1\times 10^{-4}$: the other non--gravitational forces are
time--varying with known periodicities and can be fitted and
removed from the signal as done for the tidal perturbations.

In assessing the total systematic error we have accounted for the
fact that the gravitational errors are not independent simply by
summing them up. Then, we have added them and the other
independent sources of errors in a root-sum-square fashion. Note
that the error in the LAGEOS II perigee refers to 1 year only by
assuming an error of 1 cm in its radial position. Moreover, the
estimate of the statistical formal error has been obtained by
considering the noise level in the LAGEOS data of 1998
Lense-Thirring experiment; in the near future and for longer time
spans these uncertainties will be reduced. Moreover, it is
important to emphasize that the data from the CHAMP and GRACE
missions will soon yield a notable reduction of the systematic
error due to the higher degree static even zonal harmonics of the
geopotential and of the solid and ocean tides systematic errors as
well. In particular, a very significant reduction of the
systematic errors will take place due to the static part of the
geopotential, which has turned out to be the most important source
of uncertainty.

The results obtained for the combination {\sl 1}
examined here together with those released in ref.\ct{ref:and} for
the combination $\et$ would allow to measure
$\b=\rp{2}{7}\et+\rp{3}{7}\n+\rp{4}{7}$ independently of other
measurements of $\ga$ with $ \sim 3.3\times 10^{\rm -3}$ accuracy
over 8 years; this result, which is of the same order of magnitude
of that obtained with the radar ranging technique \ct{ref:sha90},
should be compared with the most recent $\d\b=4.7\times 10^{\rm
-4}$ obtained from the LLR data \ct{ref:and}. The parameter
$\gamma$, given by
$\gamma=\frac{1}{7}\eta+\frac{12}{7}\nu-\frac{5}{7}$, could be
measured less precisely over the same time span: $\delta\gamma\sim
1.2\times 10^{\rm -2}$.
\section{The LARES mission revisited: some alternative scenarios}
\subsection{Introduction}
In order to measure the Lense-Thirring drag with an accuracy of
few percent, in \ct{ref:ciu86} it was proposed to launch a passive
geodetic laser-ranged \st- the former {\rm LAGEOS} III - with the
same orbital parameters of \lg\ apart from its inclination which
should be supplementary to that of \lg.

This orbital configuration would be able to cancel out exactly the
classical \nl\ \pc s provided that the observable to be adopted is
the sum of the residuals of the \nl\ \pc s of {\rm LAGEOS} III and
LAGEOS \eqi \delta\dt\Omega^{{\rm III}}+\delta\dt\Omega^{{\rm
I}}=62\mlt.\lb{lares}\eqf Later on the concept of the mission
slightly changed. The area-to-mass ratio of {\rm LAGEOS} III was
reduced in order to make less relevant the impact of the
non-gravitational perturbations and the eccentricity was enhanced
in order to be able to perform other \grl\ tests: the LARES was
born. The most recent error budget of the LARES Lense-Thirring
experiment claim an accuracy of the order of $3\%$. Unfortunately,
at present we do not know if the LARES mission will be approved by
any space agency.

In this section we investigate the possibility of further reducing
the error in this important proposed mission. This section is
organized as follows. In subsect. 2 we analyze in detail the
impact of the unavoidable injection \er s in the orbital
parameters of LARES on the \se\ induced by the mismodelling in the
\zh\ of the \gp\ according to the most recently released \rt\
gravity model. Moreover, in subsect. 3 we propose an alternative
configuration which should be able to reduce this error by one
order of magnitude. It adopts as observable a suitable combination
of the orbital residuals of the \nd s of \lg, \lgg\ and LARES, and
the \pg s of \lgg\ and LARES. It presents also the important
advantage that it is almost insensitive to the errors in the \ic\
of LARES, contrary to the original {\rm LAGEOS}/LARES only
configuration. The negative implications of placing the LARES in a
low-altitude polar orbits are examined in subsect. 4. Subsect. 5
is devoted to the conclusions.
\subsection{The impact of the even zonal harmonics of the geopotential on the original LARES mission}
Let us calculate the \se\ induced by the mismodelling in the even
degree zonal coefficients $J_2$, $J_4$,... of the \gp\ on the sum
of the classical precessions of the \nd s of \lg\ and LARES. It is
important to stress that it is the major source of \se\ and it
cannot be eliminated in any way. We will use the covariance matrix
of the \rt\ gravity field model EGM96 \ct{ref:lem} by summing up
in a root sum square fashion the correlated contributes up to
degree $l=20$\footnote{It is important to notice that using \st s
of the \lg\ family allows one to obtain reliable estimates with
EGM96. Indeed, the \lg' orbits are not affected by the terms of
degree $l>20$, for which the EGM96 covariance matrix elements are
determined less accurately.}. The relative error obtained by using
the nominal values amounts to \eqi\dmu_{\zs}=3\times
10^{-3}.\lb{1zon}\eqf It is not equal to zero because we have
assumed $e_{{\rm LR}}=0.04$ while $e_{{\rm {\rm LAGEOS}}}=0.0045$.
If it was $e_{{\rm LR}}= e_{{\rm {\rm LAGEOS}}}$, then the
classical \nl\ \pc s would be exactly equal in value and opposite
in sign and would cancel out. Notice that the coefficients with
which $\delta\dot\Omega^{\rm III}$ and $\delta\dot\Omega^{\rm I}$
enter the combination of \rfr{lares} do not depend on any orbital
parameters: they are constants equal to 1. Moreover, \rfr{lares}
is affected by all the classical \nl\ precessions, including those
induced by $J_2$ and $J_4$ which, instead, are cancelled out $a\
priori$ in the combination used in the \lg\ experiment
\ct{ref:ciu96}. They are the most effective in aliasing the
\leti\ precessional rates.

Now we will focus on the sensitivity of $\dmu_{\zs}$ to the
unavoidable orbital injection errors in the possible orbital
parameters of LARES. For a former analysis see ref.\ct{ref:cas90}.
It is particularly interesting to consider the impact of the
errors in the inclination and the \sa. The ranges of variation for
them have been chosen in a very conservative way in order to take
into account low-precision and low-costs injection scenarios.

\begin{figure}[ht!]
\begin{center}
\includegraphics*[width=10cm,height=7cm]{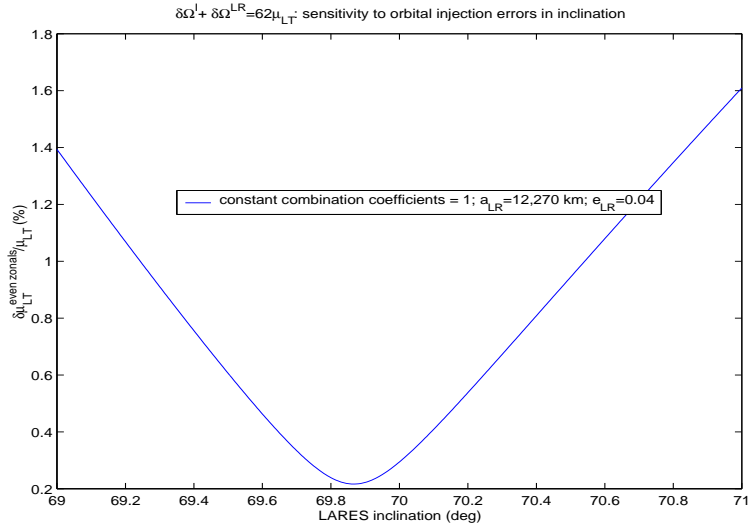}
\end{center}
\caption[Original LARES mission: sensitivity to
\ic.]{\footnotesize Influence of the injection errors in the LARES
inclination on the zonals' error.} \label{LRfigura1}
\end{figure}

\begin{figure}[ht!]
\begin{center}
\includegraphics*[width=10cm,height=7cm]{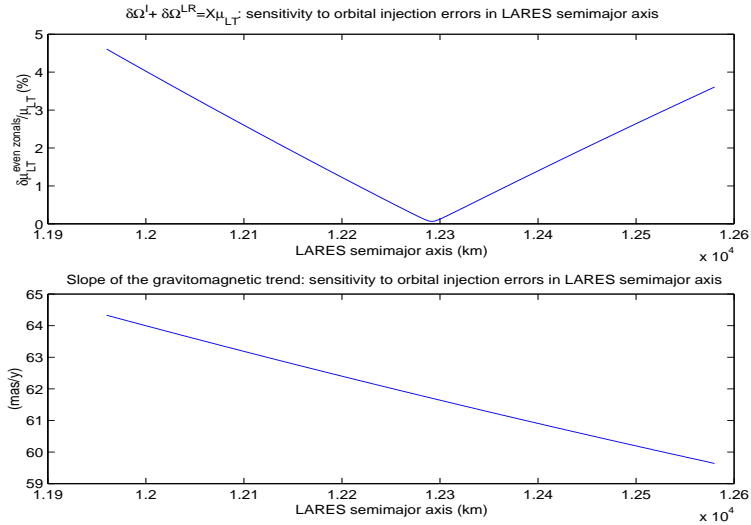}
\end{center}
\caption[Original LARES mission: sensitivity to
\sa.]{\footnotesize Influence of the injection errors in the LARES
\sa\ on the zonals' error.} \label{LRfigura2}
\end{figure}

From fig. \ref{LRfigura1} it is interesting to notice that the
minimum value of the systematic zonal error, which amounts to
$2.1\times 10^{-3}$, does not correspond to $i_{{\rm LR}}=70\
\textrm{deg}$ but it is obtained for a slightly smaller value. It
is possible to show that for $e_{{\rm LR}}=e_{{\rm {\rm LAGEOS}}}$
the minimum is 0 and that it is attained at $i_{{\rm LR}}=70\ {\rm
deg}$. The maximum error amounts to $1.6\times 10^{-2}$. This
suggests that the original LARES project is rather sensitive to
small departures of $i_{{\rm LR}}$ from its nominal value. Fig.
\ref{LRfigura2} shows that even more relevant is the sensitivity
to the LARES \sa.
Also in this case the minimum is attained at a value of $a_{{\rm
LR}}$ smaller than the nominal $a_{{\rm LR}}=12,270$ km. For
$e_{{\rm LR}}=e_{{\rm {\rm LAGEOS}}}$ the minimum error amounts to
0 and it is obtained for $a_{{\rm LR}}=12,270$ km, as expected. In
obtaining fig. \ref{LRfigura2} we have accounted for the
dependence of the \nl\
\leti\ \pc\ on $a$ by varying, accordingly, the slope of the \grl\
trend. The sensitivity to eccentricity variations is less
relevant: indeed, by varying it from 0.03 to 0.05 the relative
systematic zonal error passes from $1.6\times 10^{-3}$ to
$4.6\times 10^{-3}$
\subsection{An alternative LARES scenario}
Here we will look for an alternative observable involving the
orbital elements of LARES which\\
$\bullet$ yields a smaller value for the \se\ due to the
mismodelled \zh\ of the \gp\ than that of the simple sum of the \nd s of \lg\ and LARES\\
$\bullet$ is less sensitive to the departures of the possible
orbital elements of LARES from the nominal values. The first
requirement could be implemented by setting up a suitable orbital
combination which cancels out the contributions of as many
mismodelled even zonal harmonics as possible, following the
strategy of the \lg\ experiment outlined in \ct{ref:ciu96}. To
this aim we will consider only the \st s of the \lg\ family both
because they are the best laser-ranged tracked targets and because
the gravitational and non-gravitational perturbations affecting
their orbits have been extensively and thoroughly analyzed.

Our result is \eqi \delta\dt\Omega^{\rm {\rm
LAGEOS}}+c_1\delta\dt\Omega^{\rm {\rm LAGEOS}\
II}+c_2\delta\dt\Omega^{\rm LARES}+c_3\delta\dt\omega^{\rm {\rm
LAGEOS} \ II}+c_4\delta\dt\omega^{\rm
LARES}=61.8\mlt,\lb{LRcombi}\eqf with \eqia
c_1 & = & 6\times 10^{-3},\lb{c1}\\
c_2 & = & 9.83\times 10^{-1},\\
c_3 & = & -1\times 10^{-3},\\
c_4 & = & -2\times 10^{-3}\lb{c4}. \eqfa It is important to notice
that the coefficients given by \rfrs{c1}{c4} depend on the orbital
parameters of the \st s entering the combination and, among them,
of LARES. The values released here are calculated for the nominal
LARES parameters, as is the case for the slope in \msy\ of the
\grl\ trend. The observable of \rfr{LRcombi} allows one to cancel
out the static and dynamical contributions of the first four \zh.
The relative \se\ due to the $J_{2n},\ n\geq 5$, according to
EGM96 up to degree $l=20$, amounts to: \eqi\dmu_{\zs}=2\times
10^{-4},\lb{kazonga}\eqf which is one order of magnitude better
than the result of \rfr{1zon}.

\begin{figure}[ht!]
\begin{center}
\includegraphics*[width=10cm,height=7cm]{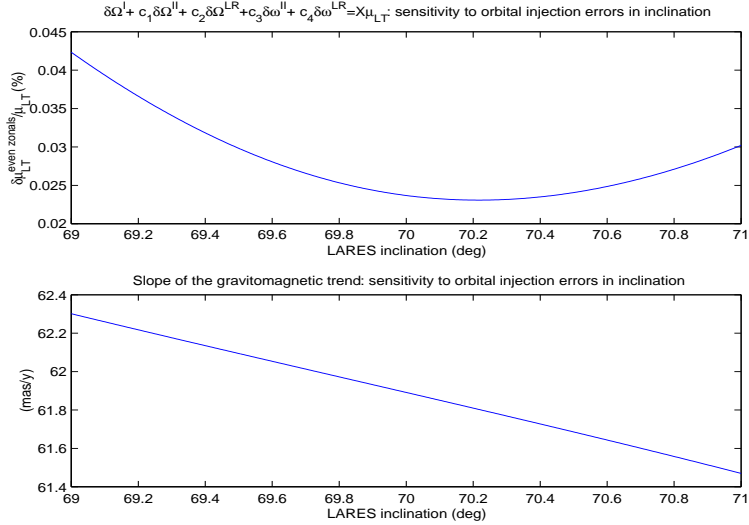}
\end{center}
\caption[Alternative LARES mission: sensitivity to
\ic.]{\footnotesize Alternative combined residuals: influence of
the injection errors in the LARES \ic\ on the zonals' error.}
\label{LRfigura3}
\end{figure}

\begin{figure}[ht!]
\begin{center}
\includegraphics*[width=10cm,height=7cm]{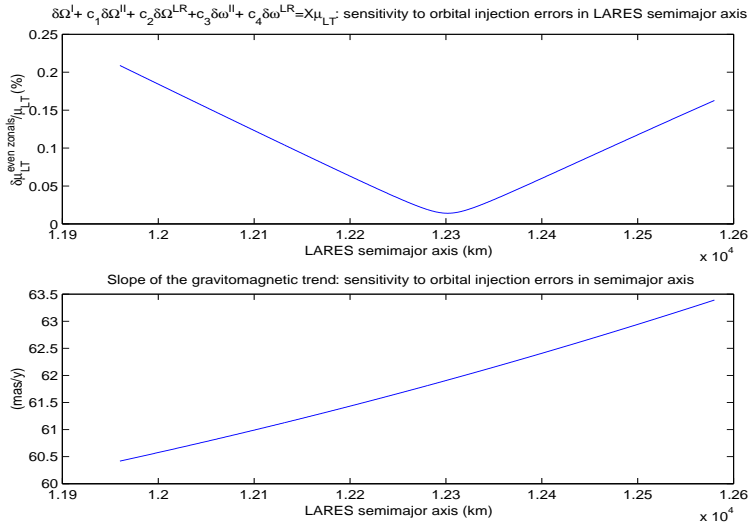}
\end{center}
\caption[Alternative LARES mission: sensitivity to
\sa.]{\footnotesize Alternative combined residuals: influence of
the injection errors in the LARES \sa\ on the zonals' error.}
\label{LRfigura4}
\end{figure}

Fig. \ref{LRfigura3} and fig. \ref{LRfigura4} show the important
achievements realized in reducing the sensitivity of the proposed
combined residuals to the orbital injection errors in the LARES
orbital elements. In obtaining fig. \ref{LRfigura3} and fig.
\ref{LRfigura4} we have accounted for the dependence on $a_{\rm
LR}$ and $i_{\rm LR}$ of both the coefficients and the
\leti\ \pc s: it turns out that the variations in the slope of the
\grl\ trend are very smooth with respect to the nominal value of
61.8 mas/y amounting to few mas/y. Now the values of the zonals'
error are much more close to the nominal one given by
\rfr{kazonga}. Also in this case, the minima are attained at
slightly different values of the LARES orbital elements with
respect to the nominal ones. It is interesting to notice in fig.
\ref{LRfigura3} that over a 3$\%$ variation of $i_{\rm LARES}$ the
error due to the mismodeled zonal harmonics remain almost
constant, while over a 5$\%$ variation of $a_{\rm LARES}$ it
changes of 1 order of magnitude, as it turns out from fig.
\ref{LRfigura4}.
It is worthwhile noting that the time-varying gravitational and
non-gravitational orbital perturbations which affect the proposed
combined residuals are depressed by the small values of the
coefficients with which some orbital elements enter the
combination. Moreover, the observational error in the {\rm LAGEOS}
II and LARES perigees, which are difficult to measure for low
eccentric \st s as \lg due to the small value of their
eccentricity, would have an impact of the order of $1\times
10^{-4}$ by assuming an uncertainty of 1 cm over 1 year in the
\st's position.
\subsection{Conclusions}
If analyzed from the point of view of the impact of the \se\
induced by the mismodelling in the \zh\ of the \gp, which is the
most important source of systematic error, the original LARES
mission seems to be affected by a certain sensitivity to the
unavoidable departures of the LARES orbital parameters from their
nominal values due to orbital injection errors. In such a refined
experiment, which would compete with the ambitious Stanford GP-B
mission and its claimed global $1\%$ accuracy level, it could be a
serious drawback. It could be enhanced if LARES will be finally
put in orbit with a low-cost launcher which, inevitably, would
induce relatively large injection errors.

The adoption of the alternative combined residuals proposed here,
including also the \nd\  of LAGEOS II and the \pg s of LAGEOS II
and LARES, would reduce by one order of magnitude the \se\ due to
the \zh\ of the \gp\, passing from $0.3\%$ to $0.02\%$, and would
reduce greatly the sensitivity of such result to errors in the
LARES orbital parameters. This would yield to less stringent
requirements on the quality and the costs of the launcher to be
adopted. It is very important to notice that the estimates
presented here are based on the most recent Earth gravity model
EGM96. When the new data on the terrestrial gravitational field
from the CHAMP and GRACE missions will be available, the
systematic zonal error will greatly reduce. The impact of the
errors related to the quality of laser data will further reduce in
the near future as well. Preliminary estimates of the standard
statistical error in the solve-for least square parameter
$\mu_{\rm LT}$, based on the present models of the time-dependent
LAGEOS perturbations worked out in section 3 and the noise level
reported in the Lense-Thirring LAGEOS experiment, yield a value of
the order of $10^{-3}$. If confirmed by further analysis of the
impact of the time-dependent gravitational and non-gravitational
perturbations, this result is very important because the total
error of this improved version of the LARES mission would be $\leq
1\%$.
\section{Conclusions and recommendations for future work}
The general relativistic features considered here are

$\bullet$ The \grc\ \lg\ experiment devoted to the measurement of
the Lense-Thirring drag of inertial frames, currently carried on
by analyzing the combined residuals of the nodes of \lg\ and \lgg\
and the \pg\ of \lgg. The assessment of the systematic error
induced by the time--dependent part of the terrestrial
gravitational field has been extensively discussed so to extend
its evaluation also to future measurements over longer time spans.
The role of the proposed LARES in a new context is also taken into
account. The proposed new observable would allow to reduce by one
order of magnitude the total systematic error

$\bullet$ The pericenter shift of a test particle due to the
Schwarzschild's gravitoelectric part of the gravitational field of
a central, static body and the possibility of measuring it at a
$10^{-3}$ level in the field of the Earth with \lg, \lgg, and,
perhaps, other SLR satellites

It is very important to stress that in the near future, when the
new and more precise data relative to the static and dynamical
parts of the  Earth's gravitational field from the CHAMP and GRACE
missions will be available, it will be possible to reduce sensibly
the uncertainties of many source of systematic errors of
gravitational origin. This is particularly true for the even zonal
harmonics of the static part of the geopotential.  In regard to
the \grc\ \lg\ experiment, this would probably allow to reduce the
total geopotential error from the present $12.9\%$ to some
percent. Moreover, the zonals error in the PPN measurement could
reach the level of $10^{-4}-10^{-5}$ according to the observable
adopted. These improvements would extend also to the LARES
project. So, it could be possible to proceed in the future by
updating the orbit determination softwares like GEODYN II with the
new data of the geopotential and, subsequently, by reanalyzing the
data of the \grc\ \lg\ experiment. In the framework of the \grc\
\lg\ experiment, it could be useful to look for alternative
combinations of orbital residuals involving also other SLR
satellites in order to reduce, on one hand, the systematic error
due to the geopotential, on the other, the systematic error due to
the non--gravitational perturbations. It would be the follow-on of
the work presented in ref.\ct{ref:cas90}. A first attempt can be
found in ref.\ct{ref:ior02}.



\appendix
\section{The classical orbital precessions}
Here we show the explicitly calculated coefficients \eqi\djn\eqf
and \eqi\djo\eqf of the satellites' classical nodal and apsidal
precessions due to the even ($l=2n$, $n=1,2,3...$) zonal ($m=0$)
harmonics of the \protect{geopotential} up to $l= 20$. Recall that
$J_{l}\ \equiv -C_{l0},\ l=2n,\ n=1,2,3...$ where the unnormalized
adimensional {Stokes} coefficients $C_{lm}$ of degree $l$ and
order $m$ can be obtained from the normalized
${\overline{C}}_{lm}$ with \eqi C_{lm}=N_{lm}{\overline{C}}_{lm}.
\eqf In it \eqi N_{lm}=
\left[\frac{(2l+1)(2-\delta_{0m})(l-m)!}{(l+m)!}\right]^{\frac{1}{2}}.\eqf
For the general expression of the classical rates of the near
\protect{Earth} satellites' Keplerian orbital elements due to the
\protect{geopotential} ${\dot a}_{\rm class},\ {\dot e}_{\rm
class},\ {\dot i}_{\rm class},\ {\dot \Omega}_{\rm class},\ {\dot
\omega}_{\rm class},\ {\dot {{\mathcal{M}}}}_{\rm class}$, see
ref.\ct{ref:kau}. The coefficients $\dot\Omega_{.2n}$ and
$\dot\omega_{.2n}$ are of crucial importance in the evaluation of
the systematic error due to the mismodelled even zonal harmonics
of the \protect{geopotential}; moreover, they enter the combined
residuals' coefficients $c_i,\ i=1,2...N$ of sect. 5 and sect. 6.
See also ref.\ct{ref:ciu96}. Since the \grl\ effects investigated
are secular perturbations, we have considered only the
perturbations averaged over one satellite' s orbital period. This
has been accomplished with the condition $l-2p+q = 0$. Since the
\protect{eccentricity} functions $G_{lpq}$ are proportional to
$e^{\vass{q}}$, for a given value of $l$ we have considered only
those values of $p$ which fulfil the condition $l-2p+q = 0$ with
$q=0$, i.e. $p=\frac{l}{2}$. This implies that in the summations
\eqi\sum_{p=0}^{l}\dert{F_{l0p}}{i}\sum_{q=-\infty}^{+\infty}G_{lpq}\eqf
and
\eqi\sum_{p=0}^{l}F_{l0p}\sum_{q=-\infty}^{+\infty}\dert{G_{lpq}}{e}\eqf
involved in the expressions of the classical rates we have
considered only $F_{l0\frac{l}{2}}$ and $G_{l\frac{l}{2}0}$.
Moreover, in working out the $G_{l\frac{l}{2}0}$ we have neglected
the terms of order ${\mathcal{O}}(e^{k})$ with $k>2$.

The nodal coefficients, proportional to \eqi\frac{1}{\sin
i}\sum_{q=-\infty}^{+\infty}G_{lpq}\sum_{p=0}^{l}\dert{F_{lmp}}{i},\eqf
are

\eqia\jn2 & = &  -\frac{3}{2}n\rtr{2}\frac{\cos i}{\ecn{2}},\\
\jn{4} & = & \jn2\left[ \frac{5}{8}\rtr{2}
\frac{(1+\frac{3}{2}e^{2})}{\ecn{2}} \left(
7\sus{2}-4\right)\right],\\
\jn{6} & = & \jn2\left[ \frac{35}{8}\rtr{4}
\frac{(1+5e^{2})}{\ecn{4}} \left(
\frac{33}{8}\sus{4}-\frac{9}{2}\sus{2}+1\right)\right],\\
\jn{8} & = & \jn2\left[ \frac{105}{16}\rtr{6}
\frac{(1+\frac{21}{2}e^{2})}{\ecn{6}} \left( \frac{715}{64}\sus{6}
-\frac{143}{8}\sus{4}+\right.\right.\nonumber\\
&+&\left.\left.\frac{33}{4}\sus{2}-1\right)\right] ,\\
\jn{10} & = & \jn2\left[  \frac{1,155}{128} \rtr{8}
\frac{(1+18e^{2})}{\ecn{8}}\left(
\frac{4,199}{128}\sus{8}-\frac{1,105}{16}\sus{6}\right.\right.\nonumber\\\nonumber\\
& +&
\left.\left.\frac{195}{4}\sus{4}-13\sus{2}+1\right)\right],\\
\jn{12} & = & \jn2\left[ \frac{3,003}{256}\rtr{10}
\frac{\ekn{55}{2}}{\ecn{10}} \left( \frac{52,003}{512}\sus{10}-
\frac{33,915}{128}\sus{8}\right.\right.\nonumber\\\nonumber\\
&+ &  \left.\left.
\frac{8,075}{32}\sus{6}-\frac{425}{4}\sus{4}+\frac{75}{4}\sus{2}-1\right)\right],\\
\jn{14} & = & \jn2\left[ \frac{15,015}{1,024}\rtr{12}
\frac{\ekn{91}{2}}{\ecn{12}} \left(
\frac{334,305}{1,024}\sus{12}-\frac{260,015}{256}\sus{10}\right.\right.\nonumber\\\nonumber\\
& +& \left.\left.
\frac{156,009}{128}\sus{8}-\frac{11,305}{16}\sus{6}+\frac{1,615}{8}\sus{4}-\frac{51}{2}\sus{2}+1\right)\right],\\
\jn{16} & = & \jn2\left[ \frac{36,465}{2,048}\rtr{14}
\frac{\ekn{105}{2}}{\ecn{14}} \left(
\frac{17,678,835}{16,384}\sus{14}-\right.\right.\nonumber\\\nonumber\\
&-&\left.\left.\frac{3,991,995}{1,024}\sus{12} +
\frac{2,890,755}{512}\sus{10}-
\frac{535,325}{128}\sus{8}+\frac{107,065}{64}\sus{6}\right.\right.\nonumber\\\nonumber\\
&-&
\left.\left.\frac{2,793}{8}\sus{4}+\frac{133}{4}\sus{2}-1\right)\right],\\
\jn{18} & = & \jn2\left[ \frac{692,835}{32,768}\rtr{16}
\frac{(1+68e^{2})}{\ecn{16}}
\left(\frac{119,409,675}{32,768}\sus{16}-\right.\right.\nonumber\\\nonumber\\&-&
\frac{30,705,345}{2,048}\sus{14}
+\frac{6,513,255}{256}\sus{12}-\frac{1,470,735}{64}\sus{10}+\nonumber\\\nonumber\\
&+&\left.\left.\frac{760,725}{64}\sus{8} -
\frac{28,175}{8}\sus{6}+\right.\right.\nonumber\\\nonumber\\
&+&\left.\left.\frac{1,127}{2}\sus{4}-
42\sus{2}+1\right)\right],\\
\jn{20} & = & \jn2\left[ \frac{1,616,615}{65,536}\rtr{18}
\frac{\ekn{171}{2}}{\ecn{18}}
\left(\frac{1,641,030,105}{131,072}\sus{18}\right.\right.\nonumber\\\nonumber\\
& -&\frac{1,893,496,275}{32,768}\sus{16}+
\frac{460,580,175}{4,096}\sus{14}-\frac{30,705,345}{256}\sus{12}\nonumber\\\nonumber\\
& +& \frac{19,539,765}{256}\sus{10}
-\frac{1,890,945}{64}\sus{8}+\frac{108,675}{16}\sus{6}-\nonumber\\\nonumber\\
&-&\left.\left.\frac{1,725}{2}\sus{4} +
\frac{207}{4}\sus{2}-1\right)\right]. \eqfa

The coefficients of the classical \protect{perigee} precession are
much more involved because they are proportional to
\eqi-\left(\frac{\cos i}{\sin
i}\right)\sum_{q=-\infty}^{+\infty}G_{lpq}\sum_{p=0}^{l}\dert{F_{lmp}}{i}+
\frac{(1-e^2)}{e}\sum_{q=-\infty}^{+\infty}\dert{G_{lpq}}{e}\sum_{p=0}^{l}F_{lmp}.\eqf
We can pose  $\jo{2n}=\jo{2n}^{a}+\jo{2n}^{b}$.

The first set is given by

\eqia\jo2^{a} & = &  \frac{3}{2}n\rtr{2}\frac{\cos^{2}
i}{\ecn{2}},\\
\jo{4}^{a} & = & \jo2^{a}\left[ \frac{5}{8}\rtr{2}
\frac{(1+\frac{3}{2}e^{2})}{\ecn{2}} \left(
7\sus{2}-4\right)\right],\\
\jo{6}^{a} & = & \jo2^{a}\left[ \frac{35}{8}\rtr{4}
\frac{(1+5e^{2})}{\ecn{4}} \left(
\frac{33}{8}\sus{4}-\frac{9}{2}\sus{2}+1\right)\right],\\
\jo{8}^{a} & = & \jo2^{a}\left[ \frac{105}{16}\rtr{6}
\frac{(1+\frac{21}{2}e^{2})}{\ecn{6}} \left( \frac{715}{64}\sus{6}-\right.\right.\nonumber\\\nonumber\\
&-&\left.\left.\frac{143}{8}\sus{4}+\frac{33}{4}\sus{2}-1\right)\right] ,\\
\jo{10}^{a} & = & \jo2^{a}\left[  \frac{1,155}{128} \rtr{8}
\frac{(1+18e^{2})}{\ecn{8}}\left(
\frac{4,199}{128}\sus{8}-\frac{1,105}{16}\sus{6}\right.\right.\nonumber\\\nonumber\\
& +&
\left.\left.\frac{195}{4}\sus{4}-13\sus{2}+1\right)\right],\\
\jo{12}^{a} & = & \jo2^{a}\left[ \frac{3,003}{256}\rtr{10}
\frac{\ekn{55}{2}}{\ecn{10}} \left( \frac{52,003}{512}\sus{10}-
\frac{33,915}{128}\sus{8}\right.\right.\nonumber\\\nonumber\\
&+ &  \left.\left.
\frac{8,075}{32}\sus{6}-\frac{425}{4}\sus{4}+\frac{75}{4}\sus{2}-1\right)\right],\\
\jo{14}^{a} & = & \jo2^{a}\left[ \frac{15,015}{1,024}\rtr{12}
\frac{\ekn{91}{2}}{\ecn{12}} \left(
\frac{334,305}{1,024}\sus{12}-\frac{260,015}{256}\sus{10}+\right.\right.\nonumber\\\nonumber\\
& +& \left.\left.
\frac{156,009}{128}\sus{8}-\frac{11,305}{16}\sus{6}+\frac{1,615}{8}\sus{4}-\right.\right.\nonumber\\\nonumber\\
&-&\left.\left.\frac{51}{2}\sus{2}+1\right)\right],\\
\jo{16}^{a} & = & \jo2^{a}\left[ \frac{36,465}{2,048}\rtr{14}
\frac{\ekn{105}{2}}{\ecn{14}} \left(
\frac{17,678,835}{16,384}\sus{14}-\frac{3,991,995}{1,024}\sus{12}\right.\right.\nonumber\\\nonumber\\
&+ &  \left.\left.\frac{2,890,755}{512}\sus{10}-
\frac{535,325}{128}\sus{8}+\frac{107,065}{64}\sus{6}\right.\right.\nonumber\\\nonumber\\
&- &
\left.\left.\frac{2,793}{8}\sus{4}+\frac{133}{4}\sus{2}-1\right)\right],\\
\jo{18}^{a} & = & \jo2^{a}\left[ \frac{692,835}{32,768}\rtr{16}
\frac{(1+68e^{2})}{\ecn{16}}
\left(\frac{119,409,675}{32,768}\sus{16}-
\right.\right.\nonumber\\\nonumber\\
&-&\frac{30,705,345}{2,048}\sus{14}+\frac{6,513,255}{256}\sus{12}-\frac{1,470,735}{64}\sus{10}+
\nonumber\\\nonumber\\
&+&\left.\left.\frac{760,725}{64}\sus{8}-\frac{28,175}{8}\sus{6}+\frac{1,127}{2}\sus{4}-\right.\right.\nonumber\\\nonumber\\
&-&\left.\left.42\sus{2}+1\right)\right],\\
\jo{20}^{a} & = & \jo2^{a}\left[ \frac{1,616,615}{65,536}\rtr{18}
\frac{\ekn{171}{2}}{\ecn{18}}
\left(\frac{1,641,030,105}{131,072}\sus{18}\right.\right.\nonumber\\\nonumber\\
& -&\frac{1,893,496,275}{32,768}\sus{16}+
\frac{460,580,175}{4,096}\sus{14}-\frac{30,705,345}{256}\sus{12}\nonumber\\\nonumber\\
& +& \left.\left.\frac{19,539,765}{256}\sus{10}
-\frac{1,890,945}{64}\sus{8}+\frac{108,675}{16}\sus{6}-\right.\right.\nonumber\\\nonumber\\
&-&\left.\left.\frac{1,725}{2}\sus{4}+\frac{207}{4}\sus{2}-1\right)\right].
\eqfa The second set is given by
\eqia \jw2 & = & -\frac{3}{2}n\rtr2,\\
\jo2^{b} & = &
\jw2\left\{\left[\frac{1}{\ecn{2}}\right]\left(\frac{3
}{2}\sin^2 i-1\right)\right\},\\
\jo4^{b} & = & \jw2\left\{\frac{5}{8}\rtr2\left[
\frac{3}{\ecn{3}}+7\frac{\ekn{3}{2}}{\ecn{4}}\right]\left(\frac{7}{4}\sus4-\right.\right.\nonumber\\
&-&\left.\left.2\sus2+\frac{2}{5}\right)\right\},\\
\jo{6}^{b} & = & \jw2 \left\{\frac{35}{8}\rtr{4}\left[
\frac{10}{\ecn{5}}+11\frac{(1+5e^2)}{\ecn{6}}\right]\left(\frac{33}{48}\sus{6}
\right.\right.\nonumber\\\nonumber\\
&-& \left.\left.\frac{9}{8}\sus{4}+\frac{1}{2}\sus{2}-\frac{1}{21}
\right)\right\},\\
\jo{8}^{b} & = & \jw2\left\{\frac{105}{16}\rtr{6}\left[
\frac{21}{\ecn{7}}+15\frac{\ekn{21}{2}}{\ecn{8}}\right]\left(\frac{715}{512}\sus{8}
\right.\right.\nonumber\\\nonumber\\
&-&
\left.\left.\frac{143}{48}\sus{6}+\frac{33}{16}\sus{4}-\frac{1}{2}\sus{2}
+\frac{1}{36}
\right)\right\},\\
\jo{10}^{b} & = & \jw2 \left\{\frac{1,155}{128}\rtr{8}\left[
\frac{36}{\ecn{9}}+19\frac{(1+18e^2)}{\ecn{10}}\right]\left(\frac{4,199}{1,280}\sus{10}
\right.\right.\nonumber\\\nonumber\\
&-&
\left.\left.\frac{1,105}{128}\sus{8}+\frac{195}{24}\sus{6}-\frac{13}{4}\sus{4}
+\frac{1}{2}\sus{2}-\frac{1}{55}
\right)\right\},\\
\jo{12}^{b} & = & \jw2 \left\{\frac{3,003}{256}\rtr{10}\left[
\frac{55}{\ecn{11}}+23\frac{\ekn{55}{2}}{\ecn{12}}\right]\left(\frac{52,003}{6,144}\sus{12}
\right.\right.\nonumber\\\nonumber\\
&-&
\frac{6,783}{256}\sus{10}+\frac{8,075}{256}\sus{8}-\frac{425}{24}\sus{6}
+\frac{75}{16}\sus{4}\nonumber\\\nonumber\\
&-&\left.\left.\frac{1}{2}\sus{2}+\frac{1}{78}
\right)\right\},\\
\jo{14}^{b} & = & \jw2 \left\{\frac{15,015}{1,024}\rtr{12}\left[
\frac{91}{\ecn{13}}+27\frac{\ekn{91}{2}}{\ecn{14}}\right]\right.\times\nonumber\\\nonumber\\
&\times&\left.\left(\frac{334,305}{14,336}\sus{14}-
\frac{260,015}{3,072}\sus{12}+\frac{156,009}{1,280}\sus{10}-\right.\right.\nonumber\\\nonumber\\
&-&\left.\left.\frac{11,305}{128}\sus{8}+\frac{1,615}{48}\sus{6}-\frac{51}{8}\sus{4}+\right.\right.\nonumber\\\nonumber\\
&+&\left.\left.\frac{1}{2}\sus{2}-\frac{1}{105}
\right)\right\},\\
\jo{16}^{b} & = & \jw2\left\{ \frac{36,465}{2,048}\rtr{14}\left[
\frac{105}{\ecn{15}}+31\frac{\ekn{105}{2}}{\ecn{16}}\right]\right.\times\nonumber\\\nonumber\\
&\times&\left.\left(\frac{17,678,835}{262,144}\sus{16}-
\frac{570,285}{2,048}\sus{14}+\frac{963,585}{2,048}\sus{12}-\right.\right.\nonumber\\\nonumber\\
&-&\left.\left.\frac{107,065}{256}\sus{10}
+\frac{107,065}{512}\sus{8}
-\frac{931}{16}\sus{6}+\right.\right.\nonumber\\\nonumber\\
&+&\left.\left.\frac{133}{16}\sus{4}-\frac{1}{2}\sus{2}
+\frac{1}{136}
\right)\right\},\\
\jo{18}^{b} & = & \jw2\left\{ \frac{692,835}{32,768}\rtr{16}\left[
\frac{136}{\ecn{17}}+35\frac{(1+68e^2)}{\ecn{18}}\right]\right.\times\nonumber\\\nonumber\\
&\times&\left.\left(\frac{39,803,225}{196,608}\sus{18}-
\frac{30,705,345}{32,768}\sus{16}+\frac{930,465}{512}\sus{14}-\right.\right.\nonumber\\\nonumber\\
&-&\left.\left.\frac{490,245}{256}\sus{12} +
\frac{152,145}{128}\sus{10} -
\frac{28,175}{64}\sus{8}+\right.\right.\nonumber\\\nonumber\\
&+&\left.\left.\frac{1,127}{12}\sus{6}-\frac{21}{2}\sus{4}
+\frac{1}{2}\sus{2}-\frac{1}{171}
\right)\right\},\\
\jo{20}^{b} & = & \jw2\left\{
\frac{1,616,615}{65,536}\rtr{18}\left[
\frac{171}{\ecn{19}}+39\frac{\ekn{171}{2}}{\ecn{20}}\right]\right.\times\nonumber\\\nonumber\\
&\times&\left.\left(\frac{328,206,021}{524,288}\sus{20}-
\frac{210,388,475}{65,536}\sus{18}+\frac{460,580,175}{65,536}\sus{16}-\right.\right.\nonumber\\\nonumber\\
&-&\left.\left.\frac{30,705,345}{3,584}\sus{14}+
\frac{6,513,255}{1,024}\sus{12}-\frac{378,189}{128}\sus{10}+\right.\right.\nonumber\\\nonumber\\
&+&\left.\left.\frac{108,675}{128}\sus{8}-\frac{575}{4}\sus{6}+\frac{207}{16}\sus{4}-\right.\right.\nonumber\\\nonumber\\
&-&\left.\left.\frac{1}{2}\sus2+ \frac{1}{210} \right)\right\}.
\eqfa The results previously obtained can be used in working out
explicitly the contributions of the mismodelled classical nodal
and apsidal precessions up to degree $l=20$ of the existing
spherical passive laser-ranged geodetic satellites and of
\protect{LARES}. They are of the form $\delta\dot
\Omega_{(2n)}=\dot \Omega _{.2n}\times \delta J_{2n}$,
$n=1,2,...10$ and $\delta\dot \omega_{(2n)}=\dot \omega
_{.2n}\times \delta J_{2n}$, $n=1,2,...10$. The values employed
for $\delta J_{2n}=\sqrt{4n+1}\times\delta{\overline{C}}_{2n\ 0}$,
$n=1,2,...10$ would be those quoted in, e.g., EGM96 model.
\section{Some useful parameters used in the text}
The data employed for the terrestrial space environment and the
LAGEOS satellites are in the following table. In it $\ve$ is the
angle between the ecliptic and the equatorial plane, $R_{\oplus}$
is the Earth mean equatorial radius, $G$ is the Newtonian
gravitational constant, $GM_{\oplus}$ and $\delta(GM_{\oplus})$
are the Newtonian gravitational constant times the Earth's mass
and its error according to IERS standard \ct{ref:iers}, $J_2,\
J_4, \delta J_2$ and $\delta J_4$ are the first two even zonal
geopotential's coefficients and their errors according to EGM96
\ct{ref:lem}, $J_{\oplus}$ is the Earth's angular momentum,
$\og_{\oplus}$ is the mean Earth angular velocity.

The conversion factor from rad/s to mas/y is 1 rad/s=$6.509\times
10^{15}$ mas/y.
\begin{table}[ht!]
\caption{Earth's and LAGEOS parameters used in the text.
}\label{parametri}
\begin{center}
\begin{tabular}{lll}
\noalign{\hrule height 1.5pt}
Parameter & Numerical value & Units \\
\hline
$\ve$ &  23.44 & \textrm{deg} \\
$R_{\oplus}  $ & $ 6,378\times 10^{5}  $ & \textrm{cm} \\
$G$ & $ 6.67259\times 10^{-8}  $ & \textrm{cm$^{3}$\ g$^{-1}$\ s$^{-2}$} \\
$GM_{\oplus}  $ & $ 3.986\times 10^{20}  $ & \textrm{cm$^{3}$\ s$^{-2}$} \\
$\delta(GM_{\oplus})$ & $ 8\times 10^{11}  $ & \textrm{cm$^{3}$\ s$^{-2}$} \\
$J_2$ & $1.0826\times 10^{-3}$ & -\\
$J_4$ & $-1.6194\times 10^{-6}$ & -\\
$\delta J_2$ & $7.9626\times 10^{-11}$ & -\\
$\delta J_4$ & $3.126\times 10^{-10}$ & -\\
$J_{\oplus}  $ & $ 5.9\times 10^{40}  $ & \textrm{g\ cm$^{2}$\ s$^{-1}$} \\
${J}_{\oplus}/(cM_{\oplus})$ & $3.3\times 10^2$ & cm\\
$\og_{\oplus}  $ & $ 7.29\times 10^{-5}  $ & \textrm{rad\
s$^{-1}$} \\
${G}/{c^{2}}$ & $7.42\times 10^{-29}$& cm g$^{-1}$ \\
$({GJ_{\oplus}})/{c^{2}}$ & $4.37\times 10^{12}$& cm$^{3}$ s$^{-1}$\\
$({GM_{\oplus}})/{c^{2}}$ & $4.43\times 10^{-1}$ & cm\\
\hline
 $a_{{\rm LAGEOS}} $ & $1.2270\times 10^{9}  $ &
\textrm{cm}\\
 $a_{{\rm LAGEOS\ II}}  $ & $1.2163\times 10^{9}  $ &
\textrm{cm}\\
$a_{{\rm LARES}}  $ & $1.2270\times 10^{9}  $ & \textrm{cm} \\
$e_{{\rm LAGEOS}}  $ & 0.0045 & -\\
$e_{{\rm LAGEOS\ II}}  $ & 0.014 & -\\
$e_{{\rm LARES}}  $ & 0.04 & -\\
$i_{{\rm LAGEOS}}  $ & 110 & \textrm{deg} \\
$i_{{\rm LAGEOS\ II}}  $ & 52.65 & \textrm{deg} \\
$i_{{\rm LARES}}  $ & 70 & \textrm{deg} \\
$n_{\rm LAGEOS}$ & $4.643\times 10^{-4}$ & s$^{-1}$\\
$n_{\rm LAGEOS\ II}$ & $4.710\times 10^{-4}$ & s$^{-1}$\\
$n_{\rm LARES}$ & $4.643\times 10^{-4}$ & s$^{-1}$\\
\noalign{\hrule height 1.5pt}
\end{tabular}
\end{center}
\end{table}
\section{WEB resources}
\begin{itemize}
\item{\bf http://wugrav.wustl.edu/People/CLIFF/tegp.html}, the
experimental gravity web page.
\item{\bf
http://www.livingreviews.org/Articles/Volume4/2001-4will/index.html},
recent review of the current status of the empirical basis of
General Relativity.
\item{\bf
http://www.laeff.esa.es/eng/laeff/activity/lageos3.html}, on
LAGEOS III project.
\item{\bf
http://w3.ing.uniroma1.it/tildespacedpt/lares.html}, on LARES
project.
\item{\bf
http://earth.agu.org/revgeophys/marsha01/node1.html}, on the force
models acting on LAGEOS satellites.
\item{\bf
http://einstein.stanford.edu/index.html}, on the GP-B mission.
\item{\bf
http://www.nas.edu/ssb/gpbexe.html}, on the GP-B mission.
\item{\bf
http://op.gfz-potsdam.de/champ/}, on the CHAMP mission.
\item{\bf
http://op.gfz-potsdam.de/grace/}, on the GRACE missions
\item{\bf http://ilrs.gsfc.nasa.gov/}, the International
Laser Ranging Service web site.
\item{\bf
http://www.ee.surrey.ac.uk/SSC/SSHP/}, small satellites web site.
\item{\bf http://cddisa.gsfc.nasa.gov/926/egm96/egm96.html}, the
EGM96 Earth gravity model.
\item{\bf
http://co-ops.nos.noaa.gov/restles1.html}, on the phenomenon of
Earth tides.
\item{\bf
http://www.astro.oma.be/D1/EARTH$\_$TIDES/wgtide.html}, home page
of the Working Group of Theoretical Tidal Model of the Earth Tide
Commission.
\item{\bf http://www.ill.fr/tas/matlab/doc/mfit.html},
useful collection of MATLAB least squares programs and other
routines.
\end{itemize}


\end{document}